\documentclass[a4paper,10pt,onecolumn,showpacs,notitlepage,floatfix,nofootinbib,superscriptaddress,prd]
{revtex4-2}
\usepackage{graphicx}
\usepackage{amsfonts}
\usepackage{amsmath}
\usepackage{hyperref}
\usepackage{float}
\usepackage{amssymb,amsbsy}
\usepackage{epstopdf}
\usepackage{color}

\newcommand{\dd}{\mathrm{d}}

\begin{document}

\title{Multicomponent relativistic dissipative fluid dynamics from the 
Boltzmann equation}
\author{Jan A.\ Fotakis}
\email{fotakis@itp.uni-frankfurt.de}
\affiliation{Institut f\"ur Theoretische Physik, 
Johann Wolfgang Goethe--Universit\"at,
Max-von-Laue-Str.\ 1, D--60438 Frankfurt am Main, Germany}

\author{Etele Moln\'ar}
\affiliation{Institut f\"ur Theoretische Physik, 
Johann Wolfgang Goethe--Universit\"at,
Max-von-Laue-Str.\ 1, D--60438 Frankfurt am Main, Germany} 
\affiliation{Incubator of Scientific Excellence--Centre for Simulations of Superdense Fluids, 
University of Wroc\l{}aw, pl. M. Borna 9, PL-50204 Wroc\l{}aw, Poland}

\author{Harri Niemi}
\affiliation{Institut f\"ur Theoretische Physik, 
Johann Wolfgang Goethe--Universit\"at,
Max-von-Laue-Str.\ 1, D--60438 Frankfurt am Main, Germany} 
\affiliation{Department of Physics, University of Jyv\"askyl\"a, P.O.\ Box
35, FI-40014 University of Jyv\"askyl\"a, Finland} 
\affiliation{Helsinki
Institute of Physics, P.O.\ Box 64, FI-00014 University of Helsinki, Finland}

\author{Carsten Greiner}
\affiliation{Institut f\"ur Theoretische Physik, 
Johann Wolfgang Goethe--Universit\"at,
Max-von-Laue-Str.\ 1, D--60438 Frankfurt am Main, Germany}

\author{Dirk H.\ Rischke}
\affiliation{Institut f\"ur Theoretische Physik, 
Johann Wolfgang Goethe--Universit\"at,
Max-von-Laue-Str.\ 1, D--60438 Frankfurt am Main, Germany} 
\affiliation{Helmholtz Research Academy Hesse for FAIR, Campus Riedberg, 
Max-von-Laue-Str.\ 12, D-60438 Frankfurt am Main, Germany}

\begin{abstract}
We derive multicomponent relativistic second-order dissipative
fluid dynamics from the Boltzmann equations for 
a reactive mixture of $N_{\text{spec}}$ particle species with $N_q$ 
intrinsic quantum numbers (e.g.\ electric charge, baryon number, and strangeness)
using the method of moments. We obtain the continuity equations for 
multiple conserved charges as well as the conservation equations for the 
total energy and momentum in the single-fluid approximation. These 
$4+N_q$ conservation laws are closed by deriving the second-order equations of motion 
for the dissipative quantities in the 
$(10+4N_q)$-moment approximation. 
The resulting fluid-dynamical equations are formally similar to those of a
single-component system, but feature different thermodynamic relations and transport 
coefficients. We derive general relations for all
transport coefficients and compute them explicitly  
in the ultrarelativistic limit.
\end{abstract}

\date{\today }
\maketitle

\section{Introduction}

Determining properties of strong-interaction matter from experimental data measured in 
high-energy 
heavy-ion experiments at BNL-RHIC and CERN-LHC is largely based on relativistic 
fluid-dynamical modeling, see, e.g.\ Refs.\ \cite{Bernhard:2019bmu, Auvinen:2020mpc, Nijs:2020roc, JETSCAPE:2020mzn, Parkkila:2021tqq}.
Consequently, relativistic fluid dynamics has become an indispensable tool in the 
description of the dynamical evolution of relativistic nuclear collisions. 

The state of the art of fluid-dynamical modeling of relativistic nuclear matter is based on 
the relativistic second-order dissipative fluid-dynamical theory of Israel and Stewart 
\cite{Israel:1979wp}. 
This theory and method is based on the pioneering works of Grad \cite{Grad} and 
M\"uller \cite{Muller_67} and was originally formulated for a simple fluid, 
i.e.\ a fluid with a single conserved charge.
On the other hand, one of the basic features of the fluid created in high-energy 
nuclear collisions is its multicomponent nature. For example, hadronic matter produced in 
nuclear collisions consists of a multitude of different types of hadrons, where each hadron 
species carries multiple intrinsic quantum numbers like baryon number $B$, 
electric charge $Q$, and strangeness $S$.
Therefore, a multicomponent extension of relativistic fluid-dynamical theories that explicitly 
accounts for multiple conserved charges is required for a proper description of heavy-ion 
collisions.

Previous attempts to derive second-order fluid-dynamical equations of motion for 
relativistic multicomponent mixtures include
the pioneering work by Prakash \textit{et al.} \cite{Prakash:1993bt}, which is an extension 
of Israel-Stewart theory to multicomponent mixtures,
and the works by Monnai and Hirano \cite{Monnai:2010qp, Monnai:2010th}, 
which generalize the equations of Ref.\ \cite{Prakash:1993bt} by including additional 
second-order 
terms as well as providing Onsager's reciprocity relations \cite{Onsager:1931jfa,Onsager:1931kxm} for the transport coefficients.
More recent developments by Kikuchi \textit{et al.} \cite{Kikuchi:2015swa}
apply the renormalization-group method to rederive the second-order 
equations of motion including additional second-order terms in the dissipative quantities
resulting from the non-linear part of the collision integral.

Furthermore it has been shown in 
Refs.~\cite{Greif:2017byw, Fotakis:2019nbq, Rose:2020sjv, Fotakis:2021diq} that many 
features of multicomponent systems depend on the detailed coupling 
between the diffusion currents associated with different conserved 
charges. In addition, the mapping between the state of the fluid and the corresponding 
momentum distribution of particles plays an important 
role~\cite{Denicol:2012yr,Wolff:2016vcm}. 

In this work, we present a derivation of multicomponent relativistic 
second-order dissipative fluid dynamics for a reactive mixture of $N_{\text{spec}}$ species 
with $N_q$ conserved quantum numbers by generalizing the method of moments 
established for single-component systems by Denicol \textit{et al.} \cite{Denicol:2012cn}.
By summing the dynamical equations of motion describing the individual particle species 
we obtain a reduced set of equations corresponding to the so-called ``single-fluid''
description of a multicomponent fluid. We derive the continuity 
equations for each conserved quantum charge as well as the conservation laws for total 
energy and momentum in this single-fluid approximation. These $4+N_q$ equations of 
motion are closed by providing second-order equations of motion 
in the $(10+4N_q)$-moment approximation for the dissipative quantities.
The latter equations are formally similar to the relaxation equations of a 
single-component system but feature different transport coefficients, which contain 
the microscopic interactions of all components.
Our approach reproduces the results of Ref.\ \cite{Denicol:2012cn}
for a single-component fluid
in the $14$-moment approximation, i.e.\ for $N_{\text{spec}} = N_q  = 1$.

This paper is organized as follows. In Sec.\ \ref{Sect:kineticfluid}, we introduce 
the Boltzmann equation, thermodynamic quantities in local equilibrium, as well as
fluid-dynamical quantities, both in and out of equilibrium.
The definition of the local rest frame, the matching conditions, and the conservation 
equations are also given.
In Sec.\ \ref{Sect:Relax} we derive the equations of motion for the irreducible moments 
from the Boltzmann equation, linearize the collision term, and discuss
the Navier-Stokes limit and the order-of-magnitude approximation.
Finally, the second-order dissipative fluid-dynamical equations of motion 
in the $(10+4N_q)$-moment approximation 
are derived and discussed. We conclude this work with a summary in the final section.
Details of the calculations are delegated to several appendices.
They also contain explicit expressions for all second-order transport coefficients, as
well as an explicit calculation of the transport coefficients in the ultrarelativistic limit.

Throughout this paper we adopt natural units, $\hbar=c=k_{B}=1$, and work in flat 
Minkowski space-time with metric tensor $g_{\mu \nu }=\text{diag}(1,-1,-1,-1)$. 
The time-like fluid four-velocity is denoted by $u^{\mu }=\gamma ( 1,\mathbf{v})^{T}$, 
with normalization $u^{\mu }u_{\mu}\equiv 1$, 
where $\mathbf{v}$ is the three-velocity and $\gamma =(1-\mathbf{v}^{2})^{-1/2}$. 
In the local rest (LR) frame of the fluid, $u_{\text{LR}}^{\mu }=( 1,\mathbf{0})^{T}$. 
The rank-two projection operator onto the three-space orthogonal to 
$u^{\mu }$ is defined as $\Delta^{\mu \nu }\equiv g^{\mu \nu}-u^{\mu}u^{\nu}$. 
We define the projection of any four-vector $A^{\mu }$ onto the 
three-dimensional subspace orthogonal to $u^{\mu}$ as 
$A^{\left\langle \mu \right\rangle}\equiv \Delta_{\nu }^{\mu}A^{\nu}$.
The generalization to projection tensors of rank $2\ell $, denoted by 
$\Delta_{\nu_{1}\cdots \nu_{\ell }}^{\mu_{1}\cdots\mu_{\ell }}$, is 
constructed using the elementary projection operator $\Delta^{\mu}_{\nu}$.
The irreducible symmetric, traceless, and orthogonal
projection of a rank-$\ell$ tensor $A^{\mu_1 \cdots \mu_\ell}$ is
denoted as $A^{\left\langle \mu_{1}\cdots \mu_{\ell}\right\rangle } 
\equiv \Delta_{\nu_{1}\cdots \nu_{\ell }}^{\mu_{1}\cdots 
\mu_{\ell }}A^{\nu_{1}\cdots \nu_{\ell }}$. 
For example, the rank-four symmetric, traceless, and orthogonal projection operator is 
defined as $\Delta_{\alpha \beta}^{\mu \nu}\equiv \frac{1}{2}( \Delta_{\alpha}^{\mu }
\Delta_{\beta}^{\nu } + \Delta_{\beta }^{\mu }\Delta_{\alpha}^{\nu }) 
-\frac{1}{3}\Delta^{\mu \nu }\Delta_{\alpha \beta }$,
hence $A^{\left\langle \mu \nu \right\rangle} \equiv \Delta_{\alpha \beta}^{\mu \nu}
A^{\alpha \beta}$.

The four-momentum of a particle of species $i$ is denoted by 
$k_{i}^{\mu }=(k_i^{0},\mathbf{k}_{i} )^{T}$, which 
is normalized to the corresponding species rest mass squared, 
$k_{i}^{\mu}k_{i,\mu }= m_{i}^{2}$. The energy of a particle of species $i$ 
is defined as $E_{i,\mathbf{k}}=k_{i}^{\mu}u_{\mu}$, 
and coincides with the on-shell energy 
$k_{i}^{0}=\sqrt{\mathbf{k}_{i}^{2}+m_{i}^{2}}$ in the LR frame of the fluid.  
The orthogonal projection of the four-momentum is 
$k_{i}^{\left\langle \mu \right\rangle }\equiv \Delta_{\nu}^{\mu} k_{i}^{\nu}$,
and in the LR frame it reduces to the three-momentum $\mathbf{k}_{i}$.

The comoving derivative $D\equiv u^{\mu}\partial_{\mu }$ of any four-vector 
$A^{\mu}=(A^0, \mathbf{A})^{T}$ is denoted by $\dot{A^\mu}
\equiv u^{\nu }\partial_{\nu}{A^\mu}= D A^\mu$, while the space-time four-gradient is 
$\nabla_{\nu}A^{\mu}\equiv \Delta_{\nu}^{\alpha}\partial_{\alpha}A^{\mu}$. 
Note that in the LR frame these relativistic space-time derivatives reduce to the usual time 
and three-space derivatives, $\partial_{t}A^{\mu}$ and 
$\mathbf{\nabla} \otimes \mathbf{A}$. 
Thus, the four-derivative is decomposed as 
$\partial_{\mu}\equiv u_{\mu }D + \nabla_{\mu}$, 
hence the relativistic Cauchy-Stokes decomposition reads, 
$\partial_{\mu} u_{\nu}\equiv u_{\mu}\dot{u}_{\nu} 
+\nabla_{\mu}u_{\nu}=u_{\mu}\dot{u}_{\nu} + \frac{1}{3}\theta \Delta_{\mu \nu }
+\sigma_{\mu\nu}+\omega_{\mu \nu}$. 
Here we have defined the expansion scalar, 
$\theta\equiv \nabla_{\mu}u^{\mu }$, the shear tensor 
$\sigma^{\mu \nu }\equiv\nabla^{\left\langle \mu \right. } 
u^{\left. \nu \right\rangle }=\frac{1}{2}(\nabla^{\mu}u^{\nu }+\nabla^{\nu}u^{\mu })
-\frac{1}{3}\theta \Delta^{\mu \nu}$, 
and the vorticity 
$\omega^{\mu \nu}\equiv \frac{1}{2}(\nabla^{\mu}u^{\nu}-\nabla^{\nu}u^{\mu})$, 
such that $\sigma^{\mu \nu}u_{\nu}=\omega^{\mu \nu}u_{\nu} = 0$.

Moreover, we label the conserved charge types in the system with the letter $q$, which will 
be treated as an index running over $B$ (baryon number), $Q$ (electric charge), and 
$S$ (strangeness) for the case a strong-interaction system. 
For notational convenience, we employ the following 
notation for the sums over charge types,
\begin{align}
	\sum_{ q }^{ \lbrace B,Q,S \rbrace } \equiv \sum_{ q\, =\, B,Q,S } \notag .
\end{align}

\section{Reactive mixtures in kinetic theory and fluid dynamics}
\label{Sect:kineticfluid}

In this section we first introduce the Boltzmann equation for a reactive mixture
with special emphasis on the collision term. Before we discuss fluid-dynamical quantities 
in local equilibrium and out of equilibrium, we study thermodynamic quantities in local
equilibrium. This is followed by a discussion of
the matching conditions and the choices for the local rest frame. Finally, we list the 
conservation equations of second-order dissipative fluid dynamics for a multicomponent 
fluid.

\subsection{The Boltzmann equation for a reactive mixture}

A mixture of $N_{\text{spec}}$ different (elementary) particle species (i.e.\ 
different chemical components) is characterized by the single-particle
distribution functions for each particle species $i$, 
$f\left( x,k_{i}\right) \equiv f_{i,\mathbf{k}}$, where we 
label the particle species by a lower index $i$. 
The space-time evolution of the distribution function of 
species $i$ is determined by the relativistic Boltzmann equation
\cite{deGroot_book,Cercignani_book}, 
\begin{equation}
k_{i}^{\mu }\partial_{\mu }f_{i,\mathbf{k}} \equiv C_{i}\left( x,k_i\right)  
=\sum_{j\,=\,1}^{N_{\text{spec}}}C_{ij}[f] \;,  
\label{BTE_i}
\end{equation}
where we neglect any external forces and assume binary collisions only for the sake of 
simplicity. For binary inelastic, i.e.\ reactive collisions, the initial and final 
particles species may be different, $i+j \rightarrow a+b$, such that the collision term reads 
\begin{equation}
C_{ij}[f] = \frac{1}{2}\sum_{a,b\,=\,1}^{N_{\text{spec}}}
\int \dd K_{j}^{\prime } \dd P_{a} \dd P_{b}^{\prime}
\left( W_{ab\rightarrow ij}^{pp^{\prime}\rightarrow kk^{\prime}} 
f_{a,\mathbf{p}}f_{b,\mathbf{p}^{\prime}} 
\tilde{f}_{i,\mathbf{k}}\tilde{f}_{j,\mathbf{k}^{\prime}} 
- W_{ij\rightarrow ab}^{kk^{\prime}\rightarrow pp^{\prime}}
f_{i,\mathbf{k}}f_{j,\mathbf{k}^{\prime}}
\tilde{f}_{a,\mathbf{p}}\tilde{f}_{b,\mathbf{p}^{\prime}}\right) \;,  
\label{COLL_INT}
\end{equation}
where $\tilde{f}_{i,\mathbf{k}}= 1-a_if_{i,\mathbf{k}}/g_i$,
with $a_i=\pm1$ for fermions/bosons, and
$a_i\rightarrow 0$ for classical particles, respectively. Here, $g_{i}$ is the 
spin degeneracy of particle species $i$, while the Lorentz-invariant 
integration measure is $\dd K_i = \dd^{3}\mathbf{k}_{i}/[ ( 2\pi)^{3}k_{i}^{0}]$.
Here, the factor $1/2$ corrects for the double counting when integrating
over the momenta of particles in the initial and final state \cite{deGroot_book}.

The transition probabilities respect certain symmetry properties under exchange
of particles in the initial and final state,
$W_{ij\rightarrow ab}^{kk^{\prime}\rightarrow pp^{\prime}} 
= W_{ji\rightarrow ba}^{k^{\prime} k\rightarrow p^{\prime}p}$, 
as well as the bilateral normalization property of the
microscopic processes \cite{deGroot_book},
\begin{align}
\sum_{a,b\,=\,1}^{N_{\text{spec}}}
\int \dd P_{a} \dd P_{b}^{\prime } \, 
W_{ab\rightarrow ij}^{pp^{\prime}\rightarrow kk^{\prime}} 
= \sum_{a,b\,=\,1}^{N_{\text{spec}}}
\int \dd P_{a} \dd P_{b}^{\prime } \, 
W_{ij\rightarrow ab}^{kk^{\prime}\rightarrow pp^{\prime}} \; .
\end{align}
In the absence of a reaction threshold, this relation is invariant under time reversal, 
$W_{ij\rightarrow ab}^{kk^{\prime}\rightarrow pp^{\prime}} 
= W_{ab\rightarrow ij}^{pp^{\prime}\rightarrow kk^{\prime}}$, 
i.e.\ microscopic reversibility or detailed balance. 
In the case of a binary process the transition probability 
is \cite{Chakraborty:2010fr, Albright:2015fpa}
\begin{align}
W_{ij \rightarrow  ab}^{kk^{\prime}\rightarrow pp^{\prime}} = \frac{(2\pi)^4}{16} 
\vert \mathcal{M}_{ij\rightarrow ab}(\sqrt{s},\Omega)\vert^2 \, \delta^{(4)}
\left( k^{\mu}_i + k^{\prime \mu}_j - p^{\mu}_a - p^{\prime \mu}_b \right),
\end{align}
where $\vert \mathcal{M}_{ij\rightarrow ab}(\sqrt{s},\Omega)\vert^2$ is the 
Lorentz-invariant 
transition probability averaged over incoming and summed over outgoing spin states. 
It only depends on the total center-of-momentum (CM) energy squared,
$s \equiv (k^\mu_i + k^{\prime \mu}_j)^2 = (p^{\mu}_a + p^{\prime \mu}_b)^2$ and 
the solid angle $\Omega$ under which outgoing particles are scattered with respect to
the direction of the incoming particles, 
while the $\delta^{(4)}$-function ensures the conservation of 
energy and momentum in each binary collision. The differential cross section in the 
CM frame, where $\mathbf{k}_i + \mathbf{k}^\prime_j 
= \mathbf{p}_a + \mathbf{p}^\prime_b \equiv 0$ and 
$k^0_i + k^{\prime 0}_j = p^{0}_a + p^{\prime 0}_b \equiv \sqrt{s} $, is defined via the 
invariant transition probability as
\begin{align}
\dd\sigma_{ij\rightarrow ab} (\sqrt{s},\Omega)= \frac{1}{64\pi^2 s} \frac{p_{ab}}{p_{ij}}  \vert 
\mathcal{M}_{ij\rightarrow ab}(\sqrt{s},\Omega)\vert^2 \, \dd\Omega\; , 
\label{eq:diff_corsssec_binary}
\end{align}
where the incoming $p_{ij}$ and outgoing $p_{ab}$ momenta in the CM frame are
\begin{align}
p_{ij} \equiv \frac{1}{2\sqrt{s}} \sqrt{\Big( s - (m_i + m_j)^2 \Big)
\Big( s - (m_i - m_j)^2 \Big) } \;.
\end{align}
We note that in the elastic limit $p_{ij} = p_{ab}$. We define the total (integrated over 
angles) cross section as
\begin{align}
\sigma_{\text{tot}, \, ij \rightarrow ab} \equiv 2\pi \gamma_{ab} \int\limits_{-1}^1 
\dd\cos\vartheta \, \frac{\dd\sigma_{ij \rightarrow ab}(\sqrt{s},\Omega)}{\dd\Omega} \;. 
\label{eq:total_crosssec}
\end{align}
Here, the symmetry factor $\gamma_{a b} \equiv 1 - \delta_{a b}/2$ accounts
for the double counting when integrating over the momenta of indistinguishable 
particles in the final state. In the case of isotropic scattering the differential cross section is
\begin{align}
\frac{\dd\sigma_{ij \rightarrow ab}(\sqrt{s}) }{\dd\Omega} \overset{\text{isotropic}}{=} 
\frac{\sigma_{\text{tot}, ij \rightarrow ab}(\sqrt{s})}{4\pi \gamma_{ab}}  \;. 
\label{eq:isotropic_crosssec}
\end{align}
In the elastic limit the transition rates in Eq.\ (\ref{COLL_INT}) 
are defined as \cite{deGroot_book}
\begin{align}
W_{ij \rightarrow ab}^{kk^{\prime}\rightarrow pp^{\prime}} \equiv \gamma_{ij} 
\left( \delta_{ia} \delta_{jb} + \delta_{ib}\delta_{ja}\right) 
W_{ij}^{kk^{\prime}\rightarrow pp^{\prime}} \;,
\label{ELASTIC_INT}
\end{align}
where
\begin{align}
W_{ij}^{kk^{\prime}\rightarrow pp^{\prime}} \equiv \frac{(2\pi)^4}{16} 
\vert \mathcal{M}_{ij\rightarrow ij}(\sqrt{s},\Omega)\vert^2 \, \delta^{(4)}
\left( k^{\mu}_i + k^{\prime \mu}_j - p^{\mu}_i - p^{\prime \mu}_j \right)\;.
\end{align}
Note that in the above expression we have already introduced the symmetry factor, 
even though we have not yet integrated over the momentum. For later convenience, 
we can use Eqs.\ \eqref{eq:diff_corsssec_binary} and 
\eqref{eq:isotropic_crosssec} to rewrite 
Eq.\ \eqref{ELASTIC_INT} in the case of isotropic elastic scattering in terms of the total 
cross section defined in Eq.\ \eqref{eq:total_crosssec},
\begin{align}
W_{ij \rightarrow ab}^{kk^{\prime}\rightarrow pp^{\prime}} \equiv \left( \delta_{ia} \delta_{jb} 
+ \delta_{ib}\delta_{ja}\right) (2\pi)^6 s \frac{\sigma_{\text{tot}, ij}}{4\pi} \, \delta^{(4)}
	\left( k^{\mu}_i + k^{\prime \mu}_j - p^{\mu}_i - p^{\prime \mu}_j \right)\; .
	 \label{eq:trans_isotrop_elastic}
\end{align}

\subsection{Local equilibrium and associated thermodynamic quantities}

In general the single-particle distribution function $f_{i,\mathbf{k}}$ for every 
species $i$ can be decomposed into an equilibrium part,
$f_{i,\mathbf{k}}^{(0)}$, and an out-of-equilibrium part, $\delta f_{i,\mathbf{k}}$, as 
\begin{equation}
f_{i,\mathbf{k}}=f_{i,\mathbf{k}}^{(0)}+\delta f_{i,\mathbf{k}}\;,
\label{f_i}
\end{equation}
where the local-equilibrium distribution function of species $i$ is given
by the J\"{u}ttner distribution function \cite{Juttner}, 
\begin{equation}
f_{i,\mathbf{k}}^{\left( 0\right) } = g_{i} \left[ \exp \left( \frac{E_{i,\mathbf{k}}
-\mu_{i}}{T}\right) + a_i \right]^{-1}\;.  \label{f0_i}
\end{equation}
Due to detailed balance, the collision integral vanishes identically for 
the local-equilibrium distribution function \cite{deGroot_book}. 
Here, $T\equiv 1/\beta $ is the 
temperature and $\mu_{i}$ is the chemical potential of species $i$ 
as defined in the local rest frame.
The exact form of the non-equilibrium part of the distribution function 
$\delta f_{i,\mathbf{k}}$ will be clarified later.

In various cases of interest such as in high-energy particle physics or relativistic 
heavy-ion collisions there are inelastic collisions where the particle number 
corresponding to a given species is not conserved due to 
particle creation and annihilation processes (i.e.\ various chemical reactions). 
Such strong-interaction matter is therefore described by a 
few conserved intrinsic quantum numbers, such as electric charge, baryon number, 
and strangeness. 

This means that in local equilibrium the chemical potential $\mu_{i}$
of a given particle $i$ may be expressed in terms of $N_{q}$ chemical potentials 
of conserved quantum "charges", 
\begin{equation}
\mu_{i}\left( \left\{ \mu_{q}\right\} \right) \equiv \sum_{q}^{\left\{B,Q,S\right\} } 
q_{i} \mu_{q}=B_{i}\mu_{B} + Q_{i}\mu_{Q}+S_{i}\mu_{S}\;,
\label{mu_i_mu_q}
\end{equation}
where $\left\{ \mu_{q}\right\}  \equiv \left\{ \mu_B, \mu_Q, \mu_S \right\}$, with
$\mu_{B}$, $\mu_{Q}$, and $\mu_{S}$, being the baryon, electric
and strangeness chemical potentials, respectively, while $B_{i}$, $Q_{i}$, and $S_{i}$ 
are the baryon number, electric charge, and strangeness of the respective particle 
species $i$. 

Now we introduce the ratio of chemical potential over temperature corresponding to 
particle species $i$ as $\alpha_{i}\equiv \mu_{i}/T$, as well as the ratio of chemical 
potential over temperature for the conserved quantum charges, 
$\alpha_{q}\equiv \mu_{q}/T$. Applying the chain rule, we
obtain from Eq.\ \eqref{mu_i_mu_q}
\begin{equation}
\dd\alpha_{i}\left( \left\{ \alpha_{q}\right\} \right) \equiv
\sum_{q}^{\left\{ B,Q,S\right\} }\frac{\partial \alpha_{i}}{\partial \alpha_{q}} \dd\alpha_{q}
= \sum_{q}^{\left\{ B,Q,S\right\} } q_{i} \dd\alpha_{q}\;,
\label{dalpha_i_alpha_q}
\end{equation}
where the intrinsic quantum number of particle species $i$ can also be obtained as
$q_{i}\equiv\partial \alpha_{i}( \lbrace \alpha_{q^{\prime }} \rbrace ) /
\partial \alpha_{q}$.
Note that the infinitesimal change in any variable $A$, here denoted by $dA$, 
can be interchangeably used for the comoving derivative, 
$D A$, as well as the space-time four-gradient, $\nabla^{\mu} A$.

In local thermodynamic equilibrium, we define the following rank-$n$ tensor
moments of given power $r\geq 0$ in
energy $E_{i,\mathbf{k}}^{r}$ for any given particle species $i$ as 
\begin{equation}
\mathcal{I}_{i,r}^{\mu_{1} \cdots \mu_{n}} 
\equiv \int \dd K_i E_{i,\mathbf{k}}^{r} k_{i}^{\mu_{1}}
\cdots k_{i}^{\mu_{n}} f_{i,\mathbf{k}}^{(0)}
= \left\langle E_{\mathbf{k}}^{r}k^{\mu_{1}}\cdots k^{\mu_{n}}\right\rangle_{i,0}\;,  
\label{I_i_r_mu1_mun}
\end{equation}
where the angular brackets are the abbreviation of the integrals,
\begin{equation}
\left\langle \cdots \right\rangle_{i,0}\equiv \int \dd K_i \left(\cdots \right)_i 
f_{i,\mathbf{k}}^{\left( 0\right) }\;.  \label{brakets_0}
\end{equation}
Following Ref.\ \cite{Israel:1979wp} we expand the equilibrium moments 
(\ref{I_i_r_mu1_mun}) in terms of the flow velocity and the associated orthogonal 
projection operator, which leads to the following expression, 
\begin{equation}
\mathcal{I}_{i,r}^{\mu_{1}\cdots \mu_{n}}=\sum_{m\,=\,0}^{\left[ n/2\right]}
\left( -1\right)^{m}\frac{n!}{2^{m}m!\left( n-2m\right) !} 
\Delta^{\left(\mu_{1}\mu_{2}\right. }\cdots \Delta^{\mu_{2m-1}\mu_{2m}}
u^{\mu_{2m+1}}\cdots u^{\left. \mu_{n}\right) } I_{i,r+n,m} \;, 
\label{I_i_r_mu1_mun_exp}
\end{equation}%
where $n$ and $m$ are natural numbers, while $\left[ n/2\right] \leq n/2$
denotes the largest integer divisible by two. The coefficient 
$n!/ [2^m m! (n - 2m)!]$ counts the number of distinct terms in the symmetrized tensor
products $\Delta^{\left( \mu_{1}\mu_{2}\right. }\cdots 
\Delta^{\mu_{2m-1}\mu_{2m}}$ $\times u^{\mu_{2m+1}}\cdots u^{\left. \mu_{n}\right) }$. 
The coefficients $I_{i,r+n,m}$ are thermodynamic integrals which only
depend on $\alpha_{i}$ and $\beta$, 
\begin{equation}
I_{i,nm}\left( \alpha_{i},\beta \right) =\frac{\left( -1\right)^{m}}
{\left( 2m+1\right) !!}\left\langle E_{\mathbf{k}}^{n-2m}\left( \Delta_{\mu
\nu }k^\mu k^\nu\right)^{m}\right\rangle_{i,0}\;,
\end{equation}
where $\left(2m+1\right) !!\equiv \left( 2m+1\right)!/(2^m m!)$ is the double
factorial of an odd integer.

The total derivative of the thermodynamic integrals with respect to the
variables $\alpha_{i}$ and $\beta$ reads
\begin{eqnarray} \nonumber
\dd I_{i,nm} &=& \left( \frac{\partial I_{i,nm}}
{\partial \alpha_{i}}\right)_{\beta } \dd\alpha_{i} 
+ \left( \frac{\partial I_{i,nm}}{\partial \beta }\right)_{\alpha_{i}} \dd\beta  \\
&\equiv& J_{i,nm} \dd\alpha_{i} - J_{i,n+1,m} \dd\beta ,
\label{dI_i_nm_0}
\end{eqnarray}
where we have defined the auxiliary thermodynamic integrals 
\begin{equation}
J_{i,nm}\left( \alpha_{i},\beta \right) \equiv \left( \frac{\partial
I_{i,nm}}{\partial \alpha_{i}}\right)_{\beta} = \frac{\left( -1\right)^{m}}
{\left( 2m+1\right) !!}\int \dd K_i E_{i,\mathbf{k}}^{n-2m}\left( \Delta_{\mu
\nu }k_{i}^\mu k_{i}^\nu\right)^{m}f_{i,\mathbf{k}}^{\left( 0\right) }
\tilde{f}_{i,\mathbf{k}}^{\left( 0\right) }\;.  \label{J_i_nm}
\end{equation}
An integration by parts with $df_{i,\mathbf{k}}^{\left( 0\right) }
/dE_{i,\mathbf{k}}=-\beta f_{i,\mathbf{k}}^{\left( 0\right) }
\tilde{f}_{i,\mathbf{k}}^{\left( 0\right) }$ leads to the following relation between the 
thermodynamic integrals, 
\begin{equation}
\beta J_{i,nm}=I_{i,n-1,m-1}+\left( n-2m\right) I_{i,n-1,m}\;.
\end{equation}
Furthermore, with Eq.\ \eqref{dalpha_i_alpha_q} we obtain from Eq.\ \eqref{dI_i_nm_0} 
an expression for the total derivative of the thermodynamic integrals of species $i$ in 
terms of the conserved quantum charges,
\begin{eqnarray} \nonumber
\dd I_{i,nm} &=& \sum_{q}^{\left\{ B,Q,S\right\} } 
\left( \frac{\partial I_{i,nm}}{\partial \alpha_{q}}\right)_{\beta } \dd\alpha_{q}
+\left( \frac{\partial I_{i,nm}}{\partial \beta }\right)_{\lbrace \alpha_{q} \rbrace} \dd\beta  \\
&\equiv& \sum_{q}^{\left\{ B,Q,S\right\} } q_{i} J_{i,nm} \dd\alpha_{q} - J_{i,n+1,m} \dd\beta \;, 
\label{dI_i_nm_q}
\end{eqnarray}
from which follows
\begin{equation}
\left( \frac{\partial I_{i,nm}}{\partial \alpha_{q}}\right)_{\beta }=q_{i}J_{i,nm}\;.
\end{equation}

For later use we define the thermodynamic integrals summed over all particle
species, which we denote by $I_{nm}$ and $J_{nm}$. Similarly, the 
thermodynamic integrals of conserved quantum charges, $I_{nm}^{q}$, $J_{nm}^{q}$, 
as well as the auxiliary thermodynamic quantities, $I_{nm}^{qq^{\prime }}$, 
$J_{nm}^{qq^{\prime }}$, are defined as follows,
\begin{align}
I_{nm} &\equiv\sum_{i\,=\,1}^{N_{\text{spec}}}I_{i,nm}\;,\quad \quad \;\;
J_{nm}\equiv \sum_{i\,=\,1}^{N_{\text{spec}}}J_{i,nm}
=\sum_{i\,=\,1}^{N_{\text{spec}}} \left(\frac{\partial I_{i,nm}}{\partial \alpha_{i}}
\right)_\beta\;,  \label{I_J_nm} \\
I_{nm}^{q} &\equiv\sum_{i\,=\,1}^{N_{\text{spec}}}q_{i}I_{i,nm}\;,
\quad \;\; \;J_{nm}^{q}\equiv \sum_{i\,=\,1}^{N_{\text{spec}}}q_{i}J_{i,nm} 
= \left(\frac{\partial I_{nm}}{\partial \alpha_{q}} \right)_\beta\;,  \label{I_J_nm_q} \\
I_{nm}^{qq^{\prime }} &\equiv\sum_{i\,=\,1}^{N_{\text{spec}}}q_{i}q_{i}^{\prime }I_{i,nm}\;,
\quad J_{nm}^{qq^{\prime }}\equiv 
\sum_{i\,=\,1}^{N_{\text{spec}}}q_{i}q_{i}^{\prime}J_{i,nm}
=\left(\frac{\partial I_{nm}^{q}}{\partial \alpha_{q^{\prime }}}\right)_\beta\;.
\label{I_J_nm_qq}
\end{align}
Now using Eq.\ \eqref{dI_i_nm_q} together with the above definitions, the
total differential of the thermodynamic integrals summed over all particle species 
as well as that of the thermodynamic integral of a specific conserved charge read
\begin{eqnarray}
\dd I_{nm}\left( \left\{ \alpha_{q}\right\}, \beta \right)
&=& \sum_{q}^{\left\{ B,Q,S\right\} } J_{nm}^{q} \dd\alpha_{q} - J_{n+1,m} \dd\beta \;,
\label{dI_nm} \\
\dd I_{nm}^{q}\left( \left\{ \alpha_{q}\right\}, \beta \right)
&=& \sum_{q^{\prime }}^{\left\{ B,Q,S\right\} } J_{nm}^{qq^{\prime }} \dd\alpha_{q^{\prime }}
- J_{n+1,m}^{q} \dd\beta \;.  \label{dI_nm_q}
\end{eqnarray}
In terms of physical quantities, we identify the thermodynamic integral of particle species 
$i$ with indices $n=1$ and $m=0$ as the particle density, $n_{i}=I_{i,10}$, 
while that with indices $n=2$ and $m=0$ is the energy density, $e_{i}=I_{i,20}$. 
From Eqs.\ \eqref{dI_i_nm_0} and \eqref{dI_i_nm_q} we then obtain the standard
thermodynamic relations
\begin{eqnarray}
\dd n_{i} &\equiv& \left( \frac{\partial n_{i}}{\partial \alpha_{i}}\right)_{\beta } \dd\alpha_{i}
+ \left( \frac{\partial n_{i}}{\partial \beta }\right)_{\alpha_{i}} \dd\beta 
= \sum_{q}^{\left\{ B,Q,S\right\} } q_{i} J_{i,10} \dd\alpha_{q} - J_{i,20} \dd\beta\; , \\
\dd e_{i} &\equiv& \left( \frac{\partial e_{i}}{\partial \alpha_{i}}\right)_{\beta } \dd\alpha_{i}
+ \left( \frac{\partial e_{i}}{\partial \beta }\right)_{\alpha_{i}} \dd\beta 
= \sum_{q}^{\left\{ B,Q,S\right\} } q_{i} J_{i,20} \dd\alpha_{q} - J_{i,30} \dd\beta\; .
\end{eqnarray}
From these results and Eqs.\ \eqref{I_J_nm} -- \eqref{I_J_nm_qq},
or directly from Eq.\ \eqref{dI_nm_q}, we may express the total 
differential of the density of conserved charge $q$ as 
\begin{eqnarray}
\dd n_{q} &\equiv& \sum_{i\,=\,1}^{N_{\text{spec}}}q_{i} \dd n_{i}
= \sum_{q^{\prime }}^{\left\{B,Q,S\right\} } \frac{\partial n_{q}}{\partial \alpha_{q^{\prime }}}
\dd\alpha_{q^{\prime }} + \frac{\partial n_{q}}{\partial \beta } \dd\beta  \notag \\
&=& \sum_{q^{\prime }}^{\left\{ B,Q,S\right\} } J_{10}^{qq^{\prime }}
\dd\alpha_{q^{\prime }} - J_{20}^{q} \dd\beta \equiv \sum_{q^{\prime }}^{\left\{B,Q,S\right\} }
\left( \mathcal{T}^{-1}\right)_{qq^{\prime }} \dd\alpha_{q^{\prime }}
+ \left( \mathcal{T}^{-1}\right)_{q0} \dd\beta \;,  \label{d_n_q}
\end{eqnarray}
and the total differential of the energy density as 
\begin{eqnarray}
\dd e &\equiv& \sum_{i\,=\,1}^{N_{\text{spec}}} \dd e_{i} = \sum_{q^{\prime }}^{\left\{
B,Q,S\right\} } \frac{\partial e}{\partial \alpha_{q^{\prime }}} \dd\alpha_{q^{\prime }} + \frac{\partial e}{\partial \beta } \dd\beta  \notag \\
&=& \sum_{q^{\prime }}^{\left\{ B,Q,S\right\} } J_{20}^{q^{\prime }} \dd\alpha_{q^{\prime }} - J_{30} \dd\beta \equiv \sum_{q^{\prime }}^{\left\{ B,Q,S\right\} } \left( \mathcal{T}^{-1}\right)_{0q^{\prime }} \dd\alpha_{q^{\prime }} + \left( \mathcal{T}^{-1}\right)_{00} \dd\beta\; ,  \label{d_e}
\end{eqnarray}
where we have defined the following inverse matrix, 
\begin{equation} \label{Inverse_Thermo}
\left( \mathcal{T}^{-1}\right)_{qq^{\prime }} \equiv \left( 
\begin{array}{cc} \displaystyle
\frac{\partial e}{\partial \beta} & \displaystyle\frac{\partial e}{\partial \alpha_{q^{\prime}}} \\ 
\displaystyle\frac{\partial n_{q}}{\partial \beta} & \displaystyle
\frac{\partial n_{q}}{\partial \alpha_{q^{\prime}}} \\ 
\end{array}
\right) 
=\left( 
\begin{array}{cccc}
-J_{30} & J_{20}^{B} & J_{20}^{Q} & J_{20}^{S} \\ 
-J_{20}^{B} & J_{10}^{BB} & J_{10}^{BQ} & J_{10}^{BS} \\ 
-J_{20}^{Q} & J_{10}^{QB} & J_{10}^{QQ} & J_{10}^{QS} \\ 
-J_{20}^{S} & J_{10}^{SB} & J_{10}^{SQ} & J_{10}^{SS}
\end{array}\right)\;.
\end{equation}
Equations \eqref{d_n_q} and \eqref{d_e} can be solved for $d\beta $ and $d\alpha_{q}$,
\begin{eqnarray}
\dd\beta &=& \mathcal{T}_{00} \dd e + \sum_{q^{\prime }}^{\left\{ B,Q,S\right\} }%
\mathcal{T}_{0q^{\prime }} \dd n_{q^{\prime }}\;,  \label{dbeta} \\
\dd\alpha_{q} &=& \mathcal{T}_{q0} \dd e + \sum_{q^{\prime }}^{\left\{ B,Q,S\right\}}
\mathcal{T}_{qq^{\prime }} \dd n_{q^{\prime }}\;.  \label{dalpha_q}
\end{eqnarray}
Note that the relations \eqref{d_n_q} and \eqref{d_e}, or equivalently \eqref{dbeta} and 
\eqref{dalpha_q}, encode the thermodynamic response of the medium to 
perturbations and contain information about the chemical composition and/or the equation 
of state. These thermodynamic relations will be used later in the equations of motion.

\subsection{Equilibrium fluid-dynamical quantities}
\label{sec:fluid_dyn_quant}

The equilibrium moments \eqref{I_i_r_mu1_mun}, for $r=0$ and for the tensor ranks 
$\ell = 1$ and $\ell = 2$, define the partial particle
four-current and energy-momentum tensor, 
\begin{eqnarray}
N_{i,0}^{\mu } &\equiv& \mathcal{I}_{i,0}^{\mu } = \int \dd K_i k_{i}^{\mu }
f_{i,\mathbf{k}}^{\left( 0\right) } 
\equiv \left\langle k^{\mu} \right\rangle_{i,0}\;, \\
T_{i,0}^{\mu \nu } &\equiv& \mathcal{I}_{i,0}^{\mu \nu }
= \int \dd K_i k_{i}^{\mu} k_{i}^{\nu } f_{i,\mathbf{k}}^{\left( 0\right) }
\equiv \left\langle k^{\mu} k^{\nu }\right\rangle_{i,0}\;.
\end{eqnarray}
The tensor decomposition (\ref{I_i_r_mu1_mun_exp})
of these quantities with respect to an arbitrary time-like 
normalized flow velocity $u^{\mu }$ and the projection operator 
$\Delta^{\mu \nu}$ reads
\begin{eqnarray}
N_{i,0}^{\mu} &\equiv& I_{i,10}u^{\mu } = n_{i}u^{\mu}\;,  \label{N_i_0_mu} \\
T_{i,0}^{\mu \nu} &\equiv& I_{i,20}u^{\mu }u^{\nu} - I_{i,21}\Delta^{\mu\nu}
= e_{i}u^{\mu}u^{\nu} - P_{i}\Delta^{\mu \nu}\;.  \label{T_i_0_munu}
\end{eqnarray}
Tensor-projecting these quantities leads to the partial particle density,
energy density, and pressure of species $i$, 
\begin{eqnarray}
n_{i} &\equiv & N_{i,0}^{\mu }u_{\mu } = \left\langle E_{\mathbf{k}}
\right\rangle_{i,0} = I_{i,10}\equiv \mathcal{I}_{i,1}\;,  \label{n_i_0} \\
e_{i} &\equiv & T_{i,0}^{\mu \nu }u_{\mu }u_{\nu } = \left\langle E_{\mathbf{k}}^{2}
\right\rangle_{i,0} = I_{i,20}\equiv \mathcal{I}_{i,2}\;,  \label{e_i_0} \\
P_{i} &\equiv & -\frac{1}{3}\,T_{i,0}^{\mu \nu }\Delta_{\mu \nu } 
= -\frac{1}{3}\left\langle \Delta_{\mu \nu }k^{\mu }k^{\nu }\right\rangle_{i,0}
= I_{i,21} \equiv -\frac{1}{3}\left( m_{i}^{2}\,\mathcal{I}_{i,0} - \mathcal{I}_{i,2}\right) \;.  
\label{P_i_0}
\end{eqnarray}
The sum over all particle species $i$ of the partial 
equilibrium moments leads to the total particle four-current, the conserved charge 
four-currents, and the energy-momentum tensor of the mixture,
\begin{eqnarray}
N_{0}^{\mu } &\equiv &\sum_{i\,=\,1}^{N_{\text{spec}}}N_{i,0}^{\mu}
=\sum_{i\,=\,1}^{N_{\text{spec}}}n_{i}u^{\mu }\equiv nu^{\mu }\;,  \label{N_mu_0_sum} \\
N_{q,0}^{\mu } &\equiv &\sum_{i\,=\,1}^{N_{\text{spec}}}q_{i}N_{i,0}^{\mu}
=\sum_{i\,=\,1}^{N_{\text{spec}}}q_{i}n_{i}u^{\mu }\equiv n_{q}u^{\mu }\;,
\label{Nq_mu_0_sum} \\
T_{0}^{\mu \nu } &\equiv &\sum_{i\,=\,1}^{N_{\text{spec}}}T_{i,0}^{\mu \nu}
=\sum_{i\,=\,1}^{N_{\text{spec}}}\left( e_{i}u^{\mu }u^{\nu }-P_{i}\Delta^{\mu \nu}\right) 
\equiv e u^{\mu }u^{\nu } - P \Delta^{\mu \nu }\;.  \label{T_munu_0_sum}
\end{eqnarray}%
Note that in our case the conserved net-charge four-currents are simply the electric, the 
baryon, and the strangeness four-current.
The primary thermodynamic quantities of the mixture, i.e.\ the total number density, 
net-charge density, total energy density, and total pressure, are obtained by
summing over all constituents
\begin{equation}
n=\sum_{i\,=\,1}^{N_{\text{spec}}}n_{i}\;,\quad 
n_{q}=\sum_{i\,=\,1}^{N_{\text{spec}}}q_{i}n_{i}\;,
\quad e=\sum_{i\,=\,1}^{N_{\text{spec}}}e_{i}\;,\quad 
P=\sum_{i\,=\,1}^{N_{\text{spec}}}P_{i}\;.
\label{equilibirum_neP}
\end{equation}
The particle and net-charge number as well as the energy are extensive thermodynamic 
quantities, while the total pressure of the mixture follows Dalton's law of partial 
pressures.\footnote{These relations hold for systems which can be described by kinetic theory, however they are violated once the Sto{\ss}zahlansatz \cite{deGroot_book} does not apply, i.e.\ when long-range interactions or multi-particle correlations become non-negligible.}
An equation of state determines these
thermodynamic quantities as functions of temperature and chemical
potentials, i.e.\ $n_{q}=n_{q}\left( T,\mu_{B},\mu_{Q},\mu_{S}\right)$, 
$e=e\left( T,\mu_{B},\mu_{Q},\mu_{S}\right)$, and 
$P=P\left( T,\mu_{B},\mu_{Q},\mu_{S}\right)$. 

Note that in local thermodynamic equilibrium the individual particle four-currents, 
$N_{0,i}^{\mu} = n_i u^\mu$, as well as the energy current of species $i$,  
$T_{0,i}^{\mu\nu}u_\nu = e_i u^{\mu}$, are parallel to each other.
Therefore, all of these currents lead to the same local rest frame of the fluid.
Out of equilibrium the fluid-dynamical four-velocity can no longer be uniquely defined.  
Nonetheless, without any loss of generality, a common flow velocity tied to a chosen local 
rest frame can still be defined. 
The difference of fluid-dynamical quantities from their local-equilibrium form will be 
discussed next.

\subsection{Out-of-equilibrium fluid-dynamical quantities}
\label{sec:ooefdq}

Out of equilibrium, the distribution function differs from its local-equilibrium 
form by $\delta f_{i,\mathbf{k}}=f_{i,\mathbf{k}} - f_{i,\mathbf{k}}^{\left( 0\right) }$. 
Introducing a similar notation for the momentum integrals as in Eq.\ \eqref{brakets_0},
\begin{equation}
\left\langle \cdots \right\rangle_{i,\delta } 
\equiv \int \dd K_i  \left(\cdots\right)_i \delta f_{i,\mathbf{k}}\; ,
\end{equation}
and
\begin{equation}
\left\langle \cdots \right\rangle_{i}\equiv \int \dd K_i \left(\cdots \right)_i
f_{i,\mathbf{k}} = \left\langle \cdots \right\rangle_{i,0} 
+ \left\langle \cdots \right\rangle_{i,\delta }\; ,
\end{equation}%
cf.\ Refs.\ \cite{Denicol:2012cn,Denicol:2012es}, we define the irreducible 
moments of tensor-rank $\ell$ and energy-rank $r$ 
of the deviation of the single-particle distribution
function from equilibrium for a given particle species $i$, 
\begin{equation}
\rho_{i,r}^{\mu_{1}\cdots \mu_{\ell}} 
\equiv \Delta_{\nu_{1}\cdots \nu_{\ell }}^{\mu_{1}\cdots \mu_{\ell}}
\int \dd K_i E_{i,\mathbf{k}}^{r} k_{i}^{\mu_{1}}\cdots k_{i}^{\mu_{\ell }}\delta f_{i,\mathbf{k}}
= \left\langle E_{\mathbf{k}}^{r} k^{\left\langle \mu_{1}\right. }\cdots 
k^{\left. \mu_{\ell }\right\rangle}\right \rangle_{i,\delta}\;.  \label{rho_i_r}
\end{equation}%
Furthermore, we expand the distribution function $f_{i,\mathbf{k}}$ around 
$f^{(0)}_{i,\mathbf{k}}$ as in Ref.\ \cite{Denicol:2012cn},
\begin{equation}
\delta f_{i,\mathbf{k}}\equiv f_{i,\mathbf{k}}^{(0) } \tilde{f}_{i,\mathbf{k}}^{(0)}
\phi_{i,\mathbf{k}} 
= f_{i,\mathbf{k}}^{(0)} \tilde{f}_{i,\mathbf{k}}^{(0)} 
\sum_{\ell\,=\,0}^{\infty} \sum_{n\,=\,0}^{N_{\ell }} 
\rho_{i,n}^{\mu_{1}\cdots \mu_{\ell}}
k_{i, \left\langle \mu_{1}\right. }\cdots k_{\left. i, \mu_{\ell}\right\rangle } 
\mathcal{H}_{i,\mathbf{k}n}^{(\ell )} \;,
\label{delta_f_i}
\end{equation}
where the irreducible tensors orthogonal to the four-flow are
$k_i^{\left\langle \mu_{1}\right. }\cdots k_i^{\left. \mu_{\ell }\right\rangle}
= \Delta^{\mu_1 \cdots \mu_{\ell}}_{\nu_1 \cdots \nu_{\ell}} 
k_i^{\nu_{1}} \cdots k_i^{\nu_{\ell }}$. 
These tensors form a complete and orthogonal basis in momentum space.
The coefficient $\mathcal{H}_{i,\mathbf{k}n}^{(\ell )}$ is a
polynomial in energy of order $N_{\ell }$ defined as 
\begin{equation}
\mathcal{H}_{i,\mathbf{k}n}^{(\ell )}=\frac{\left( -1\right)^{\ell}}{\ell !\ J_{i,2\ell ,\ell }} 
\sum_{m\,=\,n}^{N_{\ell }}a_{i,mn}^{(\ell)}
\sum_{r\,=\,0}^{m}a_{i,mr}^{\left( \ell \right) }E_{i,\mathbf{k}}^{r}\;. \label{eq:Hfunctions}
\end{equation}%
In principle, the expansion in polynomials in energy is an infinite series, i.e.~$N_\ell \rightarrow \infty$. However, 
we have already introduced parameters $N_\ell< \infty$ (for each $\ell \geq 0$) at this point since we will truncate the 
series later on in order to derive a fluid-dynamical theory. The coefficients $a_{i,mn}^{(\ell )}$ are calculated 
via the Gram-Schmidt orthogonalization procedure and can be expressed in terms of thermodynamic
integrals, see Ref.\ \cite{Denicol:2012cn} for more details.

Thus, similarly to the equilibrium moments we define the out-of-equilibrium particle 
four-current and energy-momentum tensor for particle species $i$ as 
\begin{eqnarray}
N_{i}^{\mu } &\equiv& \int \dd K_i k_{i}^{\mu} \left(f_{i,\mathbf{k}}^{(0)} + \delta f_{i,\mathbf{k}}\right)
= N_{i,0}^{\mu } + \rho_{i,0}^{\mu } = \left\langle k^{\mu}\right\rangle_{i,0} + \left\langle k^{\mu }\right\rangle_{i,\delta}
\equiv \left\langle k^{\mu }\right\rangle_{i}\;,  \label{N_mu_i} \\
T_{i}^{\mu \nu } &\equiv&  \int \dd K_i k_{i}^{\mu}k_{i}^{\nu} \left(f_{i,\mathbf{k}}^{(0)} + \delta f_{i,\mathbf{k}}\right)
= T_{i,0}^{\mu \nu } + \rho_{i,0}^{\mu \nu}
= \left\langle k^{\mu }k^{\mu }\right\rangle_{i,0}
+\left\langle k^{\mu}k^{\nu }\right\rangle_{i,\delta} 
\equiv \left\langle k^{\mu } k^{\nu}\right\rangle_{i}\;,  \label{T_munu_i}
\end{eqnarray}%
where $N_{i,0}^{\mu }$ and $T_{i,0}^{\mu \nu }$ were defined in 
Eqs.\ \eqref{N_i_0_mu} and \eqref{T_i_0_munu}, respectively.

The tensor decompositions with respect to an arbitrary time-like normalized
flow velocity $u^{\mu }$, summed over all particle species,
lead to the total fluid-dynamical quantities of the mixture,
\begin{eqnarray}
N^{\mu } &\equiv& \sum_{i\,=\,1}^{N_{\text{spec}}}N_{i}^{\mu }
=\sum_{i\,=\,1}^{N_{\text{spec}}} 
\left[ \left( n_{i}+\rho_{i,1}\right) u^{\mu }+V_{i}^{\mu }\right] \notag \\
&\equiv& \left( n+\delta n\right) u^{\mu }+V^{\mu }\;,  \label{N_mu_i_sum} \\
N_{q}^{\mu } &\equiv & \sum_{i\,=\,1}^{N_{\text{spec}}}q_{i}N_{i}^{\mu}
=\sum_{i\,=\,1}^{N_{\text{spec}}}\left[ q_{i}\left( n_{i}+\rho_{i,1}\right) u^{\mu}
+q_{i}V_{i}^{\mu }\right]  \notag \\
&\equiv &\left( n_{q}+\delta n_{q}\right) u^{\mu }+V_{q}^{\mu }\;,
\label{Nq_mu_i_sum} \\
T^{\mu \nu } &\equiv & \sum_{i\,=\,1}^{N_{\text{spec}}}T_{i}^{\mu \nu}
=\sum_{i\,=\,1}^{N_{\text{spec}}}\left[ \left( e_{i}+\rho_{i,2}\right) u^{\mu }u^{\nu}
-\left( P_{i}+\Pi_{i}\right) \Delta^{\mu \nu }+2W_{i}^{\left( \mu \right.}
u^{\left. \nu \right) }+\pi_{i}^{\mu \nu }\right]  \notag \\
&\equiv &\left( e+\delta e\right) u^{\mu }u^{\nu }-\left( P+\Pi \right)
\Delta^{\mu \nu }+2W^{\left( \mu \right. }u^{\left. \nu \right) }
+\pi^{\mu\nu }\;.  \label{T_munu_i_sum}
\end{eqnarray}
The net-particle density, the net-charge density, the energy
density, and the isotropic pressure of the out-of-equilibrium mixture are
\begin{eqnarray}
n+\delta n &\equiv &N^{\mu }u_{\mu }=\sum_{i\,=\,1}^{N_{\text{spec}}}\left\langle 
E_{\mathbf{k}}\right\rangle_{i}\equiv \sum_{i\,=\,1}^{N_{\text{spec}}}
\left( n_{i}+\rho_{i,1}\right)\; ,  \label{n+dn_def} \\
n_{q}+\delta n_{q} &\equiv &N_{q}^{\mu }u_{\mu}
=\sum_{i\,=\,1}^{N_{\text{spec}}}q_{i}\left\langle E_{\mathbf{k}}\right\rangle_{i}
\equiv \sum_{i\,=\,1}^{N_{\text{spec}}}q_{i}\left( n_{i}+\rho_{i,1}\right) \;,
\label{nq+dnq_def} \\
e+\delta e &\equiv & T^{\mu \nu }u_{\mu }u_{\nu}
=\sum_{i\,=\,1}^{N_{\text{spec}}}\left\langle E_{\mathbf{k}}^{2}\right\rangle_{i}
\equiv \sum_{i\,=\,1}^{N_{\text{spec}}}\left( e_{i}+\rho_{i,2}\right) \;,
\label{e+de_def} \\
P+\Pi &\equiv &-\frac{1}{3}T^{\mu \nu }\Delta_{\mu \nu}
=-\sum_{i\,=\,1}^{N_{\text{spec}}}\frac{1}{3}\left\langle \Delta_{\mu \nu }k^{\mu }k^{\nu }
\right\rangle_{i}=\sum_{i\,=\,1}^{N_{\text{spec}}}\left[ P_{i}-\frac{1}{3}\left(
m_{i}^{2}\rho_{i,0}-\rho_{i,2}\right) \right] \equiv
\sum_{i\,=\,1}^{N_{\text{spec}}}\left( P_{i}+\Pi_{i}\right) \;,  \label{P+Pi_def}
\end{eqnarray}%
with an equation of state relating the equilibrium quantities.
Note that the latter were defined in Eq.\ \eqref{equilibirum_neP}, 
while the partial pressure appearing in the last equation was defined in Eq.\ \eqref{P_i_0}. 
Hence, it follows that the non-equilibrium correction to the pressure, 
the so-called  bulk viscous pressure of particle species $i$, is 
\begin{equation}
\Pi_{i}\equiv -\frac{1}{3}\left\langle \Delta_{\alpha \beta}k^{\alpha}
k^{\beta }\right\rangle_{i,\delta }=-\frac{1}{3}
\left( m_{i}^{2}\rho_{i,0}-\rho_{i,2}\right)\; .  \label{Pi_i_def}
\end{equation}
The net-particle diffusion, the net-charge diffusion, and 
the energy-momentum diffusion currents are
\begin{eqnarray}
V^{\mu } &\equiv &\Delta_{\nu }^{\mu }N^{\nu}
=\sum_{i\,=\,1}^{N_{\text{spec}}}\left\langle k^{\left\langle \mu \right\rangle}\right\rangle_{i}
=\sum_{i\,=\,1}^{N_{\text{spec}}}\rho_{i,0}^{\mu }\equiv
\sum_{i\,=\,1}^{N_{\text{spec}}}V_{i}^{\mu }\;,  \label{V_mu_i_sum} \\
V_{q}^{\mu } &\equiv &\Delta_{\nu }^{\mu }N_{q}^{\nu}
=\sum_{i\,=\,1}^{N_{\text{spec}}}q_{i}\left\langle k^{\left\langle \mu \right\rangle}
\right\rangle_{i}=\sum_{i\,=\,1}^{N_{\text{spec}}}q_{i}\rho_{i,0}^{\mu }\equiv
\sum_{i\,=\,1}^{N_{\text{spec}}}q_i V_{i}^{\mu }\;,  \label{Vq_mu_i_sum} \\
W^{\mu } &\equiv &\Delta_{\alpha }^{\mu }T^{\alpha \beta }u_{\beta}
=\sum_{i\,=\,1}^{N_{\text{spec}}}\left\langle E_{\mathbf{k}}k^{\left\langle \mu
\right\rangle }\right\rangle_{i}=\sum_{i\,=\,1}^{N_{\text{spec}}}\rho_{i,1}^{\mu}
\equiv \sum_{i\,=\,1}^{N_{\text{spec}}}W_{i}^{\mu }\;.  \label{W_mu_i_sum} 
\end{eqnarray}
Finally, the shear-stress tensor of the mixture is
\begin{eqnarray}
\pi^{\mu \nu } &\equiv &\Delta_{\alpha \beta }^{\mu \nu }T^{\alpha \beta}
=\sum_{i\,=\,1}^{N_{\text{spec}}}\left\langle k^{\left\langle \mu \right. }
k^{\left.\nu \right\rangle }\right\rangle_{i}
=\sum_{i\,=\,1}^{N_{\text{spec}}}\rho_{i,0}^{\mu\nu }
\equiv \sum_{i\,=\,1}^{N_{\text{spec}}}\pi_{i}^{\mu \nu }\;.
\label{pi_munu_i_sum}
\end{eqnarray}
Equations (\ref{n+dn_def}) -- (\ref{pi_munu_i_sum}) represent the  
fluid-dynamical fields of the mixture, 
which (similarly as in chemical solutions) is a combination of multiple 
particle species and where the number of particles of an individual species may or may 
not be conserved. Originally, these fields constitute $14 N_{\text{spec}}$ variables
(10 for each energy-momentum tensor $T_i^{\mu \nu}$ and 4 for each
particle current $N_i^\mu$).
We assume that the mixture can be treated as a single fluid such that
its space-time evolution 
can be entirely determined in terms of the total energy-momentum tensor 
$T^{\mu \nu}$ and the charge four-currents $N^\mu_q$.
This approach reduces the number of unknown fluid-dynamical fields   
to $10 + 4N_{q}$.\footnote{Note that, naively counting the number of unknowns, there 
are actually $15 + 5 N_q$ degrees of freedom: 5 degrees of freedom (d.o.f.s)
for the $N_q$ charge four-currents ($n_q$, $\delta n_q$, and 3 
components of $V^\mu_q$), and 
4 scalar ($e$, $\delta e$, $P$, $\Pi$), 
6 vector ($u^\mu$ and $W^\mu$), and 5 tensor d.o.f.s ($\pi^{\mu\nu}$)
for the energy-momentum tensor.
However, one d.o.f.\ is reduced by the equation of state, $P = P(e,\{n_q\})$. 
Furthermore, $4 + N_q$ additional d.o.f.s are
reduced by the matching conditions, see Sec.\ \ref{sec:matching}.}

Further, the out-of-equilibrium part of the distribution function $\delta f_{i,\mathbf{k}}$ 
was expanded in terms of an infinite set of independent irreducible moments 
$\rho_{i,n}^{\mu_{1}\cdots \mu_{\ell}}$, each of which obeys an equation of motion 
derived from the relativistic Boltzmann equation \eqref{BTE_i} (see Sec.\ \ref{Sect:Relax}). 
The dissipative fluid-dynamical fields, Eqs.\ \eqref{Pi_i_def} -- \eqref{pi_munu_i_sum}, are
defined in terms of (some of) these moments, and thus are solutions of these equations
of motion. The crucial step in the derivation of fluid dynamics using the method of 
moments 
is to truncate the infinite set of equations of motion for the irreducible moments, and thus 
also the series in Eq.\ \eqref{delta_f_i}, in a well-defined manner. To this end, the sums
over tensor rank $\ell$ and over powers of energy $n$ in Eq.\ \eqref{delta_f_i} are 
truncated. 
The latter is already implied by the truncation parameter $N_\ell$, which depends on the 
respective tensor rank $\ell$ of the moment. The lowest possible truncation in $\ell$ and $n$ is to account 
for the lowest-order irreducible moments which explicitly appear in the energy-momentum 
tensor \eqref{T_munu_i_sum} and the charge four-flow \eqref{Nq_mu_i_sum}, namely $\rho_{i,0}$, $\rho_{i,1}$, $\rho_{i,2}$, $\rho^\mu_{i,0}$, $\rho^\mu_{i,1}$, and $\rho^{\mu\nu}_{i,0}$. This leads to the truncation $\ell \leq 2$, and $N_0 = 2$, $N_1 = 1$, and $N_2 = 0$ in the series in Eq.~\eqref{delta_f_i}.  
This is the so-called $(10 + 4N_q)$-moment approximation. As mentioned above, these 
moments, however, further depend on other moments, which may also be of higher tensor 
rank (e.g.\ $\ell > 2$). In Sec.\ \ref{sec:OofMapprox} we discuss how to further truncate the 
set of equations of motion.

\subsection{Matching conditions and local rest frames}
\label{sec:matching}

In local equilibrium the thermodynamic state of matter is completely determined 
by a few scalar fields, namely a common temperature $T$ and the chemical potentials 
of the constituent particle species $\alpha_{i}$, which are in turn given by 
the chemical potentials of the conserved charges $\alpha_{q}$ via 
Eq.~\eqref{mu_i_mu_q}. 
A common way to determine these thermodynamic variables in an arbitrary state (which
is not too far from local equilibrium) is to demand that the net-charge densities and the 
total 
energy density are the same as in some fictitious local-equilibrium reference state. 
These are the so-called Landau matching conditions \cite{Landau_book}, 
\begin{eqnarray}
\left( N_{q}^{\mu }-N_{q,0}^{\mu }\right) u_{\mu } &\equiv
&\sum_{i\,=\,1}^{N_{\text{spec}}}q_{i}\left( N_{i}^{\mu }-N_{i,0}^{\mu }\right) u_{\mu}
=\sum_{i\,=\,1}^{N_{\text{spec}}}q_{i}\rho_{i,1} \equiv \delta n_{q} = 0\;,
\label{Landau_Matching_nq} \\
\left( T^{\mu \nu }-T_{0}^{\mu \nu }\right) u_{\mu }u_{\nu } &\equiv
&\sum_{i\,=\,1}^{N_{\text{spec}}}\left( T_{i}^{\mu \nu }-T_{i,0}^{\mu \nu }\right)
u_{\mu }u_{\nu }=\sum_{i\,=\,1}^{N_{\text{spec}}}\rho_{i,2} \equiv \delta e = 0\;,
\label{Landau_Matching_e}
\end{eqnarray}
where $\left( N^{\mu }-N_{0}^{\mu }\right) u_{\mu }\equiv
\sum_{i\,=\,1}^{N_{\text{spec}}}\left( N_{i}^{\mu }-N_{i,0}^{\mu }\right) u_{\mu}
=\sum_{i\,=\,1}^{N_{\text{spec}}}\rho_{i,1} \equiv \delta n \neq 0$, since the total
number of particles is not necessarily conserved. Furthermore, Landau's
matching condition for  the energy, Eq.\ \eqref{Landau_Matching_e}, also leads to a 
simplification of the bulk viscous pressure in Eqs.\ \eqref{P+Pi_def} and \eqref{Pi_i_def}, 
\begin{equation}
\Pi \equiv -\frac{1}{3}\sum_{i\,=\,1}^{N_{\text{spec}}}m_{i}^{2}\rho_{i,0}
=\sum_{i\,=\,1}^{N_{\text{spec}}}\Pi_{i}\;.  \label{Pi_matched}
\end{equation}
We note that using the matching conditions \eqref{Landau_Matching_nq} and
\eqref{Landau_Matching_e}, we can express some
scalar moments by the others, and thus reduce the 
number of scalar moments of the multicomponent mixture by $N_{q}+1$.

The number of independent unknowns is further reduced once we choose a local rest 
frame, or equivalently a definition for the fluid-dynamical flow velocity. 
The definition of Landau and Lifshitz \cite{Landau_book} leads to the so-called 
Landau frame, or energy frame, and requires that the total energy-momentum 
diffusion current of the mixture is zero, 
\begin{equation}
W^{\mu} \equiv \sum_{i\,=\,1}^{N_{\text{spec}}} W_{i}^{\mu}=0\;.  \label{Landau_flow}
\end{equation}
This directly implies that $T^{\mu \nu} u_{\nu} = e u^{\mu}$, meaning that the flow velocity 
$u^\mu$ is the time-like eigenvector of the energy-momentum tensor with eigenvalue $e$. 
This choice reduces the total number of unknowns by three and leads to additional
constraints between the remaining species-specific vector fields $W_{i}^{\mu}$, 
i.e.\ there are only $N_{\text{spec}} - 1$ independent energy-momentum 
diffusion fluxes in the mixture of $N_{\text{spec}}$ different species. 
Note that Eq.\ (\ref{Landau_flow}) also implies that if the fluid consists of a single 
component, 
i.e.\ $N_{\text{spec}} = 1$, there is no energy-momentum diffusion present in this frame.
Unless stated otherwise, the Landau frame is our choice for the local rest frame.

More traditionally, one may use Eckart's definition \cite{Eckart:1940te} 
to fix the local rest frame by demanding that the overall diffusion of 
one of the conserved net charges, say that of charge $q^\star$, in the mixture is zero,
\begin{equation}
\tilde{V}_{q^\star}^{\mu }\equiv \sum_{i\,=\,1}^{N_{\text{spec}}}
\tilde{V}_{q^\star, i}^{\mu } 
= \sum_{i\,=\,1}^{N_{\text{spec}}}q^\star_i \tilde{V}_{i}^{\mu }=0\;, \label{qstar}
\end{equation}
where quantities in this particular $q^\star$-charge frame are denoted by a tilde.
However, in high-energy heavy-ion collisions, where there are multiple conserved
charges, which are not necessarily non-vanishing in all regions of space-time,
the definition of the rest frame according to Eckart is less suitable.

Nevertheless, in case of a single non-vanishing conserved charge $q$, the 
Landau and Eckart reference frames 
are essentially equivalent, hence choosing one over the other is a matter
of taste.
Namely, the energy-momentum diffusion current in the Eckart frame can be related
to the charge diffusion current in the Landau frame via
\begin{equation}
\tilde{W}^{\mu} \equiv - h_q  V_q^{\mu} \;,
\end{equation} 
where we introduced the enthalpy per charge, $h_q = (e+P)/n_q$, see Appendix
\ref{peculiar_flow} for more details.

\subsection{Conservation equations}

Due to the fact that in binary collisions the net charges as
well as the energy and momentum of particles are conserved, 
the equations of fluid dynamics of a mixture are derived from
the Boltzmann equation (\ref{BTE_i}) as 
\begin{eqnarray} \label{charge_cons_0}
\partial_{\mu }N_{q}^{\mu } &\equiv &\sum_{i\,=\,1}^{N_{\text{spec}}}q_{i}
\partial_{\mu }N_{i}^{\mu } = \sum_{i\,=\,1}^{N_{\text{spec}}} q_i 
\int \dd K_i C_{i}=0\;, \label{eq:charge_conservation_0}\\
\partial_{\mu }T^{\mu \nu } &\equiv &\sum_{i\,=\,1}^{N_{\text{spec}}} 
\partial_{\mu}T_{i}^{\mu \nu } = \sum_{i\,=\,1}^{N_{\text{spec}}} 
\int \dd K_i k^{\nu}_i C_{i} = 0\;,
\label{energy_momentum_cons_0}
\end{eqnarray}
where there are $N_{q}$ independent charge-conservation laws. 
Note that due to inelastic collisions the number of particles of species $i$ is 
no longer conserved and the individual particle species satisfy rate equations, 
$\partial_{\mu }N_{i}^{\mu } \neq 0$.
On the other hand, for purely elastic collisions the numbers of particles are conserved,
and the momentum integral over each partial collision term vanishes separately, 
$\partial_{\mu }N_{i}^{\mu } = 0$.

With Eq.\ \eqref{Nq_mu_i_sum} the $N_{q}$ charge-conservation equations 
(\ref{charge_cons_0}) assume the form 
\begin{eqnarray}
\partial_{\mu }N_{q}^{\mu } &\equiv &\sum_{i\,=\,1}^{N_{\text{spec}}}q_{i}
D\left(n_{i}+\rho_{i,1}\right) +\sum_{i\,=\,1}^{N_{\text{spec}}}q_{i}
\left( n_{i}+\rho_{i,1}\right) \theta 
+\sum_{i\,=\,1}^{N_{\text{spec}}}q_{i}\partial_{\mu } V_{i}^{\mu } \notag \\
&=&Dn_{q}+n_{q}\theta +\partial_{\mu }V_{q}^{\mu }=0\;,  \label{charge_cons}
\end{eqnarray}
where in the last step we used the first Landau matching condition
\eqref{Landau_Matching_nq}.

The conservation of energy of the mixture is obtained by 
projecting Eq.\ \eqref{energy_momentum_cons_0} onto $u_\nu$ and 
inserting Eq.\ \eqref{T_munu_i_sum},
\begin{align}
u_{\nu} \partial_{\mu }T^{\mu \nu } & \equiv \sum_{i\,=\,1}^{N_{\text{spec}}} 
D\left(e_{i}+\rho_{i,2}\right) +\sum_{i\,=\,1}^{N_{\text{spec}}} 
\left( e_{i}+\rho_{i,2}+P_{i}+\Pi_{i}\right) \theta  \notag \\
& + \sum_{i\,=\,1}^{N_{\text{spec}}}\partial_{\mu }W_{i}^{\mu }
-\sum_{i\,=\,1}^{N_{\text{spec}}}W_{i}^{\mu } Du_{\mu}
-\sum_{i\,=\,1}^{N_{\text{spec}}}\pi_{i}^{\mu \nu}\sigma_{\mu \nu }  \notag \\
& =De+\left( e+P+\Pi \right) \theta -\pi^{\mu \nu }\sigma_{\mu \nu }=0\; ,
\label{energy_cons}
\end{align}
where we have imposed the second Landau matching condition 
\eqref{Landau_Matching_e} and also fixed the local rest frame according to Landau's 
convention, Eq.\ \eqref{Landau_flow}. 

Using these conservation equations to
replace $dn_{q}$ and $de$ in Eqs.\ \eqref{dbeta} and \eqref{dalpha_q} leads to the
comoving derivatives of the inverse temperature and the charge chemical
potentials multiplied by the inverse temperature, 
\begin{eqnarray}
D\beta &=&-\mathcal{T}_{00}\left[ \left( e+P+\Pi \right) \theta 
-\pi^{\mu\nu }\sigma_{\mu \nu }\right] -\sum_{q^{\prime }}^{\left\{ B,Q,S\right\} }
\mathcal{T}_{0q^{\prime }}\left[ n_{q^{\prime }}\theta 
+ \partial_{\mu}V_{q^{\prime }}^{\mu }\right]\; ,  \label{D_beta_multi} \\
D\alpha_{q} &=&-\mathcal{T}_{q0}\left[ \left( e+P+\Pi \right) \theta 
-\pi^{\mu \nu }\sigma_{\mu \nu }\right] 
-\sum_{q^{\prime }}^{\left\{B,Q,S\right\} } \mathcal{T}_{qq^{\prime }}
\left[ n_{q^{\prime }}\theta+\partial_{\mu }V_{q^{\prime }}^{\mu }\right]\; .  
\label{D_alphaq_multi}
\end{eqnarray}

Finally, projecting Eq.\ (\ref{energy_momentum_cons_0}) onto $\Delta_\beta^\mu$ (which
gives the momentum conservation of the mixture), and using Eqs.\ 
\eqref{Landau_Matching_e} and \eqref{Landau_flow} leads to
\begin{align}
\Delta_{\beta }^{\mu }\partial_{\alpha } T^{\alpha \beta } &\equiv
\sum_{i\,=\,1}^{N_{\text{spec}}}\left( e_{i} + P_{i} + \Pi_{i}\right) Du^{\mu }
- \nabla^{\mu } \sum_{i\,=\,1}^{N_{\text{spec}}} \left( P_{i} + \Pi_{i} \right) 
+ \Delta_{\beta}^{\mu } \partial_{\alpha } \sum_{i\,=\,1}^{N_{\text{spec}}}\pi_{i}^{\alpha \beta } 
\notag \\
&= \left( e + P + \Pi \right) Du^{\mu } - \nabla^{\mu }\left( P + \Pi \right) + \Delta_{\beta }^{\mu} \partial_{\alpha } \pi^{\alpha \beta } = 0\;.
\label{momentum_cons}
\end{align}
This leads to an expression for the acceleration $Du^\mu$ of the fluid. 
Noting that in local equilibrium the Gibbs-Duhem relation holds in the form
\begin{equation} 
\beta \dd P = \sum_{q}^{\left\{ B,Q,S\right\} } n_{q} \dd \alpha_{q} - (e+P) \dd\beta\;,
\end{equation} 
we obtain
\begin{equation}
Du^{\mu }=-\frac{1}{\beta}\nabla^{\mu }\beta +\frac{1}{\beta (e+P)}
\sum_{q}^{\left\{ B,Q,S\right\} }n_{q}\nabla^{\mu
}\alpha_{q} -\frac{1}{e+P}\left( \Pi Du^{\mu }-\nabla
^{\mu }\Pi +\Delta_{\beta }^{\mu }\partial_{\alpha }\pi^{\alpha \beta }%
\right)\; .  \label{D_u_mu_multi}
\end{equation}

Let us recount the unknown variables and equations of the mixture. 
There are $N_{q}$ conservation equations \eqref{charge_cons} 
for $n_{q}$ and $V_{q}^{\mu}$, representing $4N_{q}$ variables. 
The conservation of energy and momentum of the mixture provides 
the four equations \eqref{energy_cons} and \eqref{momentum_cons}, entailing 
$e$, $\Pi$, $u^{\mu}$, and  $\pi^{\mu \nu}$, which represent 10 variables in total, 
since the equation of state already defines the pressure in terms of $e$ and the $n_q$'s. 
Thus, in a dissipative mixture of $N_{q}$ conserved charges we have only $4+N_{q}$
conservation equations for a total of $10+4N_{q}$ unknown fields. 
The additional equations for the dissipative fields $\Pi$, $V_{q}^{\mu }$, and 
$\pi^{\mu\nu}$ will be derived from the Boltzmann equation in the next section.

\section{Second-order dissipative fluid-dynamical equations of motion}
\label{Sect:Relax}

In this section, we first derive the equations of motion for the irreducible
tensor moments for particle species $i$. 
For a single-component fluid, these equations were first given in 
Ref.\ \cite{Denicol:2012cn}. Here, we extend them towards
multicomponent fluids. Then we linearize the collision term and discuss
the Navier-Stokes limit as well as the order-of-magnitude approximation, which
provides a simple, and yet effective way to close the set of equations 
of motion. Finally, we derive and discuss the second-order
dissipative fluid-dynamical equations of motion. 

\subsection{Equations of motion for the irreducible moments}

The equations of motion for the irreducible moments 
$\rho_{i,r}^{\mu_{1}\cdots \mu_{\ell}}$ are obtained directly from 
the Boltzmann equation \eqref{BTE_i} by multiplying it
with $E_{i,\mathbf{k}}^r k_i^{\langle \mu_1} \cdots k_i^{\mu_\ell \rangle}$,
integrating over momentum space, and taking the comoving derivative. Then,
projecting onto $\Delta^{\mu_{1}\cdots \mu_{\ell}}_{\nu_{1}\cdots \nu_{\ell}}$, we obtain
the equations of motion for the irreducible moments, i.e.\ an equation for the comoving
derivative $ \dot{\rho}_{i,r}^{\left\langle\mu_{1}\cdots \mu_{\ell} \right\rangle} 
\equiv  \Delta^{\mu_{1}\cdots \mu_{\ell}}_{\nu_{1}\cdots \nu_{\ell}} 
D\rho_{i,r}^{\nu_{1}\cdots \nu_{\ell}}$. The irreducible moments of the collision 
term \eqref{COLL_INT} are defined as 
\begin{equation}
	C_{i,r-1}^{\left\langle \mu_{1}\cdots \mu_{\ell }\right\rangle}
	\equiv \Delta_{\nu_{1}\cdots \nu_{\ell }}^{\mu_{1}\cdots \mu_{\ell}}
	\sum_{j\,=\,1}^{N_{\text{spec}}}\int \dd K_i E_{i,\mathbf{k}}^{r-1}\,k_{i}^{\nu_{1}}\cdots 
	k_{i}^{\nu_{\ell }}C_{ij}\left[ f\right]\; . \label{eq:MomentsOfCollisionMatrix}
\end{equation}
After some calculation the equation of motion for the
irreducible moments of tensor-rank zero reads 
\begin{align}
\dot{\rho}_{i,r} - C_{i,r-1} = &\; \alpha_{i,r}^{\left( 0\right) }\theta
-\sum_{q^{\prime }}^{\left\{ B,Q,S\right\} }\left( J_{i,r+1,0}\mathcal{T}_{0q^{\prime }} 
-\sum_{q}^{\left\{ B,Q,S\right\} }q_{i}J_{i,r0}\mathcal{T}_{qq^{\prime }}\right)
\left( \nabla_{\mu }V_{q^{\prime }}^{\mu }-V_{q^{\prime }}^{\mu }\dot{u}_{\mu }\right) 
\notag \\
& +\frac{\theta }{3}\left[ m_{i}^{2}\left( r-1\right) \rho_{i,r-2}-\left(r+2\right) 
\rho_{i,r}-3\left( J_{i,r+1,0}\mathcal{T}_{00}-\sum_{q}^{\left\{B,Q,S\right\} } 
q_{i}J_{i,r0}\mathcal{T}_{q0}\right) \Pi \right]  
\notag \\
& +r\rho_{i,r-1}^{\mu }\dot{u}_{\mu }-\nabla_{\mu }\rho_{i,r-1}^{\mu }
+\left[ \left( r-1\right) \rho_{i,r-2}^{\mu \nu } 
+\left( J_{i,r+1,0}\mathcal{T}_{00}-\sum_{q}^{\left\{ B,Q,S\right\} } 
q_{i}J_{i,r0}\mathcal{T}_{q0}\right) \pi^{\mu \nu }\right] \sigma_{\mu \nu }\;.  
\label{D_rho_ir}
\end{align}%
This equation is different from Eq.\ (35) of Ref.\ \cite{Denicol:2012cn}, because
thermodynamic relations are modified in mixtures 
with multiple conserved charges as compared to a single-component fluid. 
Similarly, the transport coefficient 
$\alpha_{i,r}^{\left( 0\right)}$ has additional contributions given 
by the sums $\sum_{q}^{\left\{ B,Q,S\right\}}$ when compared to 
Eq.\ (42) of Ref.\ \cite{Denicol:2012cn}, 
\begin{align}
\alpha_{i,r}^{\left( 0\right) } =& -\left[ I_{i,r0}+\left( r-1\right)
I_{i,r1}+\sum_{q^{\prime }}^{\left\{ B,Q,S\right\} }
\left( J_{i,r+1,0}\mathcal{T}_{0q^{\prime }} 
-\sum_{q}^{\left\{ B,Q,S\right\} }q_{i}J_{i,r0}\mathcal{T}_{qq^{\prime }}\right) 
n_{q^{\prime }}\right]  \notag \\
& -\left( J_{i,r+1,0}\mathcal{T}_{00}-\sum_{q}^{\left\{ B,Q,S\right\}}
q_{i}J_{i,r0}\mathcal{T}_{q0}\right) \left( e+P\right)\; .
\label{alpha_0_ir}
\end{align}

The equation of motion for the irreducible moments of tensor-rank one is very similar
to Eq.\ (36) of Ref.\ \cite{Denicol:2012cn}, except for the first-order gradient term 
$\sum_{q}^{\left\{B,Q,S\right\}}\alpha_{i,r,q}^{\left( 1\right) }\nabla^{\mu }\alpha_{q}$, 
\begin{align}
\dot{\rho}_{i,r}^{\left\langle \mu \right\rangle} 
- C_{i,r-1}^{\left\langle \mu \right\rangle } 
 =& \;\sum_{q}^{\left\{ B,Q,S\right\} } 
\alpha_{i,r,q}^{\left( 1\right) }\nabla^{\mu }\alpha_{q}+r\rho_{i,r-1}^{\mu \nu}
\dot{u}_{\nu }-\frac{1}{3}\nabla^{\mu }\left( m_{i}^{2}\rho_{i,r-1}
-\rho_{i,r+1}\right) +\alpha_{i,r}^{h}\nabla^{\mu }\Pi  
\notag \\
& -\Delta_{\lambda }^{\mu }\left( \nabla_{\nu }\rho_{i,r-1}^{\lambda \nu}
+\alpha_{i,r}^{h}\partial_{\nu }\pi^{\nu \lambda }\right) +\frac{1}{3}
\left[ m_{i}^{2}\left( r-1\right) \rho_{i,r-2}^{\mu }-\left( r+3\right)
\rho_{i,r}^{\mu }\right] \theta  
\notag \\
& +\frac{1}{3}\left[ m_{i}^{2}r\rho_{i,r-1}-\left( r+3\right) 
\rho_{i,r+1}-3\alpha_{i,r}^{h}\Pi \right] \dot{u}^{\mu }+\rho_{i,r,\nu }\omega^{\mu \nu }  
\notag \\
& +\frac{1}{5}\left[ m_{i}^{2}\left( 2r-2\right) \rho_{i,r-2,\nu } 
-\left(2r+3\right) \rho_{i,r,\nu }\right] \sigma^{\mu \nu }+\left( r-1\right)
\rho_{i,r-2}^{\mu \nu \lambda }\sigma_{\nu \lambda}\;,   
\label{D_rho_mu_ir}
\end{align}
where the transport coefficients are 
\begin{eqnarray}
\alpha_{i,r,q}^{\left( 1\right) } =  q_{i}J_{i,r+1,1} 
+ \alpha_{i,r}^{h}\frac{n_{q}}{\beta} \; , \qquad 
\alpha_{i,r}^{h} = -\beta \frac{J_{i,r+2,1}}{e+P}\;.  \label{alpha_1_irq}
\end{eqnarray}

Finally, the equation of motion for the irreducible moments of tensor-rank
two is formally unchanged from Eq.\ (37) of Ref.\ \cite{Denicol:2012cn}, only
the species-specific index $i$ is introduced here, 
\begin{align}
\dot{\rho}_{i,r}^{\left\langle \mu \nu \right\rangle }-
C_{i,r-1}^{\langle \mu \nu \rangle }= &\;2\alpha_{i,r}^{\left( 2\right) }
\sigma^{\mu \nu }+\frac{2}{15}\left[ m_{i}^{4}\left( r-1\right) 
\rho_{i,r-2}-\left( 2r+3\right) m_{i}^{2}\rho_{i,r}+\left( r+4\right) 
\rho_{i,r+2}\right] \sigma^{\mu \nu }  
\notag \\
& +\frac{2}{5}\dot{u}^{\left\langle \mu \right. }\left[ m_{i}^{2}r 
\rho_{i,r-1}^{\left. \nu \right\rangle }-\left( r+5\right) \rho_{i,r+1}^{\left.
\nu \right\rangle }\right] -\frac{2}{5}\nabla^{\left\langle \mu
\right. }\left( m_{i}^{2}\rho_{i,r-1}^{\left. \nu \right\rangle }
-\rho_{i,r+1}^{\left. \nu \right\rangle }\right) 
\notag \\
& +r\rho_{i,r-1}^{\mu \nu \lambda }\dot{u}_{\lambda }-\Delta_{\alpha \beta}^{\mu \nu }
\nabla_{\lambda }\rho_{i,r-1}^{\alpha \beta \lambda }+\left(r-1\right) 
\rho_{i,r-2}^{\mu \nu \lambda \kappa }\sigma_{\lambda \kappa}
+2\rho_{i,r}^{\lambda \left\langle \mu \right. } 
\omega_{\left. {}\right.\lambda }^{\left. \nu \right\rangle }  
\notag \\
& +\frac{1}{3}\left[ m_{i}^{2}\left( r-1\right) \rho_{i,r-2}^{\mu \nu}
-\left( r+4\right) \rho_{i,r}^{\mu \nu }\right] \theta 
+\frac{2}{7}\left[m_{i}^{2}\left( 2r-2\right) \rho_{i,r-2}^{\lambda \left\langle \mu \right.}
-\left( 2r+5\right) \rho_{i,r}^{\lambda \left\langle \mu \right. }\right]
\sigma_{\lambda }^{\left. \nu \right\rangle }\;,  
\label{D_rho_munu_ir}
\end{align}
where the coefficient $\alpha_{i,r}^{\left( 2\right) }$ is also formally
the same as in Eq.\ (44) of Ref.\ \cite{Denicol:2012cn}, 
\begin{equation}
\alpha_{i,r}^{\left( 2\right) } = \left( r-1\right) I_{i,r+2,2}+I_{i,r+2,1}\;.
\label{alpha_2_ir}
\end{equation}
These are the equations of motion for the irreducible moments
up to tensor-rank two for any particle 
species $i$. One can show that in the case of a single-component fluid they
reduce to the equations found in Ref.\ \cite{Denicol:2012cn}.

Furthermore, note that since the conserved quantities in fluid dynamics contain no 
tensors of rank higher than two, the higher-rank tensors, 
$\rho_{i,r}^{\mu_{1}\cdots \mu_{\ell }} =0$ for $\ell \ge 3$,  
in the equations of motion \eqref{D_rho_mu_ir} and \eqref{D_rho_munu_ir} 
will be neglected in the following (see Sec.\ \ref{sec:OofMapprox}).

\subsection{The linearized collision integral}

Further progress requires the linearization of the collision integral \eqref{COLL_INT} 
in the quantities  $\phi_{i,\mathbf{k}}=\delta f_{i,\mathbf{k}}/%
[f_{i,\mathbf{k}}^{(0)} \tilde{f}_{i,\mathbf{k}}^{(0)}]$, such that it simplifies to
\begin{equation}
C_{i}\left(x,k\right) \simeq \frac{1}{2}\sum_{j,a,b\,=\,1}^{N_{\text{spec}}}
\int \dd K_{j}^{\prime } \dd P_{a} \dd P_{b}^{\prime}\, 
W_{ij\rightarrow ab}^{kk^{\prime}\rightarrow pp^{\prime}} 
f_{i,\mathbf{k}}^{\left( 0\right) } f_{j,\mathbf{k}^{\prime }}^{\left( 0\right) }
\tilde{f}^{\left( 0\right) }_{a,\mathbf{p}} 
\tilde{f}^{\left( 0\right) }_{b,\mathbf{p}^{\prime }}
\left( \phi_{a,\mathbf{p}}+\phi_{b,\mathbf{p}^{\prime }}
-\phi_{i,\mathbf{k}}-\phi_{j,\mathbf{k}^{\prime }}\right) \;,
\label{C_i_A_rho}
\end{equation}%
where the bilateral normalization condition holds
and the equality
$f^{(0)}_{i,\bf{k}} f^{(0)}_{j,\bf{k}^\prime} \tilde{f}^{\left( 0\right) }_{a,\mathbf{p}} 
\tilde{f}^{\left( 0\right) }_{b,\mathbf{p}^{\prime }}
= f^{(0)}_{a,\bf{p}} f^{(0)}_{b,\bf{p}^\prime} \tilde{f}^{\left( 0\right) }_{i,\mathbf{k}} 
\tilde{f}^{\left( 0\right) }_{j,\mathbf{k}^{\prime }}$ 
was used \cite{deGroot_book,Cercignani_book}.

Using the linearized collision integral \eqref{C_i_A_rho} one can show that the 
corresponding irreducible moments (\ref{eq:MomentsOfCollisionMatrix})
of the collision integral can be expressed in terms of a linear combination of irreducible 
moments, $\rho_{i,r}^{\nu_{1}\cdots \nu_{\ell}}$,
in a similar way as in Eq.\ (50) of Ref.\ \cite{Denicol:2012cn}, 
\begin{eqnarray}
C_{i,r-1}^{\langle \mu_{1}\cdots \mu_{\ell }\rangle } &\equiv
&-\sum_{s\,=\,1}^{N_{\text{spec}}}\sum_{n\,=\,0}^{N_{\ell }}\sum_{m\,=\,0}^{\infty }
\left( \mathcal{A}_{is,rn}\right)_{\nu_{1}\cdots \nu_{m}}^{\mu_{1}\cdots \mu_{\ell }}
\rho_{s,n}^{\nu_{1}\cdots \nu_{m}},  \notag \\
&=& -\sum_{s\,=\,1}^{N_{\text{spec}}}\sum_{n\,=\,0}^{N_{\ell }} 
\left( \mathcal{A}_{is,rn}^{\left( \ell \right) }\right) \rho_{s,n}^{\mu_{1}\cdots \mu_{\ell }}\;.  
\label{Coll_int_mixture}
\end{eqnarray}%
Here, we have defined the following tensors 
\begin{align}
\left( \mathcal{A}_{is,rn}\right)^{\mu_{1}\cdots \mu_{\ell } \nu_{1}\cdots \nu_{m}}
 = &\; \frac{1}{2}\sum_{j,a,b\,=\,1}^{N_{\text{spec}}}
\int \dd K_{i} \dd K_{j}^{\prime } \dd P_{a} \dd P_{b}^{\prime } 
W_{ij\rightarrow ab}^{kk^{\prime}\rightarrow pp^{\prime}}
f_{i,\mathbf{k}}^{\left( 0\right) } f_{j,\mathbf{k}^{\prime }}^{\left( 0\right) }
\tilde{f}^{\left( 0\right) }_{a,\mathbf{p}} 
\tilde{f}^{\left( 0\right) }_{b,\mathbf{p}^{\prime }}
E_{i,\mathbf{k}}^{r-1}k_{i}^{\left\langle \mu_{1}\right. }\cdots k_{i}^{\left.
\mu_{\ell }\right\rangle }  \notag \\
& \times \left[ \delta_{si}\mathcal{H}_{s,\mathbf{k}n}^{(m)}
k_{s}^{\left\langle \nu_{1}\right. } \cdots k_{s}^{\left. \nu_{m}\right\rangle}
+\delta_{sj} \mathcal{H}_{s,\mathbf{k}^{\prime }n}^{(m)} 
k_{s}^{\prime \left\langle \nu_{1}\right. }\cdots k_{s}^{\prime
\left. \nu_{m}\right\rangle }\right.  \notag \\
& \quad \left. -\delta_{sa}\mathcal{H}_{s,\mathbf{p}n}^{(m)} 
p_{s}^{\left\langle \nu_{1}\right. } \cdots p_{s}^{\left. \nu_{m}\right\rangle } 
-\delta_{sb}\mathcal{H}_{s,\mathbf{p}^{\prime }n}^{(m)} 
p_{s}^{\prime \left\langle \nu_{1}\right. }\cdots 
p_{s}^{\prime \left. \nu_{m}\right\rangle } \right] \;,
\label{eq:CollisionMatrix_Tensor}
\end{align}
where the polynomials $\mathcal{H}_{s,\mathbf{k}n}^{(m)}$ were defined in 
Eq.\ (\ref{eq:Hfunctions}).
These tensors can be decomposed and projected, hence one finally obtains the
collision matrix 
\begin{equation}
\mathcal{A}_{is,rn}^{\left( \ell \right) }=\frac{1}{ 2\ell +1}
\Delta_{\mu_{1}\cdots \mu_{\ell }}^{\nu_{1}\cdots \nu_{\ell }}
\left(\mathcal{A}_{is,rn}\right)_{\nu_{1}\cdots \nu_{\ell }}^{\mu_{1}\cdots\mu_{\ell }}\;,
\label{eq:CollisionMatrixEntries}
\end{equation}
cf.\ Eq.\ (A18) of Ref.\ \cite{Denicol:2012cn}.
Note that in the case of a single-component system, i.e.\ 
$N_{\text{spec}}=1$, the above equation reduces to the diagonal components of 
$\mathcal{A}_{is,rn}^{\left( \ell \right) }$. Hence, the 
particle species labels $i$ and $s$ can be dropped and we simply recover the result 
of Ref.\ \cite{Denicol:2012cn}, 
$C_{i,r-1}^{\langle \mu_{1}\cdots \mu_{\ell }\rangle } 
\overset{\text{single}}{\rightarrow} C_{r-1}^{\langle \mu_{1}\cdots \mu_{\ell }\rangle }
=-\sum_{n\,=\,0}^{N_{\ell}}
\mathcal{A}_{rn}^{\left( \ell \right) }\rho_{n}^{\mu_{1}\cdots \mu_{\ell }}$ .


The inverse of the collision matrix defines the so-called relaxation-time matrix, 
\begin{equation}
\tau_{si,nr}^{\left( \ell \right) }\equiv \left( \mathcal{A}^{-1} \right)^{\left(\ell\right)}_{si,nr}\;, 
\label{MAIN_relax_time}
\end{equation}
where the matrix elements are proportional to the inverse of the mean free path between 
collisions $\lambda^{-1}_{\textrm{mfp}}$.
Therefore, multiplying both sides of Eq.\ (\ref{Coll_int_mixture}) by the relaxation-time 
matrix we obtain an important relation between the 
moments of the linearized collision integral and the irreducible moments, 
\begin{equation}
\sum_{i\,=\,1}^{N_{\text{spec}}}\sum_{r\,=\,0}^{N_{\ell }}
\tau_{si,nr}^{\left( \ell\right) } 
C_{i,r-1}^{\left\langle \mu_{1}\cdots \mu_{\ell }\right\rangle}
\equiv -\sum_{i,s^{\prime}\,=\,1}^{N_{\text{spec}}}\sum_{r,n^{\prime }\,=\,0}^{N_{\ell }}
\tau_{si,nr}^{\left( \ell\right) } 
\mathcal{A}_{is^{\prime },rn^{\prime }}^{\left( \ell \right) }
\rho_{s^{\prime },n^{\prime }}^{\mu_{1}\cdots \mu_{\ell }}
= -\rho_{s,n}^{\mu_{1}\cdots \mu_{\ell }}.  \label{Coll_int_relax_time}
\end{equation}

The infinite set of equations of motions for the irreducible moments contains 
infinitely many degrees of freedom. 
In order to close the equations of motion for the mixture treated as a single fluid, 
the number of degrees of freedom must be reduced and infinite sums must be truncated 
at some finite number.

One of the key features of transient fluid dynamics is that the 
corresponding equations of motion have a single time scale that controls the transient 
behavior, e.g.\ relaxation towards equilibrium. On the other hand,
the Boltzmann equation has infinitely many of such time scales. Even in a 
single-component gas the modes corresponding to the same tensor rank $\ell$ are  
coupled to each other and their dynamics depends on these infinitely many scales. 
In the case of a mixture the situation is even more complicated as
the modes corresponding to different particle species are also coupled. 
The reason for this is that due to interactions between particles of different species the 
corresponding moments are correlated, see Refs.~\cite{El:2011cp,El:2012ka}.

One possible way to reduce the number of degrees of freedom would be to generalize 
the approach of Ref.\ \cite{Denicol:2012cn} developed for a single-component system, 
and diagonalize the collision matrix to determine the slowest microscopic time scales, i.e.\ 
the relaxation times that are relevant in our approximation, and the corresponding modes 
that dominate the long-time dynamics of the fluid in the transient regime. This has the 
advantage that the relaxation times that appear in the equations of motion could be 
explicitly identified as real microscopic time scales. However, the downside of this method 
is the appearance of terms that are of second or higher order in gradients, denoted as 
$\mathcal{O}({\rm Kn}^2)$ in Ref.\ \cite{Denicol:2012cn}. These terms can violate stability 
and causality of the theory. In principle, this can be cured by introducing further 
independent dynamical variables, cf.\ for instance Ref.\ \cite{Denicol:2012vq}, but this is 
beyond the scope of the current work. Instead, as discussed below, we will employ a 
slightly simpler approach, where the problematic $\mathcal{O}({\rm Kn}^2)$ terms do not 
appear.

\subsection{The Navier-Stokes limit and the order-of-magnitude approximation}
\label{sec:OofMapprox}

The explicit relation between the irreducible tensor of a given rank and the corresponding 
fluid-dynamical gradients can be derived by multiplying Eqs.\ (\ref{D_rho_ir}), 
(\ref{D_rho_mu_ir}), and (\ref{D_rho_munu_ir}) by the corresponding relaxation-time
matrices, $\tau_{si,nr}^{\left( \ell \right) }$, and using Eq.\ \eqref{Coll_int_relax_time}. 
In this way, the following equations of motion for the irreducible moments of tensor rank 
$\ell =0, \,1$, and 2 are obtained,
\begin{eqnarray}
\sum_{i\,=\,1}^{N_{\text{spec}}}\sum_{r\,=\,0}^{N_{0}}
\tau_{si,nr}^{\left( 0\right) }\dot{\rho}_{i,r} + \rho_{s,n} 
&=&  -\zeta_{s,n}\theta + \mathcal{O}(2) \;, 
\label{eq:OOM1} \\
\sum_{i\,=\,1}^{N_{\text{spec}}}\sum_{r\,=\,0}^{N_{1}}\tau_{si,nr}^{\left( 1\right) }
\dot{\rho}_{i,r}^{\left\langle \mu \right\rangle } 
+\rho_{s,n}^{\mu}
&=& \sum_{q}^{\left\{ B,Q,S\right\} }\kappa_{s,n,q}\nabla^{\mu }
\alpha_{q} + \mathcal{O}(2) \;, 
\label{eq:OOM2} \\
\sum_{i\,=\,1}^{N_{\text{spec}}}\sum_{r\,=\,0}^{N_{2}} \tau_{si,nr}^{\left( 2\right)}
\dot{\rho}_{i,r}^{\left\langle \mu \nu \right\rangle }+\rho_{s,n}^{\mu \nu }
&=&2\eta_{s,n}\sigma^{\mu \nu } + \mathcal{O}(2) \;, 
\label{eq:OOM3}
\end{eqnarray}%
where $\mathcal{O}(2)$ denote all remaining second- and higher-order terms from
the corresponding equations of motion for the irreducible moments. These are
terms which are at 
least of quadratic order in the Knudsen number, $\mathcal{O}(\mathrm{Kn}^2)$, 
or in the inverse Reynolds number, $\mathcal{O}(\mathrm{Rn^{-2}})$, or of linear 
order in their product, $\mathcal{O}(\mathrm{Kn}\, \mathrm{Rn^{-1}})$. 
Here, we have defined the species-specific bulk-viscosity, diffusion, 
and shear-viscosity coefficients as
\begin{eqnarray}
\zeta_{s,n} &\equiv& -\sum_{i\,=\,1}^{N_{\text{spec}}} 
\sum_{r\,=\,0}^{N_{0}}\tau_{si,nr}^{\left(0\right) }
\alpha_{i,r}^{\left( 0\right) }\;, \label{eq:BaseCoeffScalar} \\
\kappa_{s,n,q} &\equiv& \sum_{i\,=\,1}^{N_{\text{spec}}}\sum_{r\,=\,0}^{N_{1}} 
\tau_{si,nr}^{\left( 1\right) }\alpha_{i,r,q}^{\left( 1\right) }\;, \label{eq:BaseCoeffVector} \\
\eta_{s,n} &\equiv& \sum_{i\,=\,1}^{N_{\text{spec}}}\sum_{r\,=\,0}^{N_{2}} 
\tau_{si,nr}^{\left( 2\right)}\alpha_{i,r}^{\left( 2\right) }\;. \label{eq:BaseCoeffTensor}
\end{eqnarray}
Note that the definition of the bulk-viscosity coefficient differs by a factor of
$-m^2_s/3$ compared to Eq.\ (63) of Ref.\ \cite{Denicol:2012cn}.

From here on, we will make the assumption that the irreducible moments $\rho_{i,r}$,  
$\rho_{i,r}^{\left\langle \mu \right\rangle }$, 
and $\rho_{i,r}^{\left\langle \mu \nu \right\rangle }$
are of the same order, irrespective of the particle species. 
This implies that the sum over all species of these irreducible moments,
i.e.\ $\rho_{r}$, $\rho_{r}^{\left\langle \mu \right\rangle }$, and 
$\rho_{r}^{\left\langle \mu \nu \right\rangle }$
are also of the same order as the species-specific irreducible moments.

The so-called ``order-of-magnitude approximation'' is based on
the first-order solution of the moment equations (\ref{eq:OOM1}) -- (\ref{eq:OOM3}),
which is equivalent to the Navier-Stokes limit.
In this limit, the irreducible moments are algebraically related to 
terms of first order in Knudsen number, also called thermodynamic forces, multiplied by
the corresponding transport coefficients,
\begin{eqnarray}
\rho_{s,n} &=& -\zeta_{s,n}\theta + \mathcal{O}(2)\;,
\label{rho_sn_NS} \\
\rho_{s,n}^{\mu} &=& \sum_{q}^{\left\{ B,Q,S\right\} }\kappa_{s,n,q} 
\nabla^{\mu }\alpha_{q} + \mathcal{O}(2)\; , \label{rho_mu_sn_NS} \\
\rho_{s,n}^{\mu \nu } &=& 2\eta_{s,n}\sigma^{\mu \nu} +\mathcal{O}(2)\; ,
\label{rho_munu_sn_NS}
\end{eqnarray}
while all tensor moments of rank higher than two are at least of second order, 
$\rho_{s,n}^{\mu_{1}\cdots \mu_{\ell }}\sim \mathcal{O}(2)$
for any $\ell>2$, see Ref.\ \cite{Denicol:2012cn} for details.

In principle, all irreducible moments are of first order 
in inverse Reynolds number, $\mathcal{O}(\mathrm{Rn^{-1}})$, and
thus formally independent of the power counting in Knudsen number.
The order-of-magnitude approximation, which is based on the Navier-Stokes
limit \eqref{rho_sn_NS} -- \eqref{rho_munu_sn_NS}, assumes that
the irreducible moments are of first order in Knudsen number, 
$\mathcal{O}(\mathrm{Kn})$, i.e.\ the regime where the Knudsen and the inverse 
Reynolds numbers are of the same magnitude. 
This defines a power-counting scheme, similar to the one described in 
Ref.\ \cite{Denicol:2012cn}, which helps to assign a certain order to the various
terms in the equations of motion. Then, all $\mathcal{O}(2)$ terms
on the right-hand sides of Eqs.\ (\ref{eq:OOM1}) -- (\ref{eq:OOM3}), 
as well as the comoving derivatives on the left-hand sides are of
second order in Knudsen number. 
The order-of-magnitude approximation is very similar to the ``order-of-magnitude method''
in non-relativistic fluid dynamics~\cite{Struchtrup}.

Using the approximation \eqref{rho_sn_NS} -- \eqref{rho_munu_sn_NS}
while summing over all particle species, and for the moment omitting 
$\mathcal{O}(2)$ terms, we obtain the Navier-Stokes relations for the mixture. 
From Eq.\ \eqref{Pi_matched} together with Eq.\ \eqref{rho_sn_NS}
the total bulk viscous pressure of the mixture reads 
\begin{equation}
\Pi \equiv -\sum_{s\,=\,1}^{N_{\text{spec}}}\frac{m_{s}^{2}}{3}\rho_{s,0}
= \sum_{s\,=\,1}^{N_{\text{spec}}} \frac{m_s^2}{3} \zeta_{s,0}\theta \equiv -\zeta \theta \;.
\label{Pi_NS_mixture}
\end{equation}
Similarly, we obtain from Eq.\ \eqref{Vq_mu_i_sum} 
together with Eq.\ \eqref{rho_mu_sn_NS} for the conserved charge currents 
\begin{equation}
V_{q}^{\mu} \equiv \sum_{s\,=\,1}^{N_{\text{spec}}}q_{s} \rho_{s,0}^{\mu} 
= \sum_{q^{\prime }}^{\left\{ B,Q,S\right\}} \sum_{s\,=\,1}^{N_{\text{spec}}}q_{s} 
\kappa_{s,0,q^{\prime}} \nabla^{\mu }\alpha_{q^{\prime }} 
\equiv \sum_{q^{\prime }}^{\left\{ B,Q,S\right\} } 
\kappa_{qq^{\prime }}\nabla^{\mu }\alpha_{q^{\prime }}\;.
\label{Vq_mu_NS_mixture}
\end{equation}
Finally, the shear-stress tensor of the mixture follows
from Eq.\ \eqref{pi_munu_i_sum} together with Eq.\ \eqref{rho_munu_sn_NS}, 
\begin{equation}
\pi^{\mu \nu }\equiv \sum_{s\,=\,1}^{N_{\text{spec}}} \rho_{s,0}^{\mu \nu} 
= \sum_{s\,=\,1}^{N_{\text{spec}}}2\eta_{s,0}\sigma^{\mu \nu } 
\equiv 2\eta \sigma^{\mu \nu }\;.
\label{pi_munu_NS_mixture}
\end{equation}
Note that the first-order thermodynamic forces are the same for all particle species 
and for the mixture. Due to this fact, we obtain the first-order transport coefficients of the
mixture: the bulk viscosity 
$\zeta$, the diffusion coefficients $\kappa_{qq^\prime}$, and the shear viscosity $\eta$,
\begin{equation} \label{transport_coeff_mixture}
\zeta \equiv -\sum_{s\,=\,1}^{N_{\text{spec}}} \frac{m_s^2}{3} \zeta_{s,0}\;, \qquad 
\kappa_{qq^{\prime }} 
\equiv \sum_{s\,=\,1}^{N_{\text{spec}}}q_{s} \kappa_{s,0,q^{\prime }} \;,
\qquad \eta \equiv \sum_{s\,=\,1}^{N_{\text{spec}}} \eta_{s,0}\;.
\end{equation}
Therefore, using Eqs.\ \eqref{rho_sn_NS} -- \eqref{rho_munu_sn_NS} together with the 
relativistic Navier-Stokes relations \eqref{Pi_NS_mixture} -- \eqref{pi_munu_NS_mixture} 
we readily obtain the following algebraic relations between the species-specific irreducible
moments and the primary dissipative quantities of the mixture, 
\begin{eqnarray}
\rho_{s,n} &=& \bar{\zeta}_{s,n}\Pi + \mathcal{O}(2) \;,
\label{rho_in_NS} \\
\rho_{s,n}^{\mu } &=& \sum_{q}^{\left\{ B,Q,S\right\} }
\bar{\kappa}_{s,n}^{\left(q\right) } V_{q}^{\mu } + \mathcal{O}(2)\;,  
\label{rho_mu_in_NS} \\
\rho_{s,n}^{\mu \nu } &=& \bar{\eta}_{s,n} \pi^{\mu \nu } + \mathcal{O}(2)\;,  
\label{rho_munu_in_NS}
\end{eqnarray}
where we introduced the normalized transport coefficients for each species,
\begin{equation}
\bar{\zeta}_{s,n}=\frac{\zeta_{s,n}}{\zeta }\;, \qquad 
\bar{\kappa}_{s,n}^{\left( q\right) } = \sum_{q^{\prime }}^{\left\{ B,Q,S\right\} }
\kappa_{s,n,q^{\prime }}\left( \kappa^{-1}\right)_{q^{\prime }q}\;, \qquad 
\bar{\eta}_{s,n}=\frac{\eta_{s,n}}{\eta }\;. \label{eq:normalizedcoeff}
\end{equation}
Here, $\left( \kappa^{-1}\right)_{q^{\prime }q}$ is the inverse of the diffusion-coefficient 
matrix defined in Eq.\ (\ref{transport_coeff_mixture}). We will use Eqs.\
(\ref{rho_in_NS}) -- (\ref{rho_munu_in_NS}) to close the equations of motion 
(\ref{D_rho_ir}) -- (\ref{D_rho_munu_ir}) in the next section.

\subsection{The equations of motion in $\left(10 + 4N_{q}\right)$-moment approximation}
\label{sec:EoM14mom}

In the $\left(10 + 4N_{q}\right)$-moment approximation, we truncate the infinite set of 
moment equations (\ref{D_rho_ir}), (\ref{D_rho_mu_ir}), and (\ref{D_rho_munu_ir}) in
the following way. We first multiply these equations with the corresponding 
relaxation-time matrices $\tau_{si,nr}^{\left( \ell \right) }$. Equation (\ref{D_rho_mu_ir})
is also multiplied by $q_s$, and all equations are summed over species. 
In what follows, we only consider the set of equations for $n=0$.
In all terms we then substitute $\rho_{i,r}$, $\rho_{i,r}^{\mu }$, 
and $\rho_{i,r}^{\mu \nu }$ by the dissipative quantities $\Pi,\,V_q^\mu$, 
and $\pi^{\mu \nu}$ using Eqs.\ (\ref{rho_in_NS}) -- (\ref{rho_munu_in_NS}). 
Note that in this substitution the $\mathcal{O}(2)$ terms in 
Eqs.~(\ref{rho_in_NS}) -- (\ref{rho_munu_in_NS}) become $\mathcal{O}(3)$ terms 
and can be neglected together with other higher-order terms.
In this way, we finally obtain a closed set of equations of motion for the dissipative 
quantities $\Pi,\,V_q^\mu$, and $\pi^{\mu \nu}$. Further, as discussed in Sec.~\ref{sec:ooefdq}, 
the $\left(10 + 4N_{q}\right)$-moment approximation is the lowest-order truncation of the series in Eq.~\eqref{delta_f_i}, 
where $\ell \leq 2$, and $N_0 = 2$, $N_1 = 1$, and $N_2 = 0$. It should be noted that, for $\ell \leq 2$, 
such a truncation in powers of energy (i.e.~$N_\ell<\infty$) neglects infinitely many contributions of order 
$\mathcal{O}(\text{Kn})$ in Eq.~\eqref{delta_f_i}. As shown in Ref.~\cite{Denicol:2012cn}, however, 
systematically increasing the parameters $N_0$, $N_1$, and $N_2$ the values of the 
corresponding transport coefficients exhibit rapid convergence.

The resulting equation of motion for the bulk
viscous pressure is
\begin{equation}
\tau_{\Pi }\dot{\Pi}+\Pi = -\zeta \theta -\delta_{\Pi \Pi } \,\Pi \theta
+\lambda_{\Pi \pi } \, \pi^{\mu \nu }\sigma_{\mu \nu } 
- \sum_{q}^{\left\{B,Q,S\right\} }\ell_{\Pi V}^{(q)}\,\nabla_{\mu }V_{q}^{\mu}
- \sum_{q}^{\left\{ B,Q,S\right\} }\tau_{\Pi V}^{(q)}\,V_{q}^{\mu }\dot{u}_{\mu }
- \sum_{q,q^{\prime }}^{\left\{ B,Q,S\right\} }\lambda_{\Pi V}^{(q,q^{\prime })}
\,V_{q}^{\mu }\nabla_{\mu }\alpha_{q^{\prime }}\;,
\label{bulk_relax}
\end{equation}
where we have defined the relaxation time and the 
bulk-viscosity coefficient as%
\begin{eqnarray}
\tau_{\Pi } &=& -\sum_{s,i\,=\,1}^{N_{\text{spec}}} \frac{m_{s}^{2}}{3}
\sum_{r\,=\,0}^{N_{0}} 
\tau_{si,0r}^{\left( 0\right) }\bar{\zeta}_{i,r}\;, \\
\zeta &=& -\sum_{s\,=\,1}^{N_{\text{spec}}} \frac{m_s^2}{3} \zeta_{s,0}
\equiv \sum_{s,i\,=\,1}^{N_{\text{spec}}}  \frac{m_{s}^{2}}{3}
\sum_{r\,=\,0}^{N_{0}} 
\tau_{si,0r}^{\left( 0\right) }\alpha_{i,r}^{\left( 0\right) }\;.
\end{eqnarray}
All second-order transport coefficients appearing in Eq.\ (\ref{bulk_relax})
are listed in Appendix \ref{App:bulk_coefficients}.

Similarly, the equations of motion for the charge diffusion currents read 
\begin{align}
\sum_{q}^{\left\{ B,Q,S\right\} } \tau_{q^{\prime }q} \,
\dot{V}_{q}^{\left\langle \mu \right\rangle } + V_{q^{\prime }}^{\mu}
& =\sum_{q}^{\left\{ B,Q,S\right\} }\kappa_{q^{\prime }q} \, \nabla^{\mu}\alpha_{q} 
-\sum_{q}^{\left\{ B,Q,S\right\} } \tau_{q^{\prime }q} \, V_{q,\nu }\omega^{\nu \mu } 
-\sum_{q}^{\left\{ B,Q,S\right\} }\delta_{VV}^{(q^{\prime },q)}\,V_{q}^{\mu }\theta 
-\sum_{q}^{\left\{ B,Q,S\right\}}\lambda_{VV}^{(q^{\prime },q)}\,V_{q,\nu }
\sigma^{\mu \nu }  \notag \\
& -\ell_{V\Pi }^{(q^{\prime })} \, \nabla^{\mu }\Pi 
+\ell_{V\pi }^{(q^{\prime})} \, \Delta^{\mu \nu }\nabla_{\lambda }\pi_{\nu }^{\lambda }
+\tau_{V\Pi}^{(q^{\prime })} \,\Pi \dot{u}^{\mu }  
-\tau_{V\pi }^{(q^{\prime })}\,\pi^{\mu \nu }\dot{u}_{\nu} \notag \\
&+\sum_{q}^{\left\{ B,Q,S\right\} }\lambda_{V\Pi }^{(q^{\prime },q)}\,
\Pi\nabla^{\mu }\alpha_{q} 
- \sum_{q}^{\left\{ B,Q,S\right\} }\lambda_{V\pi}^{(q^{\prime },q)}\,
\pi^{\mu \nu }\nabla_{\nu }\alpha_{q}\;,
\label{diffusion_relax}
\end{align}
where the relaxation-time matrix and the diffusion-coefficient matrix are 
\begin{eqnarray}
\tau_{q^\prime q} &=&\sum_{s,i\,=\,1}^{N_{\text{spec}}}
\sum_{r\,=\,0}^{N_{1}} q^{\prime}_{s}  \tau_{si,0r}^{\left( 1\right) } 
\bar{\kappa}^{(q)}_{i,r}\;, \label{eq:relaxtime_diff}\\
\kappa_{q^\prime q} &=& \sum_{s,i\,=\,1}^{N_{\text{spec}}} 
\sum_{r\,=\,0}^{N_{1}} q^{\prime}_{s} 
\tau_{si,0r}^{\left( 1\right)} \alpha_{i,r,q}^{\left( 1\right) }\;,
\end{eqnarray}
and the second-order transport coefficients are listed in 
Appendix \ref{App:diffusion_coefficients}. 

The diffusion-coefficient matrix $\kappa_{qq^\prime}$ has been evaluated for 
several hadronic and partonic systems in 
Refs.\ \cite{Fotakis:2019nbq,Rose:2020sjv,Fotakis:2021diq}. In general, this matrix 
couples the diffusion current of a specific charge to \emph{all} gradients of the 
charge chemical potentials via $V_{q^\prime}^\mu \sim \sum_{q}^{\left\{ B,Q,S\right\} }
\kappa_{q^{\prime }q} \, \nabla^{\mu}\alpha_{q} + \mathcal{O}(2)$. Due to this 
coupled diffusion, the density gradients in one charge could lead to the local separation 
in another charge, as demonstrated in Ref.\ \cite{Fotakis:2019nbq}. From the above 
equations of motion one can see that such a coupling is also present in 
various second-order terms.
 
The equation of motion for the shear-stress tensor follows in a similar manner,
\begin{align}
\tau_{\pi }\dot{\pi}^{\left\langle \mu \nu \right\rangle } + \pi^{\mu \nu }
&= 2\eta \sigma^{\mu \nu } 
+ 2\tau_{\pi } \,\pi_{\lambda }^{\left\langle \mu\right. } 
\omega^{\left. \nu \right\rangle\lambda } 
- \delta_{\pi \pi }\,\pi^{\mu \nu }\theta 
- \tau_{\pi \pi }\,\pi^{\lambda \left\langle \mu \right.}
\sigma_{\lambda }^{\left. \nu \right\rangle }
+\lambda_{\pi \Pi }\,\Pi \sigma^{\mu \nu }  \notag \\
& -\sum_{q}^{\left\{ B,Q,S\right\} }\tau_{\pi V}^{(q)}\,
V_{q}^{\left\langle\mu \right. }\dot{u}^{\left. \nu \right\rangle } 
+ \sum_{q}^{\left\{B,Q,S\right\} } \ell_{\pi V}^{(q)} \, 
\nabla^{\left\langle \mu \right.} V_{q}^{\left. \nu \right\rangle } 
+\sum_{q,q^{\prime }}^{\left\{B,Q,S\right\} } \lambda_{\pi V}^{(q,q^{\prime })}\,
V_{q}^{\left\langle \mu \right. }\nabla^{\left. \nu \right\rangle }\alpha_{q^{\prime }}\;,
\label{shear_relax}
\end{align}
where the relaxation time and the coefficient of the shear viscosity are given by 
\begin{eqnarray}
\tau_{\pi } &=& \sum_{s,i\,=\,1}^{N_{\text{spec}}}
\sum_{r\,=\,0}^{N_{2}} \label{eq:relaxtime_shear}
\tau_{si,0r}^{\left( 2\right) }\bar{\eta}_{i,r}\;, \\
\eta &=& \sum_{s,i\,=\,1}^{N_{\text{spec}}}\sum_{r\,=\,0}^{N_{2}} 
\tau_{si,0r}^{\left( 2\right) }\alpha_{i,r}^{\left( 2\right) }\;,
\end{eqnarray}
while the remaining second-order transport coefficients are given in Appendix 
\ref{App:shear_coefficients}. The equations of motion \eqref{bulk_relax}, 
\eqref{diffusion_relax}, and \eqref{shear_relax} are of relaxation type and
are identical to those found in 
Refs.\ \cite{Monnai:2010qp,Monnai:2010th,Kikuchi:2015swa}. 
For more details we refer to the discussion in Appendix \ref{sec:term_by_term}.

As a simple example of a relativistic multicomponent system, we discuss
an ultrarelativistic, ideal gas with elastic, isotropic hard-sphere interactions and 
multiple conserved charges in Appendix \ref{sec:ultrarelativistic_case}. 
While the transport coefficients cannot be further reduced to simple and convenient forms, 
one may easily prove that one obtains well-known results in the single-component limit 
\cite{Denicol:2012cn}.

\section{Conclusions and Outlook}

In this paper, we have presented the derivation of relativistic second-order dissipative 
fluid dynamics for multicomponent systems
in the $(10+4N_q)$-moment approximation from the relativistic Boltzmann equation 
using the method of moments.
Starting from the relativistic Boltzmann equation for a multicomponent system
we have obtained the equations of motion for the irreducible moments for 
particle species $i$. 
In the single-fluid approximation for the mixture, the sum of the 
dynamical equations of motion reduces to $4+N_q$ conservation equations 
that are closed by providing $6+3N_q$ relaxation-type equations of motion for the
dissipative quantities.
In such a mixture, where the constituents in general carry multiple quantum charges
(e.g.\ a proton carrying electric charge as well as baryon number), the equation of state 
depends on multiple chemical potentials and temperature. 
With the help of a new approximation scheme, the so-called 
order-of-magnitude approximation, we have derived a second-order dissipative 
theory that does not contain terms of second order in Knudsen number, which are known 
to render the equations of motion parabolic and thus acausal \cite{Denicol:2012cn}. 
Furthermore, the irreducible moments of the deviation of the
single-particle distribution of each particle species from equilibrium 
are directly proportional to the total bulk viscous pressure $\Pi$, the conserved 
charge-diffusion currents $V^\mu_q$, and the total shear-stress tensor $\pi^{\mu\nu}$ 
via Eqs.\ \eqref{rho_in_NS} -- \eqref{rho_munu_in_NS}. 

Similar to other works which treat multicomponent systems
\cite{Monnai:2010qp,Monnai:2010th,Kikuchi:2015swa}, in this theory the existence of 
multiple conserved charges is manifest in the equations of motion 
Eqs.\ \eqref{bulk_relax}, \eqref{diffusion_relax}, and \eqref{shear_relax}. 
As expected we obtained exactly the same equations of motion as found in 
earlier works \cite{Monnai:2010qp,Monnai:2010th,Kikuchi:2015swa}. 

Further, the coupled charge transport becomes explicit in the appearance of mixing terms 
in the equations of motions, e.g.\ a dissipative current 
($\Pi$, $V^\mu_q$, $\pi^{\mu\nu}$) is coupled to \emph{any other} gradient in chemical 
potential or diffusion current. As a prominent example, instead of a diffusive Navier-Stokes 
term with only one diffusion coefficient $\kappa$ as in a single-component system, we 
obtain a Navier-Stokes term entailing a matrix of diffusion coefficients 
$\kappa_{qq^\prime}$, 
which explicitly couples every diffusion current to all gradients in chemical potential. The 
appearance of a charge-coupled Navier-Stokes term and potential implications for the 
transport of charge was discussed in Ref.\ \cite{Fotakis:2019nbq} 
in the case of relativistic nuclear matter.

The advantage of our derivation compared to other theories is that it yields explicit 
expressions for the transport coefficients in terms of the linearized collision term. Since 
the mutual interactions of all particle species is contained in the collision term,
the multicomponent nature of the mixture is naturally encoded in the transport coefficients. 

In the future, this theory will be used to revisit the transport of coupled 
charge in heavy-ion collisions initiated in Ref.\ \cite{Fotakis:2019nbq} in a more realistic 
manner. Especially, we expect that it will be relevant for the discussion of physics of 
compressed baryonic matter at the future FAIR and NICA facilities or for the
interpretation of recent results of the isobar run at RHIC. 
We expect that the coupling of diffusion currents or the charge gradients to the bulk 
viscous pressure and the shear-stress tensor may be important in future studies.
Now that the explicit expressions of the transport coefficients have been derived, they can 
be evaluated for nuclear systems. To this end, equations of state from lattice QCD for 
non-vanishing chemical potentials may be used 
\cite{Monnai:2019hkn,Noronha-Hostler:2019ayj}. At the same time, Eq.\ \eqref{delta_f_i} 
provides an expression for the so-called $\delta f$-correction needed for the freeze-out of 
the system at the end of the fluid-dynamical phase during the simulation of a 
heavy-ion collision.

\section*{Acknowledgments}
The authors thank G.S.\ Denicol and P. Huovinen for fruitful discussions. 
They acknowledge support by the 
Deutsche Forschungsgemeinschaft (DFG, German Research Foundation) through the 
CRC-TR 211 ``Strong-interaction matter under extreme conditions'' -- project number 
315477589 -- TRR 211. J.A.F.\ acknowledges support from the Helmholtz Graduate School
for Heavy-Ion Research.
E. M.\ was also supported by the program Excellence Initiative–Research University of the
University of Wroc\l{}aw of the Ministry of Education and Science.
D.H.R.\ is supported by the State of Hesse within the Research 
Cluster ELEMENTS (Project ID 500/10.006). This project is supported by the European 
Research Council under project ERC-2018-ADG-835105 YoctoLHC. 
This publication is part of a project that has received funding from the European Union's 
Horizon 2020 research and innovation programme under grant agreement STRONG -- 2020 - No 824093. This research was funded as a part of the CoE in Quark Matter of the Academy of Finland.

\newpage

\appendix

\section{Comparison to other works}
\label{sec:term_by_term}

In this appendix we perform a comparison of the second-order 
relaxation equations found in this paper, 
Eqs.\ \eqref{bulk_relax}, \eqref{diffusion_relax}, and \eqref{shear_relax}, 
to earlier derivations of 
Monnai and Hirano \cite{Monnai:2010qp} and Kikuchi, Tsumura, and 
Kunihiro \cite{Kikuchi:2015swa}.

Our second-order relaxation equation \eqref{bulk_relax} for the bulk viscosity contains 
8 terms in total, while Eq.\ (69) of Ref.\ \cite{Monnai:2010qp} contains 13 terms.
Here, we recall this equation noting that $u^{\mu}_E =u^{\mu} $, while $J=q$ and $K=q'$,
\begin{eqnarray}
	\Pi &=& -\zeta \nabla _\mu u^\mu_E - \tau_\Pi D \Pi 
	+ \chi_{\Pi \Pi}^c \Pi \nabla _\mu u^\mu_E
	+ \chi_{\Pi \pi} \pi_{\mu \nu} \nabla^{\langle \mu} u^{\nu \rangle}_E 
	\nonumber \\
	&+& \sum_{J,K} \chi_{\Pi V_J}^{aK} V^J_\mu \nabla ^\mu \frac{\mu_{K}}{T}
	+ \sum_J \chi_{\Pi V_J}^c V^J_\mu D u ^\mu_E 
	+ \sum_J \chi_{\Pi V_J}^d \nabla ^\mu V^J_\mu 
	\nonumber \\ 
	&+& \sum_J \chi_{\Pi \Pi}^{aJ} \Pi D\frac{\mu _J}{T} 
	+ \chi_{\Pi \Pi}^b \Pi D \frac{1}{T} 
	- \zeta_{\Pi \delta e} D\frac{1}{T}
	+ \sum _J \zeta_{\Pi \delta n_J} D\frac{\mu_J}{T} 
	+ \sum_J \chi_{\Pi V_J}^b V^J_\mu \nabla ^\mu \frac{1}{T}\; .
	\label{eq:Pi_E}
\end{eqnarray}
The difference between this equation and ours is due to the 
difference in the thermodynamic forces and the 
way the comoving derivatives and space-time four-gradients are employed.
In our derivation the comoving derivatives, $D\beta=D\frac{1}{T} $ and 
$D\alpha_q = D\frac{\mu _J}{T} $, do not appear explicitly since they were replaced using
Eqs.\ \eqref{D_beta_multi} and \eqref{D_alphaq_multi}, while the space-time four-gradient 
of the inverse temperature, $\nabla^{\mu} \beta$, is given by Eq.\ \eqref{D_u_mu_multi}.
The terms that are expressed differently are in the third-line of the above equation 
(\ref{eq:Pi_E}). Now, collecting these various terms one can 
show that Eq.\ (\ref{eq:Pi_E}) reduces to Eq.\ \eqref{bulk_relax}.

Similarly, we recall Eq.\ (77) of Ref.\ \cite{Kikuchi:2015swa}, which contains 11 terms,
\begin{align}
	\label{eq:relax1}
	\Pi	&= -\zeta\theta- \tau_\Pi \frac{\partial}{\partial\tau}\Pi 
	- \sum_{A\,=\,1}^{M} \ell^A_{\Pi J}\nabla\cdot J_A
	+ \kappa_{\Pi\Pi} \Pi \theta + \kappa_{\Pi\pi}\pi_{\rho\sigma}\sigma^{\rho\sigma} 
	\nonumber\\
	&+ b_{\Pi\Pi\Pi}\Pi^2 + \sum_{A,B\,=\,1}^{M}b_{\Pi JJ}^{AB}J_A^\rho J_{B,\rho} 
	+ b_{\Pi\pi\pi}\pi^{\rho\sigma}\pi_{\rho\sigma} \nonumber\\	
	&+\sum_{A\,=\,1}^{M}\kappa^{(1)A}_{\Pi J}J_{A,\rho}\nabla^\rho T
	+\sum_{A,B\,=\,1}^{M}\kappa^{(2)BA}_{\Pi J}J_{A,\rho}\nabla^\rho \frac{\mu_B}{T} \;,
\end{align}	
where we note that $V_A = V_q$ while $A=q$ and $B=q'$. 
From these the terms in the second line are of second order in dissipative quantities, 
i.e.\  of second order in inverse Reynolds number, originating 
from the non-linear part of the collision integral. Note that such second-order terms were 
also obtained in Refs.\ \cite{Denicol:2012cn,Molnar:2013lta}, but are neglected in our 
study. The remaining 2 terms are formally the same, which can be seen using 
Eq.\ \eqref{D_u_mu_multi}.

The relaxation equation \eqref{diffusion_relax} for the conserved charge current  contains 
12 terms, while Eq.\ (70) of Ref.\ \cite{Monnai:2010qp} listed below contains 19 terms,
\begin{eqnarray}
	V_J^\mu &=& \kappa_{V_J} \nabla ^\mu \frac{\mu _J}{T} 
	+ \sum_{K\neq J} \kappa_{V_J V_K} \nabla ^\mu \frac{\mu _K}{T}
	- \tau_{V_J} \Delta^{\mu \nu} D V^{J}_\nu 	
	- \sum_{K\neq J} \tau_{V_J V_{K}} \Delta^{\mu \nu} D V^{K}_\nu \nonumber \\
	&+& \sum_{K} \chi_{V_J V_{K}}^c V_{K}^\mu \nabla _\nu u^\nu_E 
	+  \sum_{K} \chi_{V_J V_{K}}^d V_{K}^\nu \nabla _\nu u^\mu_E 
	+ \sum_{K} \chi_{V_J V_{K}}^e V_{K}^\nu \nabla ^\mu u_\nu^E \nonumber \\
	&+& \sum_{K} \chi_{V_J \pi}^{aK} \pi^{\mu \nu} \nabla_\nu \frac{\mu_{K}}{T}
	+ \chi_{V_J \pi}^c \pi^{\mu \nu} D u_\nu^E 
	+ \chi_{V_J \pi}^d \Delta^{\mu \nu} \nabla ^\rho \pi_{\nu \rho}
	+ \sum_{K} \chi_{V_J \Pi}^{aK} \Pi \nabla ^\mu \frac{\mu_{K}}{T}
	+ \chi_{V_J \Pi}^c \Pi D u^\mu_E + \chi_{V_J \Pi}^d \nabla ^\mu \Pi \nonumber \\
	&+& \chi_{V_J \pi}^b \pi^{\mu \nu} \nabla_\nu \frac{1}{T} 
	+ \chi_{V_J \Pi}^b \Pi \nabla ^\mu \frac{1}{T}
	+\sum_{K,L} \chi_{V_J V_{K}}^{aL} V_{K}^\mu D \frac{\mu_{L}}{T}
	+ \sum_{K} \chi_{V_J V_{K}}^b V_{K}^\mu D \frac{1}{T} 
	+\kappa_{V_J W} \bigg( \frac{1}{T} D u_E^\mu + \nabla^\mu \frac{1}{T} \bigg)\;.
	\label{eq:V_E}
\end{eqnarray}
After closer inspection, we observe that the first line contains two sums that are equivalent 
to our sums over charges, 
while the additional 4 terms in the last line can be incorporated into already existing terms. 
Furthermore, the last remaining term, 
$\kappa_{V_q W}(\beta Du^{\mu} + \nabla^{\mu} \beta)$, 
may be expressed using Eq.\ \eqref{D_u_mu_multi}, and hence is fully accounted for in 
our approach.

On the other hand, Eq.\ (78) of Ref.\ \cite{Kikuchi:2015swa} contains 14 terms, of which 
the last 2 are of second order in inverse Reynolds number, while the other terms are 
formally the same,
\begin{align}
\label{eq:relax2}
J_A^\mu
&= \sum_{B\,=\,1}^{M}\lambda_{AB}\frac{T^2}{h^2} \nabla^\mu \frac{\mu_{B}}{T}
- \sum_{B\,=\,1}^{M}\tau_J^{AB} \Delta^{\mu\rho}\frac{\partial}{\partial\tau}J_{B,\rho}
- \ell^A_{J\Pi}\nabla^\mu \Pi - \ell^A_{J\pi}\Delta^{\mu\rho} \nabla_\nu {\pi^\nu}_\rho
\nonumber\\
&+ \kappa^{(1)A}_{J\Pi}\Pi\nabla^\mu T 
+ \sum_{B\,=\,1}^{M}\kappa^{(2)AB}_{J\Pi}\Pi\nabla^\mu \frac{\mu_{B}}{T}
+ \sum_{B\,=\,1}^{M}\kappa^{(1)AB}_{JJ}J^\mu_B\theta
+ \sum_{B\,=\,1}^{M}\kappa^{(2)AB}_{JJ}J_{B,\rho}\sigma^{\mu\rho}
+ \kappa^{(3)AB}_{JJ}J_{B,\rho}\omega^{\mu\rho}
\nonumber\\
&+ \kappa^{(1)A}_{J\pi}\pi^{\mu\rho}\nabla_\rho T
+ \sum_{B\,=\,1}^{M}\kappa^{(2)AB}_{J\pi}\pi^{\mu\rho}\nabla_\rho \frac{\mu_{B}}{T}
+ \sum_{B\,=\,1}^{M}b^{AB}_{J\Pi J}\Pi J_B^\mu 
+  \sum_{B\,=\,1}^{M}b^{AB}_{JJ\pi}J_{B,\rho}\pi^{\rho\mu}\; .
\end{align}
The relaxation equation \eqref{shear_relax} for the shear-stress tensor  contains 
10 terms, while Eq.\ (71) of Ref.\ \cite{Monnai:2010qp} contains 12 terms
\begin{eqnarray}
\pi^{\mu \nu} &=& 2 \eta \nabla^{\langle \mu} u^{\nu \rangle}_E 
- \tau_\pi D \pi^{\langle \mu \nu \rangle} 
+ \chi_{\pi \pi}^c \pi^{\mu \nu} \nabla _\rho u^\rho_E 
+ \chi_{\pi \pi}^d \pi^{\rho \langle \mu} \nabla _\rho u^{\nu \rangle}_E 
+\chi_{\pi \Pi} \Pi \nabla^{\langle \mu} u^{\nu \rangle}_E \nonumber \\
&+& \sum_J \chi_{\pi V_J}^c V_J^{\langle \mu} D u^{\nu \rangle}_E 
+ \sum_J \chi_{\pi V_J}^d \nabla^{\langle \mu} V_J^{\nu \rangle} 
+ \sum_{J,K} \chi_{\pi V_J}^{aJ} V_J^{\langle \mu}
\nabla^{\nu \rangle} \frac{\mu_{K}}{T}  \nonumber \\
&+& \sum_J \chi_{\pi V_J}^b V_J^{\langle \mu} \nabla^{\nu \rangle} \frac{1}{T} 
+ \sum_J \chi_{\pi \pi}^{aJ} \pi^{\mu \nu} D \frac{\mu_J}{T} 
+ \chi_{\pi \pi}^b \pi^{\mu \nu} D \frac{1}{T}\;.
\label{eq:pi_E}
\end{eqnarray}
Here the last 3 terms that are expressed using $D\beta$, 
$D\alpha_q$, and $\nabla^{\mu} \beta$, may once again be incorporated into other terms.

Finally, Eq.\ (79) of Ref.\ \cite{Kikuchi:2015swa} contains 13 terms. The last 3 are of 
second order in inverse Reynolds number, while the remaining 10 terms are formally 
similar to ours,
\begin{align} 
\label{eq:relax3}
\pi^{\mu\nu}
&= 2\eta\sigma^{\mu\nu}
- \tau_\pi \Delta^{\mu\nu\rho\sigma}\frac{\partial}{\partial\tau}\pi_{\rho\sigma}
- \sum_{A\,=\,1}^{M}\ell^A_{\pi J}\nabla^{\langle\mu} J^{\nu\rangle}_A 
+\kappa_{\pi\Pi}\Pi\sigma^{\mu\nu}
+ \sum_{A\,=\,1}^{M}\kappa^{(1)A}_{\pi J}J^{\langle\mu}_A\nabla^{\nu\rangle} T
+ \sum_{A,B\,=\,1}^{M}\kappa^{(2)BA}_{\pi J}J^{\langle\mu}_A\nabla^{\nu\rangle} 
\frac{\mu_B}{T} \nonumber\\
&+ \kappa^{(1)}_{\pi\pi}\pi^{\mu\nu}\theta
+ \kappa^{(2)}_{\pi\pi}\pi^{\lambda\langle\mu} {\sigma^{\nu\rangle}}_{\lambda}
+ \kappa^{(3)}_{\pi\pi}\pi^{\lambda\langle\mu} {\omega^{\nu\rangle}}_{\lambda}
+ b_{\pi\Pi\pi} \Pi \pi^{\mu\nu}
+ \sum_{A,B\,=\,1}^{M}b^{AB}_{\pi JJ} J^{\langle\mu}_A J^{\nu\rangle}_B
+ b_{\pi\pi\pi} \pi^{\lambda\langle\mu} {\pi^{\nu\rangle}}_{\lambda}\;.
\end{align} 

\section{Eckart frame}
\label{peculiar_flow}

In most textbooks and relevant publications the local rest frame and the fluid four-velocity 
are chosen according to Eckart \cite{Eckart:1940te}, since this choice intuitively follows the 
non-relativistic interpretation of physical quantities.
On the other hand, all our results are given relative to the local rest frame of Landau.
In this appendix, we will elaborate on the differences.

We may choose to define a different time-like normalized flow vector, $\tilde{u}^{\mu}$, 
and hence a local frame of reference different from the previously chosen local rest frame 
(the Landau frame) given by $u^\mu$. The $\tilde{u}$-frame is related to the
$u$-frame by a Lorentz transformation. If we assume that the difference between
the frame vectors is small, $\tilde{u}^\mu - u^\mu \sim \mathcal{O}(1)$,
we may write
\begin{equation}
\tilde{u}^{\mu} = u^{\mu} + w^{\mu} + \mathcal{O}(2)\; . \label{tilde_u}
\end{equation} 
Computing the normalization of $\tilde{u}^{\mu}$ up to order $\mathcal{O}(1)$,
\begin{equation}
\tilde{u}^{\mu } \tilde{u}_{\mu } =  u^{\mu} u_{\mu} + 2u^{\mu }w_{\mu} + w^{\mu }w_{\mu }  
=1+ 2u^{\mu }w_{\mu} + \mathcal{O}(2)\;, \label{tilde_u_normalization}
\end{equation}
and demanding that $\tilde{u}^\mu$ is also normalized, we
conclude that  $w^\mu$ must be orthogonal to $u^\mu$, $u^{\mu }w_{\mu}=0$.
The projection operator onto the three-space orthogonal to $\tilde{u}^{\mu}$ is
\begin{equation}
\tilde{\Delta}^{\mu \nu } \equiv g^{\mu \nu } - \tilde{u}^{\mu } \tilde{u}^{\nu } 
= \Delta^{\mu \nu }- 2u^{\left( \mu \right. }w^{\left. \nu\right) }
- w^{\mu }w^{\nu }
= \Delta^{\mu \nu }- 2u^{\left( \mu \right. }w^{\left. \nu\right) }
+\mathcal{O}(2) \;. \label{tilde_Delta_munu}
\end{equation}

The tensor decomposition of the primary fluid-dynamical quantities with respect to 
$\tilde{u}^{\mu}$ leads to results similar to the tensor 
decompositions listed in Eqs.\ \eqref{Nq_mu_i_sum} -- \eqref{T_munu_i_sum},
\begin{eqnarray}
	N^{\mu}_q &\equiv & \tilde{n}_{q} \tilde{u}^{\mu} 
	+ \tilde{V}_{q}^{\mu} 
	= \tilde{n}_{q} \left(u^{\mu } + w^{\mu } \right)
	+ \tilde{V}_{q}^{\mu}\; ,  \label{N_mu_i_gen} \\ 
	T^{\mu \nu} &\equiv& \tilde{e}\, \tilde{u}^{\mu} \tilde{u}^{\nu} 
	- ( \tilde{P} + \tilde{\Pi} ) \tilde{\Delta}^{\mu \nu}
	+ \tilde{\pi}^{\mu \nu }  
	+ 2 \tilde{W}^{\left( \mu \right.} \tilde{u}^{\left. \nu \right)}\notag \\
	&=& \tilde{e} \,u^{\mu }u^{\nu } - ( \tilde{P} + \tilde{\Pi} ) \Delta^{\mu \nu } 
	+ \tilde{\pi}^{\mu \nu } + 2( \tilde{e} + \tilde{P} ) 
	w^{\left(\mu \right. } u^{\left. \nu \right) } 
	+ 2 \tilde{W}^{\left( \mu \right. } u^{\left. \nu \right)}
	+ \mathcal{O}(2)\; , \label{T_munu_i_gen}
\end{eqnarray}%
where we have explicitly 
applied Landau's matching conditions from Eqs.\ \eqref{Landau_Matching_nq},
\eqref{Landau_Matching_e}. 
We have also made use of Eq.\ (\ref{tilde_u}) and the fact
that all dissipative quantities are $\sim \mathcal{O}(1)$.
The physical quantities follow from similar projection operations as in 
Eqs.\ (\ref{n+dn_def}) -- (\ref{pi_munu_i_sum}).

Furthermore, one can show that neglecting corrections of order $\mathcal{O}(2)$, 
the net-particle density, conserved net-charge density, energy density, equilibrium and 
bulk pressures, as well the shear-stress components are equal in both local rest frames,
\begin{equation}
\tilde{n}_q = n_q\;, \quad \tilde{e} = e\;, \quad \tilde{P} = P\;, \quad
\tilde{\Pi} = \Pi\;, \quad \tilde{\pi}^{\mu \nu} = \pi^{\mu \nu}\; . \label{Eckart_Landau}
\end{equation}
Note that the quantities without tilde are taken in the Landau frame and are not  
identical with the corresponding quantities when the frame is not yet specified.

The choice of the local rest frame changes the diffusion currents orthogonal to the flow 
velocity, i.e.\ the so-called peculiar velocities. In Eckart's definition of the local rest frame, 
for \emph{one} conserved charge $q$ it is required that
\begin{eqnarray}
N_q^{\mu} \equiv n_q \tilde{u}^{\mu } \quad \textrm{and} \quad
\tilde{V}^{\mu}_q \equiv 0\;, \label{N_mu_Eckart}
\end{eqnarray}%
where $n_{q} = N_q^{\mu} \tilde{u}_{\mu }$ and $\tilde{V}_{q}^{\mu} = N_q^{\nu} 
\tilde{\Delta}^{\mu}_{\nu}$. This means that according to the 
definition of Eckart, there is no net-charge diffusion current of charge type $q$ in its own 
local rest frame.
Now, comparing this with the form of $N^\mu_q$ in the Landau frame leads to
\begin{equation}
N_q^{\mu} \equiv n_q \left(u^{\mu } + w^{\mu } \right) 
\overset{!}{=} n_q u^{\mu } + V_{q}^{\mu} \;.
\end{equation}
Thus, for an observer in Landau's local rest frame the particles are diffusing with 
peculiar four-velocities proportional to the net-charge diffusion current,
\begin{equation}
	w^{\mu} = \frac{V_q^{\mu}}{n_q}\;. \label{w_mu_Landau}
\end{equation}
Since the dissipative quantity $V_q^\mu \sim \mathcal{O}(1)$, this is consistent with our
assumption that $w^\mu \sim \mathcal{O}(1)$.
Up to terms of order $\mathcal{O}(1)$, the energy-momentum tensor 
(\ref{T_munu_i_gen}) reads
\begin{eqnarray} \nonumber
T^{\mu \nu } &\equiv& 
e \, \tilde{u}^{\mu} \tilde{u}^{\nu} - ( P+\Pi) \tilde{\Delta}^{\mu \nu} + \pi^{\mu \nu }  
+ 2 \tilde{W}^{\left( \mu \right. } \tilde{u}^{\left. \nu \right)} \\
&=& e \, u^{\mu} u^{\nu} - ( P + \Pi ) \Delta^{\mu \nu} + \pi^{\mu \nu } 
+ 2 (e + P ) w^{\left(\mu \right. } u^{\left. \nu \right)} 
+ 2 \tilde{W}^{\left( \mu \right. }u^{\left. \nu \right)}\;. \label{T_munu_Eckart}
\end{eqnarray}%
Using this result together with Eq.\ (\ref{w_mu_Landau}) and comparing it to the 
energy-momentum tensor in the Landau frame,
\begin{align}
	T^{\mu\nu} = e \, u^\mu u^\nu - (P + \Pi) \Delta^{\mu\nu} + \pi^{\mu\nu}\;, 
\end{align}
leads to the total energy-momentum diffusion current 
$\tilde{W}^{\mu} = \Delta_{\alpha }^{\mu }T^{\alpha \beta }u_{\beta}$ in the Eckart frame, 
\begin{equation}
\tilde{W}^{\mu} \equiv - \frac{e + P}{n_q} V_q^{\mu} \equiv - h_q V_q^{\mu}\;.
\label{eq:EnergyMomAndDiff}
\end{equation}%
This relates the total energy-momentum flux seen in the Eckart frame to the 
diffusion flux of charge $q$ observed in the Landau frame. 

Finally, we give the four-flow of the charges $q'$, of which the diffusion currents were 
\emph{not} chosen to vanish by Eckart's choice \eqref{N_mu_Eckart} (i.e.\ $q' \neq q$).
Using the fact that the charge densities $n_{q'}$ are equal in both frames, we obtain
the condition
\begin{align}
	N^\mu_{q'} = n_{q'} \left( u^\mu + w^\mu \right) + \tilde{V}^\mu_{q'} 
	\overset{!}{=} n_{q'} u^\mu + V^\mu_{q'} \;.
\end{align}
Employing Eq.\ \eqref{w_mu_Landau}, we arrive at an expression for the 
diffusion currents as observed in the Eckart frame:
\begin{align}
	\tilde{V}^\mu_{q'} = V^\mu_{q'} - \frac{n_{q'}}{n_q} V^\mu_q\;.
\end{align}
Note that for the case $q = q'$, we again recover the requirement by the Eckart frame 
definition for the charge $q$, Eq.\ \eqref{N_mu_Eckart}.

\section{Transport Coefficients}
\label{app:transportcoefficients}

In this appendix, we list all second-order transport coefficients in the
equations of motion \eqref{bulk_relax}, \eqref{diffusion_relax}, and \eqref{shear_relax}.

\subsection{The coefficients in the bulk viscosity equation}
\label{App:bulk_coefficients} 

The second-order transport coefficients in the equation of motion \eqref{bulk_relax} for
the bulk viscous pressure are 
\begin{align}
	\delta_{\Pi \Pi }&
	=\sum_{s,i\,=\,1}^{N_{\text{spec}}}
	\sum_{r\,=\,0}^{N_{0}} \frac{m_{s}^{2}}{9} 
	\tau_{si,0r}^{\left( 0\right) }
	\left[m_{i}^{2}\left( r-1\right) \bar{\zeta}_{i,r-2}-\left( r+2\right) 
	\bar{\zeta}_{i,r}-3\left( J_{i,r+1,0}\mathcal{T}_{00}-\sum_{q}^{\left\{ B,Q,S\right\}}
	q_{i}J_{i,r,0}\mathcal{T}_{q0}\right) \right]  \notag \\
	&+ \sum_{s,i\,=\,1}^{N_{\text{spec}}}
	\sum_{r\,=\,0}^{N_{0}} \frac{%
		m_{s}^{2}}{3}\tau_{si,nr}^{\left( 0\right) } \sum_{q}^{\left\{ B,Q,S\right\}}
	  \frac{\partial \bar{\zeta}_{i,r}}{\partial \alpha_{q}} \left[ \mathcal{T}_{q0} 
	\left( e+P\right) +\sum_{q^{\prime }}^{\left\{ B,Q,S\right\} } 
	\mathcal{T}_{qq^{\prime }}n_{q^{\prime }}\right]   \notag \\
	&+ \sum_{s,i\,=\,1}^{N_{\text{spec}}}
	\sum_{r\,=\,0}^{N_{0}} \frac{m_{s}^{2}}{3} 
	\tau_{si,nr}^{\left( 0\right) } \frac{\partial \bar{\zeta}_{i,r}}{\partial \beta} 
	\left[ \mathcal{T}_{00}\left( e+P\right) 
	+ \sum_{q^{\prime }}^{\left\{ B,Q,S\right\} }\mathcal{T}_{0q^{\prime}} 
	n_{q^{\prime }}\right]\; , \\
	\lambda_{\Pi \pi} & = -\sum_{s,i\,=\,1}^{N_{\text{spec}}}
	\sum_{r\,=\,0}^{N_{0}}\frac{m_{s}^{2}}{3}\tau_{si,0r}^{\left( 0\right) }
	\left[ \left(r-1\right) \bar{\eta}_{i,r-2} + J_{i,r+1,0}\mathcal{T}_{00}
	- \sum_{q}^{\left\{ B,Q,S\right\} }q_{i}J_{i,r,0}\mathcal{T}_{q0}\right]\; , \\
	\ell_{\Pi V}^{(q)} & = -\sum_{s,i\,=\,1}^{N_{\text{spec}}}
	\sum_{r\,=\,0}^{N_{0}}\frac{m_{s}^{2}}{3} 
	\tau_{si,0r}^{\left( 0\right) } \left( \bar{\kappa}_{i,r-1}^{\left( q\right) } 
	+ J_{i,r+1,0}\mathcal{T}_{0q}-\sum_{q^{\prime }}^{\left\{ B,Q,S\right\} } 
	q_{i}^{\prime }J_{i,r,0} \mathcal{T}_{q^{\prime }q} \right)\; , \\
	\tau_{\Pi V}^{(q)} & =\sum_{s,i\,=\,1}^{N_{\text{spec}}}
	\sum_{r\,=\,0}^{N_{0}} \frac{m_{s}^{2}}{3} 
	\tau_{si,0r}^{\left( 0\right) } \left( r\bar{\kappa}_{i,r-1}^{\left( q\right) } 
	+ \beta \frac{\partial \bar{\kappa}_{i,r-1}^{\left( q\right) }}{\partial \beta }
	+ J_{i,r+1,0}\mathcal{T}_{0q}-\sum_{q^{\prime }}^{\left\{ B,Q,S\right\} }
	q_{i}^{\prime }J_{i,r,0} \mathcal{T}_{q^{\prime }q}\right)  \;, \\
	\lambda_{\Pi V}^{(q,q^{\prime})} & = -\sum_{s,i\,=\,1}^{N_{\text{spec}}}
	\sum_{r\,=\,0}^{N_{0}}\frac{m_{s}^{2}}{3} \tau_{si,0r}^{\left( 0\right) }
	\left( \frac{\partial \bar{\kappa}_{i,r-1}^{\left( q\right) }}{\partial \alpha_{q^{\prime}}} 
	+\frac{n_{q^{\prime }}}{e+P}\frac{\partial \bar{\kappa}_{i,r-1}^{\left(q\right) }}{\partial 
	\beta }\right) \;.
\end{align}

\subsection{The coefficients in the charge diffusion equations}
\label{App:diffusion_coefficients} 

The coefficients in the equations of motion \eqref{diffusion_relax} for the charge 
diffusion currents are:
\begin{align}
	\delta_{VV}^{(q^{\prime },q)}& = -\sum_{s,i\,=\,1}^{N_{\text{spec}}}q_{s}^{\prime}
	\sum_{r\,=\,0}^{N_{1}}\tau_{si,0r}^{\left( 1\right) } 
	\frac{1}{3}\left[ m_{i}^{2}\left( r-1\right) \bar{\kappa}_{i,r-2}^{\left( q\right)}
	-\left( r+3\right) \bar{\kappa}_{i,r}^{\left( q\right) }\right]   \notag \\
	& -\sum_{s,i\,=\,1}^{N_{\text{spec}}}q_{s}^{\prime }\sum_{r\,=\,0}^{N_{1}}
	\tau_{si,0r}^{\left( 1\right) }\sum_{q^{\prime \prime }}^{\left\{ B,Q,S\right\} }
	\frac{\partial \bar{\kappa}_{i,r}^{\left( q\right) }}{\partial \alpha_{q^{\prime \prime}}}
	\left[ \mathcal{T}_{q^{\prime \prime }0}\left(e+P\right) 
	+\sum_{q^{\prime \prime \prime }}^{\left\{ B,Q,S\right\} }
	\mathcal{T}_{q^{\prime \prime }q^{\prime \prime \prime }}n_{q^{\prime \prime
			\prime }}\right]   \notag \\
	& -\sum_{s,i\,=\,1}^{N_{\text{spec}}}q_{s}^{\prime }\sum_{r\,=\,0}^{N_{1}}
	\tau_{si,0r}^{\left( 1\right) }\frac{\partial \bar{\kappa}_{i,r}^{\left(
			q\right) }}{\partial \beta }\left[ \mathcal{T}_{00}\left( e+P\right)
	+\sum_{q^{\prime \prime }}^{\left\{ B,Q,S\right\} }\mathcal{T}_{0q^{\prime
			\prime }}n_{q^{\prime \prime }}\right] \;, \\
	\lambda_{VV}^{(q^{\prime },q)} & =-\sum_{s,i\,=\,1}^{N_{\text{spec}}}q_{s}^{\prime }
	\sum_{r\,=\,0}^{N_{1}}\tau_{si,0r}^{\left( 1\right) }\frac{1}{5} 
	\left[m_{i}^{2}\left( 2r-2\right) \bar{\kappa}_{i,r-2}^{(q)} -\left( 2r+3\right) 
	\bar{\kappa}_{i,r}^{(q)}\right] \;, \\
	\ell_{V\Pi }^{(q^{\prime })}& =\sum_{s,i\,=\,1}^{N_{\text{spec}}}q_{s}^{\prime}
	\sum_{r\,=\,0}^{N_{1}}\tau_{si,0r}^{\left( 1\right) }
	\left[ \frac{1}{3}\left( m_{i}^{2}\bar{\zeta}_{i,r-1}
	-\bar{\zeta}_{i,r+1}\right)-\alpha_{i,r}^{h}\right]\; , \\
	\ell_{V\pi }^{(q^{\prime })}& =-\sum_{s,i\,=\,1}^{N_{\text{spec}}}q_{s}^{\prime}
	\sum_{r\,=\,0}^{N_{1}}
	\tau_{si,0r}^{\left( 1\right) }\left( \bar{\eta}_{i,r-1}+\alpha_{i,r}^{h}\right) \;, 
\end{align}
\begin{align}
	\tau_{V\Pi }^{(q^{\prime })}& =\sum_{s,i\,=\,1}^{N_{\text{spec}}}q_{s}^{\prime}
	\sum_{r\,=\,0}^{N_{1}}\tau_{si,0r}^{\left( 1\right) }
	\frac{1}{3}\left[ m_{i}^{2}r\bar{\zeta}_{i,r-1}-\left( r+3\right) \bar{\zeta}_{i,r+1}
	-3\alpha_{i,r}^{h}+m_{i}^{2} \beta \frac{\partial \bar{\zeta}_{i,r-1}}{\partial \beta }
	 - \beta \frac{\partial \bar{\zeta}_{i,r+1}}{\partial \beta }\right] \;, \\
	\tau_{V\pi }^{(q^{\prime })} &= -\sum_{s,i\,=\,1}^{N_{\text{spec}}}q_{s}^{\prime}
	\sum_{r\,=\,0}^{N_{1}}
	\tau_{si,0r}^{\left( 1\right) }\left(r\bar{\eta}_{i,r-1} 
	+ \beta \frac{\partial \bar{\eta}_{i,r-1}}{\partial \beta}+\alpha_{i,r}^{h} \right)\; , \\
	\lambda_{V\Pi }^{(q^{\prime },q)}
	& =-\sum_{s,i\,=\,1}^{N_{\text{spec}}}q_{s}^{\prime }\sum_{r\,=\,0}^{N_{1}}
	\tau_{si,0r}^{\left( 1\right) }\frac{1}{3}\left[ m_{i}^{2}\left(
	\frac{\partial \bar{\zeta}_{i,r-1}}{\partial \alpha_{q}}
	+\frac{n_{q}}{e+P} \frac{\partial \bar{\zeta}_{i,r-1}}{\partial
		\beta }\right) - \frac{\partial \bar{\zeta}_{i,r+1}}{\partial \alpha_{q}}
	-\frac{n_{q}}{e+P}\frac{\partial \bar{\zeta}_{i,r+1}}{\partial \beta } \right]\; , \\
	\lambda_{V\pi}^{(q^{\prime },q)}
	& =\sum_{s,i\,=\,1}^{N_{\text{spec}}}q_{s}^{\prime }\sum_{r\,=\,0}^{N_{1}}
	\tau_{si,0r}^{\left( 1\right) }
	\left( \frac{\partial \bar{\eta}_{i,r-1}}{\partial \alpha_{q}}
	+\frac{n_{q}}{e+P}\frac{\partial \bar{\eta}_{i,r-1}}{\partial \beta }\right)\; .
\end{align}

\subsection{The coefficients in the shear-stress equation}
\label{App:shear_coefficients} 

The coefficients in the equation of motion \eqref{shear_relax}
for the shear-stress tensor are:
\begin{align}
	\delta_{\pi \pi } & =-\frac{1}{3}\sum_{s,i\,=\,1}^{N_{\text{spec}}}	
	\sum_{r\,=\,0}^{N_{2}}\tau_{si,0r}^{\left( 2\right) }\left[ m_{i}^{2}\left( r-1\right) 
	\bar{\eta}_{i,r-2}-\left( r+4\right) \bar{\eta}_{i,r}\right] 
	\notag \\
	& -\sum_{s,i\,=\,1}^{N_{\text{spec}}}
	\sum_{r\,=\,0}^{N_{2}}
	\tau_{si,0r}^{\left(2\right) }\left\{\sum_{q}^{\left\{B,Q,S\right\}}
	 \frac{\partial \bar{\eta}_{i,r}}{\partial \alpha_{q}} 
	\left[ \mathcal{T}_{q0}\left( e+P\right) 
	+ \sum_{q^{\prime }}^{\left\{B,Q,S\right\}}\mathcal{T}_{qq^{\prime }}n_{q^{\prime }}
	\right] 
	+\frac{\partial \bar{\eta}_{i,r}}{\partial \beta }
	\left[ \mathcal{T}_{00}\left(e+P\right) +\sum_{q^{\prime }}^{\left\{ B,Q,S\right\}}
	\mathcal{T}_{0q^{\prime }}n_{q^{\prime }}\right] \right\} \;, 
\end{align}
\begin{align}	
	\tau_{\pi \pi } &
	=-\frac{2}{7}\sum_{s,i\,=\,1}^{N_{\text{spec}}}
	\sum_{r\,=\,0}^{N_{2}}\tau_{si,0r}^{\left( 2\right) }\left[
	m_{i}^{2}\left( 2r-2\right) \bar{\eta}_{i,r-2}-\left( 2r+5\right) \bar{\eta}_{i,r}\right] \;, \\
	\lambda_{\pi \Pi }& =\sum_{s,i\,=\,1}^{N_{\text{spec}}}
	\sum_{r\,=\,0}^{N_{2}}\tau_{si,0r}^{\left( 2\right) }\frac{2}{15}\left[ m_{i}^{4}
	\left( r-1\right) \bar{\zeta}_{i,r-2}-\left( 2r+3\right) m_{i}^{2}\bar{\zeta}_{i,r}
	+\left(r+4\right) \bar{\zeta}_{i,r+2}\right]\; , \\
	\tau_{\pi V}^{(q)} & =-\frac{2}{5}\sum_{s,i\,=\,1}^{N_{\text{spec}}}
	\sum_{r\,=\,0}^{N_{2}}\tau_{si,0r}^{\left( 2\right) }
	\left[ m_{i}^{2}r\ \bar{\kappa}_{i,r-1}^{\left(q\right) } 
	-\left( r+5\right) \bar{\kappa}_{i,r+1}^{\left( q\right)}
	+m_{i}^{2}\beta \frac{\partial \bar{\kappa}_{i,r-1}^{\left( q\right) }}{\partial \beta }
	-\beta \frac{\partial \bar{\kappa}_{i,r+1}^{\left( q\right) }}{\partial \beta }\right]\;, \\
	\ell_{\pi V}^{(q)} &=-\frac{2}{5}\sum_{s,i\,=\,1}^{N_{\text{spec}}}
	\sum_{r\,=\,0}^{N_{2}}\tau_{si,0r}^{\left( 2\right) }
	\left(m_{i}^{2}\bar{\kappa}_{i,r-1}^{\left( q\right) } 
	- \bar{\kappa}_{i,r+1}^{\left( q\right) }\right) \;, \\
	\lambda_{\pi V}^{(q,q^{\prime })} & =-\frac{2}{5}\sum_{s,i\,=\,1}^{N_{\text{spec}}}
	\sum_{r\,=\,0}^{N_{2}}\tau_{si,0r}^{\left( 2\right) }\left[m_{i}^{2}
	\left( \frac{\partial \bar{\kappa}_{i,r-1}^{\left(q\right) }}{\partial \alpha_{q^{\prime}}} 
	+\frac{n_{q^{\prime }}}{e+P}\frac{\partial \bar{\kappa}_{i,r-1}^{\left( q\right) }}{\partial 
	\beta }\right) -
	\frac{\partial \bar{\kappa}_{i,r+1}^{\left( q\right) }}{\partial \alpha_{q^{\prime}}} 
	-\frac{n_{q^{\prime }}}{e+P} 
	\frac{\partial \bar{\kappa}_{i,r+1}^{\left( q\right) }}{\partial \beta } \right]\; .
\end{align}

\section{Comparison to a single-component fluid}

In previous works the transport coefficients in the equations of motion
were calculated in a similar manner as explained in this paper. In order to facilitate a 
comparison and for the sake of completeness we recall the 
notational convention and provide some basic relations. 

For an arbitrary function of energy,  $F(E_{i,\mathbf{k}})$, the irreducible tensors satisfy 
the following orthogonality condition \cite{deGroot_book},
\begin{align}
	\int \dd K_i F(E_{i,\mathbf{k}}) k^{\langle \mu_1}_i \cdots k^{\mu_\ell \rangle}_i 
	k^{\langle \nu_1}_i \cdots k^{\nu_n \rangle}_i  
	= \frac{\ell!\, \delta_{\ell n}}{(2\ell + 1)!!} 
	\Delta^{\mu_1 \cdots \mu_\ell \nu_1 \cdots \nu_n} \int \dd K_i F(E_{i,\mathbf{k}})
	\left( \Delta_{\alpha\beta} k^\alpha_i k^\beta_i \right)^\ell\; .
\end{align}
Therefore, for a given species $i$, any irreducible moment of tensor 
rank $\ell$ of arbitrary order $r\leq 0$ may be expressed as a linear combination 
of irreducible moments of the same tensor rank $\ell$, but with different
power of energy $n$ as
\begin{equation}
	\rho_{i,r}^{\mu_{1}\cdots \mu_{\ell }}=\sum_{n\,=\,0}^{N_{\ell }}
	\rho_{i,n}^{\mu_{1}\cdots \mu_{\ell }} \mathcal{F}_{i,-r,n}^{\left( \ell \right) } \;.
	\label{useful}
\end{equation}
where for $r,n\geq 0$, $\mathcal{F}_{-r,n}^{\left( \ell \right) }=\delta_{r n}$. 
Therefore, for  $r \rightarrow -r$, we obtain
\begin{equation}
	\rho_{i,-r}^{\mu_{1}\cdots \mu_{\ell }}=\sum_{n\,=\,0}^{N_{\ell }} 
	\rho_{i,n}^{\mu_{1}\cdots \mu_{\ell }}
	\mathcal{F}_{i,r,n}^{\left( \ell \right) }\;,
	\label{rho_negative_r}
\end{equation}
where using Eqs.\ \eqref{J_i_nm} and \eqref{eq:Hfunctions} we defined the following 
coefficient similar to Eq.\ (66) of Ref.\ \cite{Denicol:2012cn},
\begin{eqnarray} \label{F_irn} \nonumber
	\mathcal{F}_{i, \pm r,n}^{\left( \ell \right) }
	&\equiv& \frac{\ell !}{\left( 2\ell +1\right)!!}
	\int \dd K_i E_{i,\mathbf{k}}^{\mp r}\mathcal{H}_{i,\mathbf{k}n}^{\left( \ell \right)}
	\left( \Delta_{\alpha \beta }k_{i}^{\alpha }k_{i}^{\beta }\right)^{\ell} 
	f^{(0)}_{i,\mathbf{k}} \tilde{f}^{(0)}_{i,\mathbf{k}} \\ 
	&=& \sum_{n^{\prime}\,=\,n}^{N_{\ell }} \sum_{m\,=\,0}^{n^{\prime}}
	\frac{J_{i,\mp r+m +2 \ell, \ell}}{J_{i,2 \ell, \ell}} 
	a_{i,n^{\prime}n}^{\left( \ell \right)} a^{(\ell)}_{i,n^{\prime}m}\; .
\end{eqnarray}
Therefore, using these results one can also show that the expansion coefficients are 
related as
\begin{align}
	\sum_{m\,=\,0}^{n^{\prime}} \frac{J_{i,r+m +2 \ell, \ell}}{J_{i,2 \ell, \ell}} 
	a_{i,n^{\prime}n}^{\left( \ell \right)} a^{(\ell)}_{i,n^{\prime}m} 
	= \delta_{r n^{\prime}} \delta_{r n} \;, \label{eq:expansion_coeff}
\end{align}
which in the case of a single-component system is equivalent to the matrix equation 
provided in Appendix E of Ref.\ \cite{Denicol:2012cn}.

Truncating these expressions in the $(10+4N_q)$-moment approximation, hence using 
Eq.\ \eqref{rho_negative_r} with the summation limits
$N_{0}=2,\,N_{1}=1,\,N_{2}=0$ for the various tensor ranks, we obtain the
following relations,
\begin{align}
	\rho_{i,-r}& \equiv \sum_{n\,=\,0}^{N_{0}}\rho_{i,n}\mathcal{F}_{i,r,n}^{\left( 0\right) }
	\approx -\frac{3}{m_{i}^{2}}\Pi_i \mathcal{F}_{i,r,0}^{\left( 0\right)} 
	+\rho_{i,1} \mathcal{F}_{i,r,1}^{\left( 0\right)}
	+\rho_{i,2} \left(\frac{1}{m_{i}^{2}} \mathcal{F}_{i,r,0}^{\left( 0\right)} 
	+ \mathcal{F}_{i,r,2}^{\left( 0\right)} \right)\;,  \label{OMG_rho} \\
	\rho_{i,-r}^{\mu }& \equiv \sum_{n\,=\,0}^{N_{1}}\rho_{i,n}^{\mu }
	\mathcal{F}_{i,r,n}^{\left( 1\right) }
	\approx V_{i}^{\mu }\mathcal{F}_{i,r,0}^{\left(1\right)} + 
	W_{i}^{\mu }\mathcal{F}_{i,r,1}^{\left( 1\right) } \;,  \label{OMG_rho_mu} \\
	\rho_{i,-r}^{\mu \nu }& \equiv \sum_{n\,=\,0}^{N_{2}}\rho_{i,n}^{\mu \nu } 
	\mathcal{F}_{i,r,n}^{\left( 2\right) }
	\approx \pi_i^{\mu \nu }\mathcal{F}_{i,r,0}^{\left(2\right) }\;,
	\label{OMG_rho_mu_nu}
\end{align}
Furthermore, using Eqs.\ \eqref{OMG_rho} -- \eqref{OMG_rho_mu_nu} for irreducible 
moments with positive $r$ we obtain similar relations by replacing $-r \rightarrow r$ .

On the other hand, summing Eqs.\ \eqref{OMG_rho} -- \eqref{OMG_rho_mu_nu}
over species, the irreducible moments of the mixture also lead to the expressions
for a single-component fluid in the Landau frame,
\begin{align}
	\rho_{-r} &\equiv \sum_{i\,=\,1}^{N_{\text{spec}}} \rho_{i,-r} 
	= \sum_{i\,=\,1}^{N_{\text{spec}}}\sum_{n\,=\,0}^{N_{0}}\rho_{i,n}
	\mathcal{F}_{i,r,n}^{\left( 0\right) } \overset{\text{single}}{\rightarrow} -\frac{3}{m^{2}}\Pi \gamma_{r}^{(0)} 
	+\mathcal{O}(1) \;,  \label{rho_mix} \\
	\rho_{-r}^{\mu } &\equiv \sum_{i\,=\,1}^{N_{\text{spec}}} \rho_{i,-r}^{\mu } 
	=  \sum_{i\,=\,1}^{N_{\text{spec}}}\sum_{n\,=\,0}^{N_{1}}\rho_{i,n}^{\mu }
	\mathcal{F}_{i,r,n}^{\left( 1\right) } 
	\overset{\text{single}}{\rightarrow} V^{\mu } \gamma_{r}^{(1)} + \mathcal{O}(1)\;, 
	\label{rho_mix_mu} \\
	\rho_{-r}^{\mu \nu} &\equiv \sum_{i\,=\,1}^{N_{\text{spec}}}\rho_{i,-r}^{\mu \nu } 
	= \sum_{i\,=\,1}^{N_{\text{spec}}}\sum_{n\,=\,0}^{N_{2}}\rho_{i,n}^{\mu \nu} 
	\mathcal{F}_{i,r,n}^{\left( 2\right) }
	\overset{\text{single}}{\rightarrow} \pi^{\mu \nu } \gamma_{r}^{(2)}
	+\mathcal{O}(\mathrm{1})\;,
	\label{rho_mix_mu_nu}
\end{align}
where the coefficients $\gamma^{(\ell)}_r$ are listed in 
Ref.\ \cite{Denicol:2012cn} for the case of a single-component fluid, $i=N_{\text{spec}}=1$. 

Notice that moments with negative power of energy are expressed as a linear combination 
of moments with positive $r$ which represent the coupling between moments even for 
simple fluids with a single conserved charge.
On the other hand in mixtures the summations over all particle species lead to further 
couplings, which renders the above expressions rather difficult to use.
To circumvent this we have introduced the order-of-magnitude approximation in Sec.\ 
\ref{sec:OofMapprox} to express the negative moments and their sums.

Furthermore, in order to compare the irreducible moments to 
Eqs.\ \eqref{rho_in_NS} -- \eqref{rho_munu_in_NS}, we have to demand
\begin{equation}
	\Pi \bar{\zeta}_{1,-r} = \Pi \gamma_{r}^{(0)}\; , \quad
	\pi^{\mu \nu }  \bar{\eta}_{1,-r} = \pi^{\mu \nu } \gamma_{r}^{(2)}\;,
\end{equation}
and since we are dealing with a single charge, say $q$, then $V_{q}^{\mu } = q V^{\mu }$ 
and hence
\begin{equation}
	V_{q}^{\mu } \bar{\kappa}_{1,-r}^{\left( q\right)}
	= q V^{\mu } \gamma_{r}^{(1)}\;.
\end{equation}
\section{Performing the collision integrals}
\label{sec:perform_coll_int}

In order to evaluate the transport coefficients first we need to calculate the irreducible 
moments (\ref{Coll_int_mixture}) of the collision term (\ref{COLL_INT}).
These moments are related to the entries of the collision matrix defined in 
Eqs.\ (\ref{eq:CollisionMatrix_Tensor}) -- (\ref{eq:CollisionMatrixEntries}). 
For the sake of convenience we define the following tensor, 
similarly as in Ref.\ \cite{Denicol:2012cn}, 
\begin{align}
	&\mathcal{X}_{s ijab,r}^{\mu_1 \cdots \mu_\ell \nu_1 \cdots \nu_{\ell + m}} 
	\equiv \mathcal{L}_{s ijab,r}^{\mu_1 \cdots \mu_\ell \nu_1 \cdots \nu_{\ell + m}} 
	- \mathcal{G}_{s ijab,r}^{\mu_1 \cdots \mu_\ell \nu_1 \cdots \nu_{\ell + m}}\;, 
\end{align}
where the loss term is
\begin{align}  \nonumber
	\mathcal{L}_{s ijab,r}^{\mu_1 \cdots \mu_\ell \nu_1 \cdots \nu_{\ell + m}} 
	&\equiv \int \dd K_i \dd K^\prime_j f^{(0)}_{i,\mathbf{k}} f^{(0)}_{j,\mathbf{k^\prime}} 
	E^{r-1}_{i,\mathbf{k}} k^{\mu_1}_i \cdots  k^{\mu_\ell}_i \Big(  k_i^{\nu_1} \cdots 
	k_i^{\nu_{\ell+m}} \delta_{is} + {k^\prime}_j^{\nu_1} \cdots {k^\prime}_j^{\nu_{\ell + m}} 
	\delta_{js} \Big)  \\ 
	&\times \int \dd P_a \dd P^\prime_b  \, W_{ij\rightarrow ab}^{kk^{\prime}\rightarrow 
	pp^{\prime}} \, \tilde{f}^{(0)}_{a,\mathbf{p}} \tilde{f}^{(0)}_{b,\mathbf{p^\prime}} \;,
	 \label{Loss_term}
\end{align}
while the gain term is
\begin{align}  \nonumber
	\mathcal{G}_{s ijab,r}^{\mu_1 \cdots \mu_\ell \nu_1 \cdots \nu_{\ell + m}} 
	&\equiv \int \dd K_i \dd K^\prime_j f^{(0)}_{i,\mathbf{k}} f^{(0)}_{j,\mathbf{k^\prime}} 
	E^{r-1}_{i,\mathbf{k}} k^{\mu_1}_i \cdots  k^{\mu_\ell}_i \\
	&\times \int \dd P_a \dd P^\prime_b  \, W_{ij\rightarrow ab}^{kk^{\prime}\rightarrow 
	pp^{\prime}} \tilde{f}^{(0)}_{a,\mathbf{p}} \tilde{f}^{(0)}_{b,\mathbf{p^\prime}} 
	\Big( p_a^{\nu_1} \cdots p_a^{\nu_{\ell + m}} \delta_{as} + {p^\prime}_b^{\nu_1} 
	\cdots {p^\prime}_b^{\nu_{\ell + m}} \delta_{b s} \Big) \;. \label{Gain_term} 
\end{align}
Therefore, once we have evaluated the corresponding $\mathcal{X}$-tensors, the 
elements of the collision matrix $\mathcal{A}^{(\ell)}_{is,rn}$ in Eq.\ 
\eqref{eq:CollisionMatrixEntries} can be obtained by calculating the following projections:
\begin{align}
	\mathcal{A}^{(\ell)}_{is,rn} &= \frac{1}{2} \sum_{n^\prime\,=\,n}^{N_{\ell}} 
	\sum_{m\,=\,0}^{n^\prime} 
	\frac{ (-1)^\ell \, a^{(\ell)}_{s,n^\prime n} a^{(\ell)}_{s,n^\prime m}}{\ell! \,(2\ell + 1) \,
	J_{s,2\ell,\ell}} 
	\sum_{j,a,b\,=\,1}^{N_{\text{spec}}} u_{\nu_{\ell + 1}} \cdots u_{\nu_{\ell + m}} 
	\Delta^{\mu_1 \cdots \mu_\ell }_{\nu_1 \cdots \nu_\ell } 
	\left(\mathcal{X}_{s ijab,r}\right)_{\mu_1 \cdots \mu_\ell }^{\nu_1 \cdots \nu_{\ell + m}}
	\; . 
	\label{eq:CollisionMatrixContraction}
\end{align}
In order to evaluate the $\dd P_a \dd P^\prime_b$ integrals of the loss term it is useful 
to choose the 
center-of-momentum (CM) frame to perform the integration over the transition rate 
$ W_{ij \, \rightarrow \, ab}^{kk^{\prime}\rightarrow pp^{\prime}}$. 
The  total momentum involved in binary collisions,  
$P^{\mu}_T \equiv k^\mu_i + k^{\prime \mu}_j$, defines the Mandelstam variable
\begin{align}
	s &\equiv \left(k^\mu_i + k^{\prime \mu}_j\right)^2 
	= \left( p^{\mu}_a + p^{\prime \mu}_b\right)^2 \equiv P^{\mu}_T P_{T, \mu} \;. 
	\label{Mandelstam_s}
\end{align}
The CM frame is defined such that 
\begin{align}
	\sqrt{s} &\equiv k^0_i + k^{\prime 0}_j = p^0_a + p^{\prime 0}_b \equiv P^0_T, \\ 
	0 &\equiv \mathbf{k}_i + \mathbf{k}^\prime_j = \mathbf{p}_a 
	+ \mathbf{p}^{\prime}_b \equiv \mathbf{P}_T\;.
\end{align}
In the following, we use the following substitutions,
\begin{align}
	x \equiv p^0_a + p^{\prime 0}_b, \qquad \frac{\dd x}{x} 
	= \frac{|\mathbf{p}|\,\dd|\mathbf{p}|}{p^0_a p^{\prime 0}_b}\;,
\end{align}
and 
\begin{equation}
	|\mathbf{p}| = \frac{1}{2x} \sqrt{\big(x^2 - m_+\big)\big(x^2 - m_-\big)}\;,
\end{equation}
where $|\mathbf{p}| = |\mathbf{p}_a| =|\mathbf{p}^\prime_b|$, and 
$m_\pm \equiv \big(m_a \pm m_b \big)^2$. 
Therefore, the second integral in Eq.\ \eqref{Loss_term} leads to
\begin{align}
	\mathcal{P}_{ab} \equiv \int \dd P_a \dd P^\prime_b  \, 
	W_{ij\rightarrow ab}^{kk^{\prime}\rightarrow pp^{\prime}} 
	\tilde{f}^{(0)}_{a,\mathbf{p}} \tilde{f}^{(0)}_{b,\mathbf{p^\prime}} 
	= \frac{1}{2} \frac{1}{16(2\pi)^2} |\bar{\mathcal{M}}_{ij \, \rightarrow \, ab}|^2 
	\tilde{f}^{(0)}_{a,\sqrt{s}} \tilde{f}^{(0)}_{b,\sqrt{s}} 
	\sqrt{\left(1 - \frac{m_+}{s}\right) \left(1 - \frac{m_-}{s}\right)}\;, 
	\label{eq:FirstMomentumIntegral}
\end{align}
where we introduced the notation
\begin{align}
	\tilde{f}^{(0)}_{i,\sqrt{s}} \equiv 1 - \frac{a_i}{\exp\left( \beta \sqrt{ m_i^2 
	+ \frac{s}{4} \left(1 - \frac{m_+}{s} \right) \left(1 - \frac{m_-}{s} \right) } - \alpha_i \right) 
	+ a_i}\; .
\end{align}
Now using this result the remaining integral in Eq.\ \eqref{Loss_term} can be calculated.
For later use we define the angle-integrated transition probability:
\begin{align}
	\vert \bar{\mathcal{M}}_{ij\,\rightarrow\,ab}(\sqrt{s})\vert^2 
	\equiv \int_{S^2} \dd\Omega \, 
	\vert \mathcal{M}_{ij\,\rightarrow\,ab}(\sqrt{s},\Omega) \vert^2
	= 2\pi \int_{0}^{\pi} \dd\vartheta \, \sin\vartheta 
	\vert \mathcal{M}_{ij\,\rightarrow\,ab}(\sqrt{s},\vartheta)\vert^2\;,
\end{align}
where $\vartheta$ is the scattering angle in the reaction plane defined as
\begin{equation}
\cos \vartheta  = \frac{ \left(k^{\mu} - k^{\prime \mu}\right) \left(p_{\mu} 
- p^{\prime}_{\mu}\right)}{ \left(k^{\mu} - k^{\prime \mu}\right)^2} \;.
\end{equation}

The integral in Eq.\ \eqref{Gain_term} is more tedious. 
Here, we restrict ourselves to isotropic scattering processes, hence all 
integrals only depend on the normalized total momentum $\tilde{P}^\mu_T \equiv 
\left( k^\mu_i + {k^\prime}^\mu_j \right)/\sqrt{s}$. 
We also introduce the corresponding projection operator 
$\Delta^{\mu\nu}_P = g^{\mu\nu} - \tilde{P}^\mu_T \tilde{P}^\nu_T$ orthogonal to the 
total momentum, $\Delta^{\mu\nu}_P  \tilde{P}_{T,\nu} = 0$. Therefore, similarly to the 
thermodynamic integrals in Eq.\ \eqref{I_i_r_mu1_mun_exp}, we decompose the integrals 
in terms of the normalized total momentum and the associated orthogonal projection 
operator,  
\begin{align}
	\Theta^{\nu_1 \cdots \nu_n}_{a} &\equiv \int \dd P_a \dd P^\prime_b  \,
	W_{ij\rightarrow ab}^{kk^{\prime}\rightarrow pp^{\prime}}
	\tilde{f}^{(0)}_{a,\mathbf{p}} \tilde{f}^{(0)}_{b,\mathbf{p^\prime}} \, p^{\nu_1}_a 
	\cdots p^{\nu_n}_a \nonumber \\
	&= \sum_{m\,=\,0}^{[n/2]} (-1)^m  \frac{2^{-m} n!}{m!(n-2m)!}  \Delta^{(\nu_1\nu_2}_P 
	\cdots \Delta^{\nu_{2m-1}\nu_{2m}}_P \tilde{P}^{\nu_{2m+1}}_T \cdots 
	\tilde{P}^{\nu_n)}_T  C_{a,nm}\; ,
\end{align}
where the coefficients are
\begin{align}
	C_{a,nm} &\equiv \frac{(-1)^m}{(2m+1)!!} \int \dd P_a \dd P^\prime_b  \,
	W_{ij\rightarrow ab}^{kk^{\prime}\rightarrow pp^{\prime}} \tilde{f}^{(0)}_{a,\mathbf{p}} 
	\tilde{f}^{(0)}_{b,\mathbf{p^\prime}} \, \left( \Delta_{P,\,\alpha\beta}\,
	p^\alpha_a p^\beta_a 
	\right)^m \left( \tilde{P}_{T,\,\alpha}\,p^\alpha_a \right)^{n-2m} 
	\label{eq:MomentumDecompositionCoefficients} \nonumber \\
	&= \frac{(-1)^m}{(2m+1)!!} \,\frac{1}{16(2\pi)^2} 
	|\bar{\mathcal{M}}_{ij \, \rightarrow \, ab}|^2 \tilde{f}^{(0)}_{a,\sqrt{s}} 
	\tilde{f}^{(0)}_{b,\sqrt{s}} \, \frac{s^{n/2}}{2^{n+1}}  
	 \left[\left(1 - \frac{m_+}{s}\right)\left(1 - \frac{m_-}{s}\right) \right]^\frac{2m+1}{2} 
	 \nonumber \\
	&\qquad \times \left[ \frac{4m_a^2}{s} + \left(1 - \frac{m_+}{s}\right)
	\left(1 - \frac{m_-}{s}\right) \right]^{\frac{n - 2m}{2}} \; . 
\end{align}


\section{Classical, ultrarelativistic system with hard-sphere interactions in 
$(10+4N_q)$-moment approximation}
\label{sec:ultrarelativistic_case}

In this appendix we evaluate the transport coefficients of the theory in the 
$(10+4N_q)$-moment approximation ($N_0 = 2$, $N_1 = 1$, and $N_2 = 0$) for a 
classical ($a_i \rightarrow 0$), ultrarelativistic ($m_i/T \rightarrow 0$) multicomponent gas
with elastic hard-sphere interactions, for which the transition rate is given by 
Eq.\ \eqref{eq:trans_isotrop_elastic}.
Equivalently, in Eqs.\ \eqref{eq:FirstMomentumIntegral} and 
\eqref{eq:MomentumDecompositionCoefficients} we can just replace
\begin{align}
	\frac{1}{16(2\pi)^2} |\bar{\mathcal{M}}_{ij \, \rightarrow \, ab}|^2 
	= \left( \delta_{ia} \delta_{jb} + \delta_{ib}\delta_{ja}\right) s \sigma_{\text{tot}}\; ,
\end{align}
with $\sigma_{\text{tot},ij} \equiv \sigma_{\text{tot}} = \text{const}$. 
With this replacement, we obtain for Eqs.\ \eqref{eq:FirstMomentumIntegral} and 
\eqref{eq:MomentumDecompositionCoefficients}
\begin{align}
	\mathcal{P}_{ab} &= \frac{1}{2} \Big( \delta_{ia} \delta_{jb} 
	+ \delta_{ib} \delta_{ja} \Big) 
	s \, \sigma_{\text{tot}}\;, \\
	C_{a,nm} &= \frac{1}{2} \Big( \delta_{ia} \delta_{jb} + \delta_{ib} \delta_{ja} \Big) 
	\frac{(-1)^m}{(2m+1)!!}  \frac{s^{(n+2)/2}}{2^{n}}  \sigma_{\text{tot}} \;.
\end{align}
Furthermore, according to Eqs.\ (\ref{equilibirum_neP}), 
$P = \sum\limits_{i\,=\,1}^{N_{\text{spec}}} P_i$ is the total pressure of the system, and it 
fulfills the ideal gas laws $e = 3 P$ and $n_i = \beta P_i$.

\subsection{Collision matrix and its inverse}

In the following we evaluate the collision matrix \eqref{eq:CollisionMatrixContraction} 
for the vector and tensor moments. The scalar moments play no role since their transport 
coefficients are proportional to mass, and therefore vanish in the ultrarelativistic case.
In these calculations we make use of the ultrarelativistic limit of Eq.\ (\ref{Mandelstam_s}), 
leading to $s = 2 k_{i,\mu} {k^\prime}^{\mu}_j$. 

Furthermore, in the classical limit, we express the thermodynamic integrals in terms of 
the partial pressure of the respective species,
\begin{align}
	J_{i,nm} = I_{i,nm} = \frac{1}{2} \frac{(n+1)!}{(2m+1)!!} \frac{1}{\beta^{n-2}} P_i \;,
\end{align}
where the thermodynamic integrals are related by
\begin{equation}
	I_{i,n+2,m}=m_{i}^{2}I_{i,nm}+\left( 2m+3\right) I_{i,n+2,m+1}\;.
\end{equation}
Following Eq.\ \eqref{eq:expansion_coeff} the relevant expansion coefficients in the 
$(10+4N_q)$-moment approximation are:
\begin{align}
	a^{(\ell)}_{i,00} &= 1, \quad \forall \, \ell \geq 0\;, \\ 
	a^{(1)}_{i,11} &= \frac{\beta}{2}\;, \quad 
	a^{(1)}_{i,10} = -\frac{4}{\beta} a^{(1)}_{i,11} = -2\; .
\end{align}

\paragraph{Tensor moments ($\ell = 2$):}

The only relevant contraction of the $\mathcal{X}$-tensor reads
\begin{align}
	\sum_{j,a,b\,=\,1}^{N_{\text{spec}}} \Delta_{\mu_1 \mu_2 \nu_1 \nu_2} 
	\mathcal{X}^{\mu_1\mu_2\nu_1\nu_2}_{s ijab,r} = \frac{4}{9} \sigma_{\text{tot}} 
	\sum_{j\,=\,1}^{N_{\text{spec}}} \left[ \big( 2\delta_{is} - \delta_{js} \big)  
	I_{i,r+4,0} I_{j,10} + \frac{2}{3} \big( \delta_{is} + \delta_{js} \big) I_{i,r+3,0} I_{j,20}  
	\right]\;,
\end{align}
and therefore the elements of the collision matrix (\ref{eq:CollisionMatrixContraction})
for the tensor moments read in $(10+4N_q)$-moment approximation 
(note that $N_2 = 0$)
\begin{align}
	\mathcal{A}^{(2)}_{is,00} &= \frac{1}{2} \frac{ a^{(2)}_{s,00} a^{(2)}_{s,00}}{10 \,
	I_{s,42}} 
	\sum_{j,a,b\,=\,1}^{N_{\text{spec}}} \Delta_{\mu_1 \mu_2 \nu_1 \nu_2} 
	\mathcal{X}^{\mu_1\mu_2\nu_1\nu_2}_{s ijab,0} \nonumber \\
	&= \frac{\sigma_{\text{tot}}}{45 I_{s,42}} \sum_{j\,=\,1}^{N_{\text{spec}}} 
	\left[ \big( 2\delta_{is} - \delta_{js} \big)  I_{i,40} I_{j,10} + \frac{2}{3} \big( \delta_{is} 
	+ \delta_{js} \big) I_{i,30} I_{j,20}  \right]\; .
\end{align}
Expressing this in terms of the pressure, we obtain
\begin{align}
	\mathcal{A}^{(2)}_{is,00} =  \frac{\sigma_{\text{tot}} \beta}{5} 
	\sum_{j\,=\,1}^{N_{\text{spec}}} \frac{P_i P_j}{P_s} 
	\big( 4\delta_{is} - \delta_{js} \big)  
	= \frac{\sigma_{\text{tot}} \beta}{5} \left( 4 \delta_{is} P - P_i \right)\;, 
	\label{eq:CollMatrix14MomShearUltra}
\end{align}
and in the single-component limit ($N_{\text{spec}} \rightarrow 1$), where 
$P_i \equiv P_0 = n_0/\beta$, we reproduce the result from Ref.\ \cite{Denicol:2012cn}:
\begin{align}
\mathcal{A}^{(2)}_{11,00} \rightarrow \mathcal{A}^{(2)}_{00} 
= \frac{3}{5} \sigma_{\text{tot}} n_0 \;.
\end{align}
Here, Eq.\ (\ref{eq:CollMatrix14MomShearUltra}) defines the entries of an 
$N_{\text{spec}}$-dimensional rectangular, regular matrix. The elements of its inverse are:
\begin{align}
	\tau^{(2)}_{si,00} = \frac{5}{\sigma_{\text{tot}} \beta} \frac{1}{12 P^2} 
	\left( 3 \delta_{si} P + P_s \right) \;.
\end{align}

\paragraph{Vector moments ($\ell = 1$):}

The relevant contractions of the $\mathcal{X}$-tensor for the vector moments are
\begin{align}
	\sum_{j,a,b\,=\,1}^{N_{\text{spec}}} \Delta_{\mu_1 \nu_1} 
	\mathcal{X}_{s ijab,r}^{\mu_1 \nu_1} = & \;\sigma_{\text{tot}} 
	\sum_{j\,=\,1}^{N_{\text{spec}}} \left( \delta_{js} - \delta_{is} \right) 
	\Bigg( \frac{1}{3} I_{i,r+1,0} I_{j,20} +  I_{i,r+2,0} I_{j,10} \Bigg)\;, \\
	\sum_{j,a,b\,=\,1}^{N_{\text{spec}}} \Delta_{\mu_1 \nu_1} u_{\nu_2} 
	\mathcal{X}_{s ijab,r}^{\mu_1 \nu_1 \nu_2} = & \; \sigma_{\text{tot}} 
	\sum_{j\,=\,1}^{N_{\text{spec}}} \left[ \frac{2}{9} 
	\left( 2\delta_{js} - \delta_{is} \right) I_{i,r+1,0} I_{j,30} \right. \nonumber \\
	& \left. + \frac{4}{9} \left( \delta_{js} + \delta_{is} \right) I_{i,r+2,0} I_{j,20} 
	+ \frac{6}{9} \left( \delta_{js} - 2\delta_{is} \right) I_{i,r+3,0} I_{j,10} \right] \;,
\end{align}
thus the elements of the collision matrix (\ref{eq:CollisionMatrixContraction}) 
read in the $(10+4N_q)$-moment approximation (where $N_1 = 1$):
\begin{align}
	\mathcal{A}^{(1)}_{is,r0} &= -\frac{1}{2} \sum_{n^\prime\,=\,0}^{1} 
	\sum_{m\,=\,0}^{n^\prime} \frac{  a^{(1)}_{s,n^\prime0} a^{(1)}_{s,n^\prime m}}{3 \,
	I_{s,21}}  \sum_{j,a,b\,=\,1}^{N_{\text{spec}}} u_{\nu_{2}} \cdots u_{\nu_{1 + m}} 
	\Delta^{\mu_1 \nu_1} \mathcal{X}_{s ijab,r}^{\mu_1 \nu_1 \hdots \nu_{1 + m}} 
	\nonumber \\
	&= -\frac{\sigma_{\text{tot}}}{6 \, I_{s,21}}\sum_{j\,=\,1}^{N_{\text{spec}}} 
	\Bigg\{ \left( a^{(1)}_{s,00} a^{(1)}_{s,00} + a^{(1)}_{s,10} a^{(1)}_{s,10} \right) 
	\left( \delta_{js} - \delta_{is} \right) \Bigg( \frac{1}{3} I_{i,r+1,0} I_{j,20} 
	+  I_{i,r+2,0} I_{j,10} \Bigg) \nonumber \\ 
	& + a^{(1)}_{s,10} a^{(1)}_{s,11}  \left[ \frac{2}{9} \left( 2\delta_{js} - \delta_{is} 
	\right) I_{i,r+1,0} I_{j,30} + \frac{4}{9} \left( \delta_{js} + \delta_{is} \right) 
	I_{i,r+2,0} I_{j,20} + \frac{6}{9} \left( \delta_{js} - 2\delta_{is} \right) I_{i,r+3,0} I_{j,10} 
	\right] \Bigg\}\;, 
\end{align}
and
\begin{align}
	\mathcal{A}^{(1)}_{is,r1} = &\;-\frac{1}{2} \sum_{m\,=\,0}^{1} \frac{  a^{(1)}_{s,11} 
	a^{(1)}_{s,1m}}{3 \,I_{s,21}} \sum_{j,a,b\,=\,1}^{N_{\text{spec}}} u_{\nu_{2}} \cdots 
	u_{\nu_{1 + m}} \Delta^{\mu_1 \nu_1} 
	\mathcal{X}_{s ijab,m}^{\mu_1 \nu_1 \hdots \nu_{1 + m}} \nonumber \\
	= & \; -\frac{\sigma_{\text{tot}}}{6 \, I_{s,21}}\sum_{j\,=\,1}^{N_{\text{spec}}} 
	\Bigg\{ a^{(1)}_{s,11} a^{(1)}_{s,10} \left( \delta_{js} - \delta_{is} \right) 
	\Bigg( \frac{1}{3} I_{i,r+1,0} I_{j,20} +  I_{i,r+2,0} I_{j,10} \Bigg) \nonumber \\ 
	& + a^{(1)}_{s,11} a^{(1)}_{s,11}  \left[ \frac{2}{9} \left( 2\delta_{js} - \delta_{is} \right) 
	I_{i,r+1,0} I_{j,30} + \frac{4}{9} \left( \delta_{js} + \delta_{is} \right) I_{i,r+2,0} I_{j,20} 
	+ \frac{6}{9} \left( \delta_{js} - 2\delta_{is} \right) I_{i,r+3,0} I_{j,10} \right] \Bigg\}\; .
\end{align}
After some calculations, the relevant matrix elements read in terms of pressure:
\begin{align}
	\mathcal{A}^{(1)}_{is,00} &= -\frac{4}{9} \sigma_{\text{tot}} \beta 
	\sum_{j\,=\,1}^{N_{\text{spec}}} \frac{P_i P_j}{P_s} \left( \delta_{js} - 2\delta_{is} \right) 
	= \frac{4}{9} \sigma_{\text{tot}} \beta \left(2\delta_{is}P - P_i \right)\;, 
	\label{eq:CollMatrix14MomDiffUltra1} \\ 
	\mathcal{A}^{(1)}_{is,10} &= -\frac{1}{2} \sigma_{\text{tot}} 
	\sum_{j\,=\,1}^{N_{\text{spec}}} \frac{P_i P_j}{P_s} 
	\left( \delta_{js} - \delta_{is} \right) 
	= \frac{1}{2} \sigma_{\text{tot}} \left(\delta_{is}P - P_i \right)\;, 
	\label{eq:CollMatrix14MomDiffUltra2} \\
	\mathcal{A}^{(1)}_{is,01} &= -\frac{1}{18} \sigma_{\text{tot}} \beta^2 
	\sum_{j\,=\,1}^{N_{\text{spec}}} 
	\frac{P_i P_j}{P_s} \left( \delta_{js} + \delta_{is} \right) 
	= -\frac{1}{18} \sigma_{\text{tot}} \beta^2 \left(\delta_{is}P + P_i \right)\;, 
	\label{eq:CollMatrix14MomDiffUltra3} \\ 
	\mathcal{A}^{(1)}_{is,11} &= -\frac{1}{2} \sigma_{\text{tot}} \beta 
	\sum_{j\,=\,1}^{N_{\text{spec}}} \frac{P_i P_j}{P_s} \left( \delta_{js} - \delta_{is} \right) 
	= \frac{1}{2} \sigma_{\text{tot}} \beta \left(\delta_{is}P - P_i \right)  \;. 
	\label{eq:CollMatrix14MomDiffUltra4} 
\end{align}
Equations \eqref{eq:CollMatrix14MomDiffUltra1} -- \eqref{eq:CollMatrix14MomDiffUltra4} 
are the elements of the four $N_{\text{spec}}$-dimensional rectangular block matrices 
of the $(2N_{\text{spec}} \times 2N_{\text{spec}})$-matrix $\mathcal{A}^{(1)}$. 
Its single-component limit ($N_{\text{spec}} = 1$) is consistent with 
Ref.\ \cite{Denicol:2012cn}:
\begin{align}
	\mathcal{A}^{(1)} \equiv \begin{pmatrix}
		\mathcal{A}^{(1)}_{11,00} & \mathcal{A}^{(1)}_{11,01} \\ 
		\mathcal{A}^{(1)}_{11,10} & \mathcal{A}^{(1)}_{11,11}
	\end{pmatrix} = \frac{1}{9} \sigma_{\text{tot}} \beta P_0 \begin{pmatrix}
		4 & - \beta \\ 0 & 0 
	\end{pmatrix} \;. \label{eq:CollMatrixSingleCompDiff}
\end{align}
We observe that the $\mathcal{A}^{(1)}$ matrix is singular even in the single-component 
limit. This is due to the momentum-conservation equation.
In order to construct the inverse matrix $\tau^{(1)}$, we follow the steps presented in 
Appendix \ref{sec:InverseCollisionMatrix} by introducing the reduced matrix 
$\tilde{\mathcal{A}}^{(1)}$ (therefore effectively removing the irreducible moment 
$\rho^\mu_{1,1}$), inverting it, and adding zero elements corresponding to the originally 
removed rows and columns to that inverse, yielding $\tau^{(1)}$. 
We illustrate this procedure in the single-component limit. The reduced matrix and its 
inverse then just consist of one entry,
\begin{align}
	\tilde{\mathcal{A}}^{(1)} = \frac{4}{9}  \sigma_{\text{tot}} \beta P_0\;, 
	\quad \text{and} \quad \tilde{\mathcal{\tau}}^{(1)} 
	= \frac{9}{4}  \frac{1}{\sigma_{\text{tot}} \beta P_0}\;.
\end{align}
Adding zero elements yields the final inverse
\begin{align}
	\mathcal{\tau}^{(1)} = \frac{9}{4}  \frac{1}{\sigma_{\text{tot}} \beta P_0} 
	\begin{pmatrix} 1 & 0 \\ 0 & 0 \end{pmatrix} \;.
\end{align}
In the multicomponent case, we find the following entries of the 
$2N_{\text{spec}}$-dimensional rectangular inverse matrix:
\begin{align}
\tau^{(1)}_{si,00} &= \frac{9}{68 \beta \sigma_{\text{tot}} P^2} ( 8 \delta_{si} P + 9 P_s )\;,	
\\ 
\tau^{(1)}_{si,01} &= \frac{2}{17 \sigma_{\text{tot}} P} 
( \delta_{si} - \delta_{s N_{\text{spec}}} )\;, \\
\tau^{(1)}_{si,10} &= \frac{18}{17 \sigma_{\text{tot}} \beta^2 P^2} 
(1 - \delta_{s N_{\text{spec}}}) \left( P_s - \delta_{si} P \right)\;, \\ 
\tau^{(1)}_{si,11} &= \frac{32}{17 \sigma_{\text{tot}} \beta P} \delta_{si} 
(1 - \delta_{s N_{\text{spec}}}) \;.
\end{align}
We note that the elements $\tau^{(1)}_{si,nr}$ are indeed constructed in a way that 
they vanish in the cases $n = 1$ and $s = N_{\text{spec}}$, or 
$r = 1$ and $i = N_{\text{spec}}$ (i.e.\ the row and column which was originally removed 
from $\mathcal{A}^{(1)}$). We remind the reader that adding these zeros simplifies our 
notation in this work.

\subsection{Transport coefficients}

Now that the collision matrix has been determined, we can proceed to 
calculate the transport coefficients of the theory. We remark that the scalar moments have 
not been discussed since the bulk viscous pressure vanishes identically in the 
ultrarelativistic (massless) case, $\Pi \equiv \sum\limits_{s\,=\,1}^{N_{\text{spec}}} 
\frac{m_s^2}{3} \rho_{s,0} = 0$. The coefficients in the $(10+4N_q)$-moment 
approximation ($N_0 = 2$, $N_1 = 1$, 
$N_2 = 0$) have been defined in Sec.\ \ref{sec:EoM14mom} and 
Appendix \ref{app:transportcoefficients}. We evaluate them in the ultrarelativistic scenario, 
where all mass terms vanish. Since the transport coefficients are defined via the 
coefficients listed in Eqs.\ \eqref{eq:BaseCoeffScalar} -- \eqref{eq:BaseCoeffTensor}, these 
have to be evaluated first. For this, we introduce short-hand notations for the charge 
concentration of type $q$, and the concentration of the charge combination 
$q q^\prime$ in the system, respectively: 
\begin{align}
	\mathbf{c}_q \equiv \sum_{j\,=\,1}^{N_{\text{spec}}} q_j \frac{P_j}{P}\;, \quad 
	\text{and} \quad \mathbf{c}_{qq^\prime} 
	\equiv \sum_{j\,=\,1}^{N_{\text{spec}}} q_j q^\prime_j \frac{P_j}{P} \;.
\end{align}
Further, the derivatives in temperature and chemical potential of the weighted partial 
pressures of a classical gas read
\begin{align}
	\frac{\partial}{\partial \beta} \left( \frac{P_i}{P} \right) = 0 \quad 
	\text{and} \quad \frac{\partial}{\partial \alpha_q} \left( \frac{P_i}{P} \right) 
	= \frac{P_i}{P} \left( q_i - \mathbf{c}_q \right),
\end{align}
and from this the derivatives in the charge concentration follow
\begin{align}
	\frac{\partial \mathbf{c}_q}{\partial \beta} = 0 \quad \text{and} 
	\quad \frac{\partial \mathbf{c}_q}{\partial \alpha_{q^\prime}} 
	= \mathbf{c}_{qq^\prime} - \mathbf{c}_q \mathbf{c}_{q^\prime}\;.
\end{align}
respectively.
The relevant expressions for the vector and tensor moments are then obtained from 
Eqs.\ \eqref{eq:BaseCoeffScalar} -- \eqref{eq:BaseCoeffTensor} as:
\begin{align}
	\eta_{s,0} &\equiv \sum_{i\,=\,1}^{N_{\text{spec}}}\sum_{r\,=\,0}^{N_{2}} 
	\tau_{si,0r}^{\left( 2\right)}\alpha_{s,r}^{\left( 2\right) } 
	= \frac{4}{3} \frac{P_s}{ P} \frac{1}{\sigma_{\text{tot}} \beta}\;, \\
	\kappa_{s,0,q} &\equiv \sum_{i\,=\,1}^{N_{\text{spec}}}\sum_{r\,=\,0}^{N_{1}} 
	\tau_{si,0r}^{\left( 1\right) }\alpha_{i,r,q}^{\left( 1\right) } 
	= \frac{8}{17\, \sigma_{\text{tot}}} \frac{P_s}{P} \left( q_s  - \frac{77}{128} 
	\mathbf{c}_q \right)\;, \\
	\kappa_{s,1,q} &\equiv \sum_{i\,=\,1}^{N_{\text{spec}}}\sum_{r\,=\,0}^{N_{1}} 
	\tau_{si,1r}^{\left( 1\right) }\alpha_{i,r,q}^{\left( 1\right) } 
	= \frac{26}{17 \sigma_{\text{tot}} \beta} (1 - \delta_{s N_{\text{spec}}}) 
	\frac{P_s}{P} \left( q_s - \mathbf{c}_q \right) \;.
\end{align}
From this, the shear viscosity and the diffusion-coefficient matrix immediately follow:
\begin{align}
	\eta &\equiv \sum_{s\,=\,1}^{N_{\text{spec}}} \eta_{s,0} 
	= \frac{4}{3} \frac{1}{\sigma_{\text{tot}} \beta}\;, \\
	\kappa_{qq^\prime} &\equiv \sum_{s\,=\,1}^{N_{\text{spec}}}q_{s} 
	\kappa_{s,0,q^{\prime }} = \frac{8}{17\, \sigma_{\text{tot}}} 
	\left( \mathbf{c}_{qq^\prime} 
	 - \frac{77}{128} \mathbf{c}_{q} \mathbf{c}_{q^\prime} \right) \;.
\end{align}
It should be noted that the derivative of the above diffusion coefficients in inverse 
temperature, and therefore also from its inverse, vanishes,
\begin{align}
	\frac{\partial}{\partial\beta} \kappa_{qq^\prime} = 0 \quad 
	\text{and} \quad \frac{\partial}{\partial\beta} \left(\kappa^{-1}\right)_{qq^\prime} = 0\; .
\end{align} 
We note that the result for the shear viscosity is the same as in Ref.\ \cite{Denicol:2012cn}, 
while the diffusion coefficients are different since they depend on the various 
charges in the system, which are not taken into account in a single-component approach. 
However, in the limit where there is only one conserved particle species 
in the system (i.e.\ $N_q = 1$ and $q_s = 1$ and therefore $\mathbf{c}_q = 
\mathbf{c}_{qq^\prime} = 1$), the obtained expression is equivalent to the value derived in 
Ref.\ \cite{Denicol:2012cn}, $\kappa = 3 /( 16 \sigma_{\text{tot}} )$. We also remark 
that from the above equation we read off that the diffusion-coefficient matrix is symmetric, 
$\kappa_{qq^\prime} = \kappa_{q^\prime q}$ as shown by the Onsager reciprocal relations\cite{Onsager:1931jfa,Onsager:1931kxm}. The relevant weighted base coefficients then 
read
\begin{align}
	\bar{\eta}_{i,0} &\equiv \frac{\eta_{i,0}}{\eta} = \frac{P_i}{P}\;, \\ 
	\bar{\eta}_{i,-1} &\equiv \mathcal{F}^{(2)}_{i,1,0} \bar{\eta}_{i,0} 
	= \frac{\beta}{5} \frac{P_i}{P}\;, \\
	\bar{\kappa}^{(q)}_{i,0} &\equiv \sum_{q^{\prime}}^{\left\{ B,Q,S\right\} } 
	\kappa_{i,0,q^\prime} \left( \kappa^{-1} \right)_{q^\prime q} 
	= \frac{8}{17 \sigma_{\text{tot}}} \sum_{q^{\prime}}^{\left\{ B,Q,S\right\} } 
	\left( \kappa^{-1} \right)_{q^\prime q} \frac{P_i}{P} \left( q^\prime_i 
	- \frac{77}{128} \mathbf{c}_{q^\prime} \right)\;, \\
	\bar{\kappa}^{(q)}_{i,1} &= \frac{26}{17 \beta \sigma_{\text{tot}}} 
	\sum_{q^{\prime}}^{\left\{ B,Q,S\right\} } \left( \kappa^{-1} \right)_{q^\prime q} 
	\frac{P_i}{P} \left( q^\prime_i - \mathbf{c}_{q^\prime} \right) \left( 1 - 
	\delta_{iN_{\text{spec}}} \right)\;. 
\end{align} 
For the relaxation times defined in Eqs.\ \eqref{eq:relaxtime_diff} and 
\eqref{eq:relaxtime_shear} we get:
\begin{align}
	\tau_{\pi } &= \frac{5}{3} \frac{1}{\sigma_{\text{tot}} \beta P}\;, \\
	\tau_{q^\prime q} &= \frac{9}{68 \beta \sigma^2_{\text{tot}} P } 
	\sum_{q^{\prime\prime}}^{\left\{ B,Q,S\right\} } 
	\left( \kappa^{-1}\right)_{q^{\prime\prime} q} \left( \frac{784}{153} 
	\mathbf{c}_{q^{\prime\prime}q^\prime} 
	- \frac{4741}{2448}\mathbf{c}_{q^{\prime\prime}} \mathbf{c}_{q^\prime} \right) 
	\overset{\text{single}}{\rightarrow} \frac{9}{4} \frac{1}{\beta \sigma_{\text{tot}} P }\;.
\end{align}
We again note that also the relaxation-time matrix is symmetric, 
$\tau_{qq^\prime} = \tau_{q^\prime q}$. It is further apparent that the shear relaxation time 
is equivalent to the value derived in Ref.\ \cite{Denicol:2012cn}. In the case where only 
one conserved particle species is present (see above), such an equivalence is also 
recovered for the diffusive relaxation time, $\tau_q = 27/(64 \beta \sigma_{\text{tot}}^2 
P \kappa) = 9/(4 \beta \sigma_{\text{tot}} P)$, as indicated with the notation 
``$\overset{\text{single}}{\rightarrow}$''. In the following we will keep this notation, and 
continue with the second-order coefficients. For the coefficients
in the shear-stress tensor equation, defined in Appendix 
\ref{App:shear_coefficients}, we get:
\begin{align}
	\delta_{\pi \pi } &= \frac{4}{3} \tau_\pi\;, \\
	\tau_{\pi\pi} &= \frac{10}{7} \tau_\pi \;, \\
	\ell_{\pi V}^{(q)} &= \frac{52}{51} \frac{1}{\beta^2 \sigma_{\text{tot}}^2 P} 
	\sum_{q^\prime}^{\lbrace B, Q, S\rbrace} (\kappa^{-1})_{q^\prime q} 
	\left( \mathbf{c}_{q^\prime} - q^\prime_{N_{\text{spec}}} \right) 
	\frac{P_{N_{\text{spec}}}}{P} \overset{\text{single}}{\rightarrow} 0 \;, \\
	\tau_{\pi V}^{(q)} &= 4\beta \frac{\partial}{\partial \beta} \ell^{(q)}_{\pi V} 
	- 7 \ell^{(q)}_{\pi V}  \overset{\text{single}}{\rightarrow} 0\;,  \\
	\lambda_{\pi V}^{(q,q^{\prime })} &= \frac{\partial}{\partial\alpha_{q^\prime}} 
	\ell^{(q)}_{\pi V} + \frac{\beta}{4} \mathbf{c}_{q^\prime} \frac{\partial}{\partial \beta} 
	\ell^{(q)}_{\pi V} - 2 \mathbf{c}_{q^\prime} \ell^{(q)}_{\pi V} \overset{\text{single}}
	{\rightarrow} 0\;. 
\end{align}
For the coefficients in the vector equations of motion (see Appendix
\ref{App:diffusion_coefficients}) we derive:
\begin{align}
	\delta_{VV}^{(q^{\prime },q)} = & \; \frac{1}{\beta \sigma^2_{\text{tot}} P} 
	\sum_{q^{\prime\prime}}^{\lbrace B,Q,S\rbrace} 
	\left(\kappa^{-1}\right)_{q^{\prime\prime}q} \left[ 
	\frac{640}{867} \mathbf{c}_{q^\prime q^{\prime\prime}} 
	- \frac{17551}{55488} \mathbf{c}_{q^\prime} \mathbf{c}_{q^{\prime\prime}} 
	+ \frac{52}{289} \left( \mathbf{c}_{q^\prime q^{\prime\prime}} 
	- \mathbf{c}_{q^\prime} \mathbf{c}_{q^{\prime\prime}} \right) 
	\frac{4P}{\beta} \left( \mathcal{T}_{00} + \frac{\beta}{4} 
	\sum_{q^{\prime\prime\prime}}^{\lbrace B,Q,S\rbrace} 
	\mathcal{T}_{0q^{\prime\prime\prime}} \mathbf{c}_{q^{\prime\prime\prime}} \right) 
	\right] \notag \\
	& \quad - \frac{4}{\beta\sigma^2_{\text{tot}}} 
	\sum_{q^{\prime\prime}}^{\lbrace B,Q,S\rbrace} 
	\left\{ \frac{\partial}{\partial \alpha_{q^{\prime\prime}}} 
	\left[ \sum_{q^{\prime\prime\prime}}^{\lbrace B,Q,S\rbrace} 
	\left(\kappa^{-1}\right)_{q^{\prime\prime\prime}q} \left( 
	\frac{196}{289} \mathbf{c}_{q^\prime q^{\prime\prime\prime}} 
	- \frac{1109}{2312} \mathbf{c}_{q^\prime} \mathbf{c}_{q^{\prime\prime\prime}} \right) 
	\right] \right. \notag \\
	& \left. \quad + \frac{162}{289} \mathbf{c}_{q^\prime} 
	\frac{\partial}{\partial \alpha_{q^{\prime\prime}}} \left[ 
	\sum_{q^{\prime\prime\prime}}^{\lbrace B,Q,S\rbrace} 
	\left(\kappa^{-1}\right)_{q^{\prime\prime\prime}q} 
	\left( \mathbf{c}_{q^\prime q^{\prime\prime\prime}} 
	- \mathbf{c}_{q^\prime} \mathbf{c}_{q^{\prime\prime\prime}} \right) \right] \right\} 
	\left( \mathcal{T}_{q^{\prime\prime}0} + \frac{\beta}{4} 
	\sum_{q^{\prime\prime\prime}}^{\lbrace B,Q,S\rbrace} 
	\mathcal{T}_{q^{\prime\prime}q^{\prime\prime\prime}} 
	\mathbf{c}_{q^{\prime\prime\prime}} \right)
	\overset{\text{single}}{\rightarrow} \tau_q\;.
\end{align}
\begin{align}
	\lambda_{VV}^{(q^{\prime},q)} &= \frac{9}{68 \beta \sigma^2_{\text{tot}} P} 
	\sum_{q^{\prime\prime}}^{\left\{ B,Q,S\right\} } 
	\left( \kappa^{-1}\right)_{q^{\prime\prime} q} \Bigg( \frac{2768}{765}
	\mathbf{c}_{q^{\prime\prime}q^\prime} 
	- \frac{20879}{12240}\mathbf{c}_{q^{\prime\prime}} \mathbf{c}_{q^\prime} \Bigg)
	 \overset{\text{single}}{\rightarrow} \frac{3}{5} \tau_q\;, \\
	\ell_{V\pi }^{(q^{\prime })} &= \frac{9}{80 \sigma_{\text{tot}} P} \mathbf{c}_{q^\prime} 
	\overset{\text{single}}{\rightarrow} \frac{\beta}{20} \tau_q\;, \\
	\tau_{V\pi}^{(q^{\prime })} &= \ell_{V\pi }^{(q^{\prime })} 
	\overset{\text{single}}{\rightarrow} \frac{\beta}{20} \tau_q\;, \\
	\lambda_{V\pi}^{(q^{\prime },q)} &= \frac{28}{85}\frac{1}{\sigma_{\text{tot}} P} 
	\left( \mathbf{c}_{qq^\prime} - \frac{295}{448} \mathbf{c}_{q} 
	\mathbf{c}_{q^\prime}  \right) \overset{\text{single}}{\rightarrow} 
	\frac{\beta}{20} \tau_q\; .
\end{align}
Note that the coefficient $\tau_{n\pi}$ from Denicol (2012) \cite{Denicol:2012cn} 
was defined as $\tau_{n\pi} \overset{\text{single}}{=} \frac{1}{\epsilon + P} 
\tau_{V\pi }^{(q^{\prime })}$, which then yields $\tau_{n\pi} = \frac{\beta}{80} \tau_q$. 
Therefore, in the single-component limit we retrieve the same coefficients as in 
Ref.\ \cite{Denicol:2012cn}.

\section{Inverting the collision matrix}
\label{sec:InverseCollisionMatrix}

For the calculation of the transport coefficients, the inverse of the
collision matrix $\mathcal{A}^{(\ell)} \equiv \left(\mathcal{A}^{(\ell)}_{ij,rn}\right)$ from Eq.\
\eqref{Coll_int_mixture} must be calculated. In the tradition of Ref.\ \cite{Denicol:2012cn}, 
we provide a detailed discussion of the derivation of the linearized collision term 
(see Appendix \ref{sec:perform_coll_int}) and its inverse. In this section, we show that the 
collision matrix is singular in the cases $\ell = 0$ and $\ell = 1$ due to the conservation of 
energy-momentum and charge. While in the single-component system the construction of 
the inverse was immediately clear \cite{Denicol:2012cn}, in the case of a multicomponent 
system such a construction is not obvious.

The conservation equations \eqref{eq:charge_conservation_0} for the various charges 
imply that certain moments of the Boltzmann equation vanish,
\begin{align}
	\sum_{i\,=\,1}^{N_{\text{spec}}} q_i C_{i,0} 
	&= \sum_{i\,=\,1}^{N_{\text{spec}}} 
	q_i \int \dd K_i  k^\nu_i \partial_\nu f_{i,\mathbf{k}} \equiv 0\;. 
\end{align}
Similarly, projections of the conservation law \eqref{energy_momentum_cons_0}
for energy and momentum give
\begin{align}
	\sum_{i\,=\,1}^{N_{\text{spec}}} C_{i,1} &= \sum_{i\,=\,1}^{N_{\text{spec}}} 
	\int \dd K_i E_{i,\mathbf{k}} k^\nu_i \partial_\nu f_{i,\mathbf{k}} \equiv 0, \\
	\sum_{i\,=\,1}^{N_{\text{spec}}} C^{\langle\mu\rangle}_{i,0} 
	&= \sum_{i\,=\,1}^{N_{\text{spec}}} \int \dd K_i k^{\langle\mu\rangle}_i 
	k^\nu_i \partial_\nu f_{i,\mathbf{k}} \equiv 0 \;.
\end{align}
These relations imply that a subset of row (or column) vectors of the collision matrices 
$\mathcal{A}^{(\ell)}$ are linearly dependent. Since the irreducible moments 
$\rho^{\mu_1 \cdots \mu_\ell}_{i,r}$ are in principle independent of each other, 
Eq.\ \eqref{Coll_int_mixture} implies that the following linear combinations of the elements 
of the collision matrix must vanish:
\begin{align}
	0 &= \sum_{i\,=\,1}^{N_{\text{spec}}} q_i \mathcal{A}^{(0)}_{is,1n}\;, 
	\label{constraint_q0}\\
	0 &= \sum_{i\,=\,1}^{N_{\text{spec}}} \mathcal{A}^{(0)}_{is,2n}\; , \label{constraint_0}\\
	0 &= \sum_{i\,=\,1}^{N_{\text{spec}}} \mathcal{A}^{(1)}_{is,1n} \;.
\end{align}
This means that the conservation laws render the collision matrix for the scalar moments, 
$\mathcal{A}^{(0)}$, and the vector moments, $\mathcal{A}^{(1)}$, singular. We note that
the tensor moments ($\ell = 2$) are not affected, and thus $\mathcal{A}^{(2)}$ is in 
principle regular. For a meaningful fluid-dynamical theory, an equivalent description of the 
above discussed collision matrices has to be found such that they are rendered invertible, 
and at the same time their microscopic information is not altered. The linear dependence 
further implies that $N_q + 4$ equations need to be removed from the set 
\eqref{D_rho_ir}, \eqref{D_rho_mu_ir}, and \eqref{D_rho_munu_ir} of equations of motion, 
i.e.\ $N_q+1$ scalar moments $\rho_{i,r}$ and one vector moment $\rho^\mu_{i,r}$ 
(three equations). In the case of the vector moments, the choice of the frame provides a 
relationship between the vector moments, 
and it allows us to eliminate one of them from the equations of motion entirely.

For the Landau frame, from Eq.\ \eqref{Landau_flow} we have 
\begin{equation} \label{special_relation}
\rho^\mu_{N_{\text{spec}},1} = -\sum\limits_{i\,=\,1}^{N_{\text{spec}} - 1} \rho^\mu_{i,1}\;,
\end{equation} 
while in the Eckart frame we could impose via Eq.\ \eqref{eq:EnergyMomAndDiff} 
$\rho^{\mu}_{N_{\text{spec}},1} = - h_q \sum\limits_{i\,=\,1}^{N_{\text{spec}}} q_i 
\rho^{\mu}_{i,0} - \sum\limits_{i\,=\,1}^{N_{\text{spec}}-1} \rho^{\mu}_{i,1}$. As before, we 
proceed in the Landau frame. With the help of Eq.\ (\ref{special_relation}), we can write:
\begin{align}
	\sum_{s\,=\,1}^{N_{\text{spec}}} \sum_{n\,=\,0}^{N_1} \mathcal{A}^{(1)}_{is,rn} 
	\rho^\mu_{s,n} &= \sum_{n\,=\,0,\neq 1}^{N_1} \sum_{s\,=\,1}^{N_{\text{spec}}} 
	\mathcal{A}^{(1)}_{is,r0} \rho^\mu_{s,0} + \sum_{s\,=\,1}^{N_{\text{spec}} - 1} 
	\mathcal{A}^{(1)}_{is,r1} \rho^\mu_{s,1} + \mathcal{A}^{(1)}_{iN_{\text{spec}},r1} 
	\rho^\mu_{N_{\text{spec}},1} \label{eq:pseudo_derivation} \nonumber \\
	&= \sum_{n\,=\,0,\neq 1}^{N_1} \sum_{s\,=\,1}^{N_{\text{spec}}} 
	\mathcal{A}^{(1)}_{is,r0} 
	\rho^\mu_{s,0} + \sum_{s\,=\,1}^{N_{\text{spec}} - 1} \underbrace{ 
	\left( \mathcal{A}^{(1)}_{is,r1} - \mathcal{A}^{(1)}_{iN_{\text{spec}},r1} \right) }_{(\star)} 
	\rho^\mu_{s,1} \overset{!}{=} \sum_{n\,=\,0}^{N_1} \sum_{s\,=\,1}^{N_{\text{spec}}} 
	\tilde{\mathcal{A}}^{(1)}_{is,rn} \rho^\mu_{s,n}\; .
\end{align}
First, we note that the term marked with $(\star)$ vanishes for $s = N_{\text{spec}}$. 
Therefore, we can extend the last sum to run up to $s = N_{\text{spec}}$. 
In the last step of the above equation, we introduced the matrix $\tilde{\mathcal{A}}^{(1)}$,
which is a reduced version of the matrix $\mathcal{A}^{(1)}$ with 
the row corresponding to 
$r = 1$ and $i = N_{\text{spec}}$ and the column
corresponding to $n = 1$ and $s = N_{\text{spec}}$, respectively, 
removed, i.e.\ its elements read:
\begin{align}
\tilde{\mathcal{A}}^{(1)}_{is,rn} = \mathcal{A}^{(1)}_{is,rn} ~~ \text{for} ~~ n \neq 1 \;, 
\quad \text{and} \quad \tilde{\mathcal{A}}^{(1)}_{is,r1} = \mathcal{A}^{(1)}_{is,r1} 
- \mathcal{A}^{(1)}_{iN_{\text{spec}},r1} \;.
\end{align}
Then, $\tilde{\mathcal{A}}^{(1)}$ is an $(N_{\text{spec}} \cdot N_1 -1)$-dimensional, 
rectangular, regular matrix. 

An explicit example for the construction of the corrected matrix is given in Appendix 
\ref{sec:ultrarelativistic_case} in the case of an ultrarelativistic gas mixture. 
We further note that we are free to choose which irreducible moment $\rho^\mu_{i,1}$ we 
remove from the set of equations of motion, and thus we could have chosen any 
line and associated column (corresponding to that particular moment) to be removed 
(e.g.\ $\rho^\mu_{\lambda,1}$ instead of $\rho^\mu_{N_{\text{spec}},1}$).

Once the reduced collision matrix $\tilde{\mathcal{A}}^{(1)}$ is obtained it can be inverted. 
This yields the reduced inverse $\tilde{\mathcal{\tau}}^{(1)}$ of dimension 
$N_{\text{spec}} \cdot N_1 -1$. In order to make the inverse $\mathcal{\tau}^{(1)}$ 
equivalent to the one  of dimension 
$N_{\text{spec}} \cdot N_1$ introduced in Eq.\ \eqref{MAIN_relax_time}, 
and in order to keep a simple notation regarding the 
summations over the indices $r$ and $s$ in all equations following that definition 
(e.g.\ Eqs.\ \eqref{eq:OOM1} ff.), we add zero-element row(s) and column(s), which 
correspond to the ones originally removed from the matrix $\tilde{\mathcal{A}}^{(1)}$.
For instance, in the case of the vector moments this means that we add a zero row for 
$r = 1$ and $i = N_{\text{spec}}$, and a zero column for $n = 1$ and 
$s = N_{\text{spec}}$. 
Due to the zero-element row and column, the removed irreducible moment 
$\rho^{\mu}_{N_{\text{spec}},1}$, even though it formally still appears in the equations 
following Eq.\ \eqref{eq:OOM1}, effectively does not contribute anymore.

The procedure is analogous for the removal of the $N_q + 1$ scalar moments. The 
energy- and charge-conservation laws provide $N_q+1$ relations for the linear 
dependence of the row vectors of matrix $\mathcal{A}^{(0)}$. These are given by 
Eqs.\ (\ref{constraint_q0}) and (\ref{constraint_0}).
In order to remove the corresponding moments, we impose the Landau 
matching conditions,
\begin{align}
	\sum_{i\,=\,1}^{N_{\text{spec}}} \rho_{i,2} = 0\;, \quad \text{and} \quad 
	\sum_{i\,=\,1}^{N_{\text{spec}}} q_i \rho_{i,1} = 0\; .
\end{align}
The corrected matrix $\tilde{\mathcal{A}}^{(0)}$ is then an 
$(N_{\text{spec}} \cdot N_0 - N_q -1)$-dimensional, rectangular, regular matrix and its 
elements can be written as:
\begin{align}
	\tilde{\mathcal{A}}^{(0)}_{is,rn} = \mathcal{A}^{(0)}_{is,rn} ~~ 
	\text{for} ~~ n \neq 1,2 ~, 
	\quad \tilde{\mathcal{A}}^{(0)}_{is,r1} = \mathcal{A}^{(0)}_{is,r1} 
	- \frac{q_s}{q_\lambda} 
	\mathcal{A}^{(0)}_{i\lambda,r1}, \quad \text{and} 
	\quad \tilde{\mathcal{A}}^{(0)}_{is,r2} = \mathcal{A}^{(0)}_{is,r2} 
	- \mathcal{A}^{(0)}_{i\lambda,r2} \;, \label{eq:ScalarMatrixEntries}
\end{align}
where again $i \neq \lambda$ for $r = 1,2$, and $s \neq \lambda$ for $n = 1,2$. 
It should be noted that the $N_q$ charge-conservation laws and the $N_q$ matching 
conditions associated to the net-charge densities only allow for the removal of moments 
corresponding to species $\lambda$ with non-vanishing charge of type $q$, 
$q_\lambda \neq 0$. This can be understood by noting that the above relations 
\eqref{eq:ScalarMatrixEntries} are not well-defined when $q_\lambda = 0$. 
In order to simplify the notation in this work, we construct 
the inverse $\tau^{(0)}$ as an $(N_{\text{spec}} \cdot N_0)$-dimensional, rectangular 
matrix, which is the inverse of $\tilde{\mathcal{A}}^{\left(0\right)}$ and contains zero
elements for the rows and columns which were removed from
$\mathcal{A}^{\left(0\right)}$.


\end{document}